\documentclass[12pt]{article}
\usepackage{cancel}
\usepackage{autobreak}
\usepackage{slashed}
\usepackage{babel}
\usepackage{lipsum}
\usepackage{feynmp}
\usepackage{comment}
\usepackage{subcaption}
\usepackage{physics}
\usepackage{fancyhdr}
\usepackage{amsmath}
\usepackage{amssymb}
\usepackage{slashed}
\usepackage{graphicx}
\usepackage{tikz}
\usepackage[compat=1.1.0]{tikz-feynman}
\usepackage[compat=1.1.0]{tikz-feynhand}
\usepackage{braket}
\newcommand{\ii}{i}
\newcommand{\md}{d}
\newcommand{\im}{\mathrm{Im}}
\newcommand{\me}{\mathrm{e}}
\newcommand{\q}{\quad}
\newcommand{\OPE}{\mathrm{OPE}}

\usepackage[a4paper,text={16.8cm,22.4cm}]{geometry}
\usepackage{amsmath,amsfonts,slashed,amssymb,tikz,bm,psfrag,graphicx,color,dsfont}
\usepackage{multicol}
\usepackage{float}
\usepackage[figuresright]{rotating}
\usepackage{epstopdf}
\usepackage[colorlinks,linkcolor=blue,citecolor=blue,linktoc=page]{hyperref} 
\RequirePackage[sort&compress,square,comma,numbers]{natbib}
\allowdisplaybreaks
\begin{document}
\begin{titlepage}
\fancyhf{}

\fancyfoot[L]{\rule{0.4\textwidth}{0.35pt}\\ \small  Email: hanxy@ihep.ac.cn, lulongshun@ihep.ac.cn, lucd@ihep.ac.cn, shenylmeteor@ouc.edu.cn, shibx@mail.nankai.edu.cn}
\renewcommand{\headrulewidth}{0pt} 
\pagestyle{fancy} 
\thispagestyle{fancy}

\begin{flushright}
\normalsize
\today
\end{flushright}
\vspace{0.1cm}
\begin{center}
\Large\bf
\end{center}
\vspace{0.5cm}
\begin{center}
\Large {\bf Next-to-leading order QCD corrections to the form factors of $B$ to scalar meson decays
}
\end{center}
\vspace{0.5cm}
\begin{center}
{\bf Xue-Ying Han\textsuperscript{\textit{a,b}}, Long-Shun Lu\textsuperscript{\textit{a,b}}, Cai-Dian L\"u\textsuperscript{\textit{a,b}}, Yue-Long Shen\textsuperscript{\textit{c}}, \\Bo-Xuan Shi\textsuperscript{\textit{d}}\\}
\vspace{0.7cm}
\textit{ ${}^a$ Institute of High Energy Physics, CAS, Beijing 100049, China  
\\ ${}^b$ School of Physics, University of Chinese Academy of Sciences, Beijing 100049, China
\\ ${}^c$ College of Physics and Photoelectric Engineering, Ocean University of China, Qingdao 266100,
China
\\ ${}^d$ School of Physics, Nankai University, 300071 Tianjin, China\\}
\end{center}

\vfil
\begin{abstract}
We calculate the next-to-leading order QCD corrections to $B$ to scalar meson form factors from QCD light-cone sum rules with $B$ meson light-cone distribution amplitudes. We demonstrate that the $B$ meson-to-vacuum correlation functions can be factorized into the convolution of short-distance coefficients and light-cone distribution amplitudes at the one-loop level and find that only $\phi_B^+(\omega,\mu)$ contributes to the form factors. We then employ the $z$-parameterization combined with constraints from strong coupling constants to reconstruct the $q^2$ dependence of the form factors in the whole kinematic allowed regions. Due to the large cancellations between the hard functions and the jet functions,  the next-to-leading order results show a modest increase of approximately 5\% compared to the leading order results. Based on the results of form factors, we predict the branching ratios of semi-leptonic $B\to S\ell\bar{\nu}_\ell$ and $B\to S\nu_\ell\bar{\nu}_\ell$ processes, as well as several angular observables, such as forward-backward asymmetries, “flat terms” and lepton polarization asymmetries. We compare these results with calculations from other methods. Experimental verification of these results is required in future experiments.
\end{abstract}
\end{titlepage}
\tableofcontents

\section{Introduction}

The semileptonic decays of $B$ meson play an indispensable role in the determinations of the Cabibbo-Kobayashi-Maskawa (CKM) matrix elements and also provide an ideal place to test the universality of the lepton couplings. The semileptonic $B$ decays with the hadronic final state being a pseudo-scalar or a vector meson have been extensively studied, meanwhile, there are still very few investigations on the $B$ meson decaying into an orbital excited state or other excited states, which can serve as a good supplement other than the $B \to P, V$ ($P$ and $V$ stand for the pseudo-scalar and the vector meson respectively) transitions. The scalar meson is among the most important excited light meson states, thus it is of significant importance to study the  $B \to S$ ($S$ denotes a scalar meson) transition processes. 

Different from the ground states, the internal structures of the scalar meson have not been well understood. It has been suggested that the scalar mesons below $1.7\,\mathrm{GeV}$ can be described with two nonets \cite{Close:2000yk, Close:2002zu, Close:2005vf},  which correspond to scalar mesons below $1\,\mathrm{GeV}$ and above $1\,\mathrm{GeV}$, respectively. There is currently no widespread consensus on the understanding of scalar mesons below $1\,\mathrm{GeV}$, whose possible structures include $q\bar{q}$ states \cite{Cheng:2005nb, Cheng:2019tgh, Cheng:2023knr}, meson-meson molecule states \cite{Weinstein:1982gc}, tetraquark states \cite{PhysRevD.15.267, PhysRevD.15.281}, and the superposition of these states \cite{Amsler:1995tu, Amsler:1995td, Amsler:2002ey}. 
On the other hand, the nonet above $1\,\mathrm{GeV}$ are commonly regarded as $q\bar{q}$ states, and the existing controversy is that they might be either the low-lying p-wave state of $q\bar q$  or the first radial excited states with respect to the low-lying p-wave state,  and the former viewpoint and the latter one are referred to  Scenario 2 and Scenario 1 respectively \cite{PhysRevD.73.014017, Lu:2006fr}. Most of the studies on the structure of the scalar mesons prefer to support  Scenario 2 \cite{Du:2004ki, Lu:2006fr, Chen:2021oul}, thus we set   Scenario 2 to be our default choice in the present paper. For the flavor-neutral particles $f_0(1370)$, $f_0(1500)$ and $f_0(1710)$, recent studies suggest that they are mixing of scalar glueball and scalar quarkonia \cite{Cheng:2006hu, Yuan:2011xz, Close:2000yk, Close:2002zu, Close:2005vf, Fariborz:2015era}, and the mixing matrix is required to be further determined. In this work, we follow the assumption  in \cite{Cheng:2006hu, Lu:2006fr} where the main component  of $f_0(1500)$ is $s\bar{s}$. 

The essential task in studying the semileptonic decays of  $B$ mesons into scalar mesons is to compute  $B\to S$ transition form factors. It is generally agreed that the heavy-to-light transition form factors are nonperturbative in nature,  one has to take advantage of the nonperturbative QCD methods to evaluate them. 
Several approaches have been employed in the existing literature to investigate the $B\to S$ transition form factors, including the light-front approach \cite{Chen:2007na, Cheung:1995ub, Zhang:1994hg, Choi:1999nu}, QCD sum rules (QCDSR) \cite{PhysRevD.73.034027, Aliev:2007rq, Shifman:1978by, NOVIKOV1981301}, light-cone sum rules (LCSR) \cite{Wang:2008da, PhysRevD.83.025024, Han:2013zg, Wang:2014vra, Huang:2022xny, Khosravi:2022fzo, Braun:1988qv, CHERNYAK1990137} etc. 
Since the correlation function employed in the sum rules is dominated by the light-cone region in the coordinate space,  the heavy quark decay is not suitable to be treated using short-distance expansion in QCDSR. The LCSR approach has been developed, because of its unique advantage of studying the heavy-to-light form factors at large recoil region. An alternative approach to study the heavy-to-light form factors is the perturbative QCD approach \cite{Li:2008tk, Keum:2000ph, Keum:2000wi, Lu:2000em} where the transition form factors are calculable in the perturbation theory since the endpoint singularity is removed.

In this study, we will employ the LCSR method with LCDA of  $B$ mesons \cite{Khodjamirian:2005ea, DeFazio:2005dx, Khodjamirian:2006st, DeFazio:2007hw} to compute the $B\to S$ transition form factors, and also study the semileptonic $B \to S\ell\nu$ and $B \to S\nu\bar \nu$ decays. 
Both higher-order perturbative corrections to the correlation functions and the contributions stemming from subleading-power effects in the context of heavy-to-light  $B$ meson decay form factors \cite{Wang:2015vgv, Shen:2016hyv, Lu:2018cfc, Gubernari:2018wyi, Gao:2019lta, Shen:2021yhe}, heavy-to-heavy  $B$ meson decay form factors \cite{Gubernari:2018wyi, Wang:2017jow, Gao:2021sav, Cui:2023bzr}, and semileptonic heavy-baryon decay form factors \cite{Wang:2009hra, Feldmann:2011xf, Wang:2015ndk}, have been systematically computed utilizing this approach. The main idea of the calculation of  $B\to S$ transition form factors is similar to \cite{Wang:2015vgv, Shen:2016hyv, Lu:2018cfc,  Gao:2019lta, Shen:2021yhe}, and we will take advantage of method of regions to calculate the QCD corrections to the correlation function to extract the hard functions and jet functions directly. 
Compared with the calculation of $B \to P, V$ transition form factors, the main difference in our calculation is the scalar density (rather than the vector current) should be employed to interpolate the scalar meson state,  which is power-suppressed relative to the vector current in the aspect of soft-collinear effective theory (SCET), as a result, the LCDA $\phi^+_B(\omega,\mu)$  rather than  LCDA $\phi^-_B(\omega,\mu)$ appears at the sum rules of $B\to S$ transition form factors. In addition, one also has to consider the renormalization of the scalar density in the calculation. 

The structure of this paper is as follows:
In Section 2, we provide the definition of form factors and utilize LCSR at the tree level to calculate the form factors of $B\rightarrow S$ transitions.
In Section 3, we establish the factorization formulae and compute the hard scattering kernels at NLO, followed by a detailed analysis and discussion of the results.
In Section 4, we present the NLO calculation results of the form factors using LCSR.
In Section 5, we perform a numerical analysis to obtain the results of the form factors in the large recoil region. We extrapolate these results to the entire kinematic allowed region using the BCL parameterization and constraints from the strong coupling constant. Furthermore, we provide some results for observables that can be tested in future experiments.

\section{The LCSR of form factors at tree level}

\subsection{Definition of \texorpdfstring{$B \to S$}{B -> S}  form factors}

According to the standard Lorentz decomposition of the bilinear quark currents we define the heavy-to-light form factors as following \cite{Beneke:2000wa, Wang:2008da}
\begin{align}
    \bra{S(p)} \bar{q}' i\sigma_{\mu\nu} \gamma_5 q^\nu b(0) \ket{B(p_B)} &= - i\left[2 q^2 p_\mu - (m^2_B - m^2_S-q^2) q_\mu \right] \dfrac{f^T_{BS} (q^2)}{m_B+m_S},\\
    \bra{S(p)} \bar{q}' \gamma_\mu \gamma_5 b(0) \ket{B(p_B)}& = -i \left[p_\mu f_{BS}^+(q^2)+q_\mu f_{BS}^-(q^2)\right],\notag\\
    & = -i \left[(p_B+p-\dfrac{m_B^2-m_S^2}{q^2}q )_\mu f_{BS}^A(q^2)+\dfrac{m_B^2-m_S^2}{q^2} q_\mu f_{BS}^P\right],\notag\\
    & = -i\left[(2p_\mu+q_\mu)f_+(q^2)+q_\mu f_-(q^2)\right],
\end{align}
where $m_S$ and $p$ denote the mass and momentum of light scalar meson, and $q$ stands for the transfer momentum of weak transition current. We presented three different parameterizations for the axial-vector current matrix element, where the superscripts $A$ and $P$ in the second parameterization indicate the spin-parity of the virtual $W$ bosons, and in sum rules, they also correspond to the spin-parity of the intermediate states which contribute. The other two parameterizations are chosen for the purpose of convenience in calculation and comparison with other known results. The interrelations between these three parameterizations of the axial-vector current matrix element are as follows
\begin{align}
    f_{BS}^A(q^2)&=\dfrac{1}{2}f_{BS}^+(q^2)=f_+(q^2),\\
    f_{BS}^P(q^2)&=\dfrac{1}{2}\left(1-\dfrac{q^2}{m_B^2-m_S^2}\right)f_{BS}^+(q^2)+\dfrac{q^2}{m_B^2-m_S^2}f_{BS}^-(q^2)\notag\\
    &=f_+(q^2)+\dfrac{q^2}{m_B^2-m_S^2}f_-(q^2).
\end{align}

\subsection{The LCSR for \texorpdfstring{$B\to S$}{B -> S} form factors at tree level}

Following the process given in \cite{Wang:2015vgv,Lu:2018cfc}, we construct the LCSR of form factors with the $B$-to-vacuum correlation functions at  $B$ meson static frame
\begin{align}
	\Pi_\mu^A(n\cdot p,\bar{n}\cdot p) =\,& \int \md^4 x\ \me^{\ii p \cdot x} \bra{0} T\{\bar{q}(x) q'(x), \bar{q}'(0) \gamma_\mu \gamma_5 b(0) \} \ket{B(p_B)} \notag\\
	=\,& \Pi^A(n\cdot p,\bar{n}\cdot p) n_\mu +\widetilde{\Pi}^A(n\cdot p,\bar{n}\cdot p) \bar{n}_\mu, \\
        \Pi_\mu^T(n\cdot p,\bar{n}\cdot p) =\,& \int \md^4 x \ \me^{\ii p \cdot x} \bra{0} T\{\bar{q}(x) q'(x), \bar{q}'(0) \ii\sigma_{\mu\nu}q^\nu\gamma_5 b(0) \} \ket{B(p_B)} \notag\\
	=\,& \Pi^T(n\cdot p,\bar{n}\cdot p)[\bar{n}\cdot q\,n_{\mu}-n\cdot q\,\bar{n}_\mu],  
     \label{eq:3}
\end{align}
where we have introduced the light-cone coordinate
\begin{align}
   &n\cdot\bar{n}=2,\quad n\cdot n=0,\quad \Bar{n}\cdot \Bar{n}=0,\notag\\
   &v^\mu=\dfrac{n^\mu+\Bar{n}^\mu}{2},\notag\\
   &\gamma_{\perp}^\mu=\gamma^\mu-\dfrac{\slashed{n}}{2}\Bar{n}^\mu-\dfrac{\slashed{\Bar{n}}}{2}n^\mu.
\end{align}

In the region of $\bar{n}\cdot p<0$ we apply the light-cone operator product expansion (OPE) to calculate correlation functions, and we will demonstrate the factorization of the correlation function at one-loop level. The hard scattering part in the factorization function does not depend on the external state, thus we can replace $\ket{B(p_B)}$ with $\ket{b(p_B-k)\bar{q}(k)}$ in our calculation. The partonic level correlation function is then written by
\begin{equation}
  \begin{aligned}
      \Pi_{\mu,b\bar{q}}^{(0)}(n\cdot p, \bar{n}\cdot p)=\int \md\omega'\ T_{\mu,\alpha\beta}^{(0)}(n\cdot p,\bar{n}\cdot p,\omega')\Phi_{b,\bar{q}}^{(0)\alpha\beta}(\omega'),
  \end{aligned}  
\end{equation}
where $T_{\alpha\beta}^{(0)}$ is the hard scattering kernel at tree-level  
\begin{align}
    T^{(0),A}_{\mu,\alpha\beta}=\,&\dfrac{i}{2}\dfrac{1}{\bar{n}\cdot p-\omega'+i0}\left[\slashed{\bar{n}}\gamma_\mu\gamma_5\right]_{\alpha\beta},\\
    T^{(0),T}_{\mu,\alpha\beta}=\,&\dfrac{i}{2}\dfrac{1}{\bar{n}\cdot p-\omega'+i0}\left[\slashed{\bar{n}}i\sigma_{\mu\nu}q^\nu\gamma_5\right]_{\alpha\beta}.
\end{align}
The partonic distribution amplitude (DA) is given by
\begin{equation}
    \Phi^{\alpha\beta}_{b\bar{q}}(\omega')=\int\dfrac{\md \tau}{2\pi}\ \mathrm{e}^{i\omega'\tau}\bra{0}\bar{q}_\beta(\tau\bar{n})[\tau\bar{n},0]b_\alpha(0)\ket{b(p_B-k)\bar{d}(k)}, 
\end{equation}
and at the tree level, it is obtained straightforwardly
\begin{equation}
    \Phi^{(0)\alpha\beta}_{b\bar{q}}(\omega')=\delta(\bar{n}\cdot k-\omega')\bar{d}_\beta(k) b_\alpha(p_B-k).
\end{equation}
To arrive at the final expression of the partonic correlation function, we employ the light-cone projector of  $B$ meson in momentum space in dimension-$D$ \cite{DeFazio:2007hw, Beneke:2000wa}
\begin{equation}
    M_{\beta\alpha}=-\dfrac{\ii \Tilde{f}_B (\mu)m_B}{4}\left\{\dfrac{1+\slashed{v}}{2}\left[\phi_B^+(\omega)\slashed{n}+\phi_B^-(\omega)\slashed{\bar{n}}-\dfrac{2\omega}{D-2}\phi_B^-(\omega)\gamma_\perp^\rho\dfrac{\partial}{\partial k_\perp^{\rho}}\right]\gamma_5\right\}_{\alpha\beta}
\end{equation}
with replacement $\phi^{\pm}_B(\omega')\rightarrow\phi^{\pm}_{b\bar{q}}(\omega')$.
Here $\tilde{f}_B$ is the  $B$ meson decay constant under the static limit and it can be related to the QCD decay constant by
\begin{equation}
f_B=\tilde{f}_B(\mu)\left[1+\frac{\alpha_s C_F}{4 \pi}\left(-3 \ln \frac{\mu}{m_b}-2\right)\right].\label{eq: fBdef}
\end{equation}
The partonic correlation function at the tree level then reads
\begin{align}
    \Pi_{\mu,b\bar{q}}^{A,(0)}&=-\dfrac{1}{2}\tilde{f}_B m_B \dfrac{\phi^{+}_{b\bar{q}}}{\omega-\bar{n}\cdot p-i0}n_\mu,\\
    \Pi_{\mu,b\bar{q}}^{T,(0)}&=\dfrac{1}{4}\tilde{f}_B m_B \dfrac{\phi^{+}_{b\bar{q}}}{\omega-\bar{n}\cdot p-i0}[\bar{n}\cdot q\,n_{\mu}-n\cdot q\,\bar{n}_\mu].
\end{align}
Taking the physical $B$ state into account, the factorization formulae of the correlation function at tree level yields
\begin{align}
    \Pi_\mu^{A,(0)}
    &=-\dfrac{1}{2}\Tilde{f}_B m_B\int_0^\infty \md\omega\dfrac{\phi_B^+(\omega)}{\omega-\bar{n}\cdot p-\ii 0}n_\mu,\\
    \Pi_\mu^{T,(0)}
    &=\dfrac{1}{4}\Tilde{f}_B m_B \int_0^\infty \md\omega\dfrac{\phi_B^+(\omega)}{\omega-\bar{n}\cdot p-\ii 0}[\bar{n}\cdot q\,n_{\mu}-n\cdot q\,\bar{n}_\mu].
\end{align}
We should note that the DA $\phi_B^+$ rather than $\phi_B^-$ appears in the factorization formulae, which is different from the conditions in the $B\to \pi$ form factors.
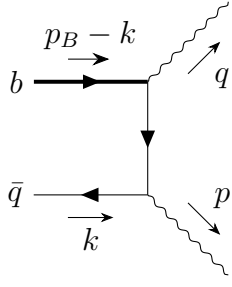
\begin{figure}[t]
    \centering
    \begin{tikzpicture}
        \begin{feynman}
            \vertex (a) {\(b\)};
            \vertex [right=1.73cm of a] (b);
            \vertex [below=of b] (c);
            \vertex [left=of c] (d) {\(\bar{q}\)};
            \vertex [above right=of b] (e);
            \vertex [below right=of c] (f) ;
            \vertex [above right=of b] (e) ;
            \diagram* {
                (a) -- [fermion, line width=1.5pt,momentum={[arrow shorten=0.3]\(p_B-k\)}] (b),
                (b) -- [fermion] (c),
                (c) -- [fermion,reversed momentum={[arrow shorten=0.3]\(k\)}] (d),
                (b) -- [photon,momentum'={[arrow shorten=0.3]\(q\)}] (e),
                (c) -- [photon,momentum={[arrow shorten=0.3]\(p\)}] (f),
            };
        \end{feynman}
    \end{tikzpicture}
    \caption{Perturbative part at tree level.}
    \label{fig:feynman1}
\end{figure}

On the hadronic side,  we insert the unitarity relation \cite{Colangelo:2000dp} between weak transition currents and scalar meson interpolation currents. Finally, the correlation functions are expressed as
\begin{align}
    \Pi_\mu^A(p_B,q) =\,& \dfrac{\bra{0} \bar{q} q' \ket{S(p)} \bra{S(p)} \bar{q}' \gamma_\mu \gamma_5 \ket{B(p_B)} }{m^2_S-p^2}+\dots \notag\\ 
    =\,&-\dfrac{i m_S \bar{f}_S}{2(m_S^2-p^2)} \left[ \bar{n}_\mu (n\cdot p\, f^+_{BS}(q^2) - (n\cdot p-m_B) f^-_{BS}(q^2) )+n_\mu m_B f^-_{BS}(q^2) \right]+\dots \label{polemodel1},\\
    \Pi_\mu^T(p_B,q) =\,& \dfrac{\bra{0} \bar{q} q' \ket{S(p)} \bra{S(p)} \bar{q}' i\sigma_{\mu\nu} q^\nu \gamma_5  \ket{B(p_B)} }{m^2_S-p^2}+\dots \notag\\ 
    =\,&\dfrac{i m_S \bar{f}_S\,n\cdot p\,m_B}{2(m_S^2-p^2)} [\bar{n}\cdot q\,n_{\mu}-n\cdot q\,\bar{n}_\mu]             \dfrac{f^T_{BS}(q^2)}{m_B+m_S} +\dots,\label{polemodel2}
\end{align}
where the ellipsis represents contributions from the excited states and the continuum. 

Here we have treated the scalar meson as a pole, resulting in the phase space integral
of the scalar meson being represented as a Dirac delta function which is a simplified model. In order to make our descriptions more faithful to reality, it is necessary to retain the width information of resonances. As a result, eq. (\ref{polemodel1}) and eq. (\ref{polemodel2}) will become integrals over the form factors. The current approach to handling this has been extensively discussed for other processes and is insightful for us \cite{Descotes-Genon:2023ukb, VonDetten:2021rax, Descotes-Genon:2019bud, Cheng:2017smj}. To adopt a similar methodology in our own processes, we will need to identify appropriate resonance models for describing form factors. This is a challenging task due to the intricate structures of scalar hadron spectra, which will be the focus of our ongoing research.

Applying Borel transformation $p^2\rightarrow M^2$ on both sides of partonic level and hadronic level correlation functions, and equalizing the higher state and continuum contribution with the partonic dispersion integral above the threshold,  we derive the tree level sum rules for $B\rightarrow S$ form factors

\begin{align}
        f^+_{BS}(q^2) & = \dfrac{\tilde{f}_B m_B}{m_S \bar{f}_S} \me^{m^2_{S} /(n\cdot p\, \omega_M)} \int_{0}^{\omega_s} \md \omega\, \me^{-\omega/\omega_M} \phi^+_B (\omega ) + \mathcal{O}(\alpha_s) ,  \notag\\
        f^-_{BS}(q^2) & = 0 + \mathcal{O}(\alpha_s), \notag\\
        f^T_{BS}(q^2) &=\dfrac{m_B+m_S}{2m_B}f^+_{BS}(q^2),
\end{align}
where $\omega_M=M^2/n\cdot p,\, \omega_s=s_0/n\cdot p$, and $M^2,\,s_0$ is the Borel parameter and the effective threshold in the QCD part for sum rule separately.

\section{Factorization of correlation function at \texorpdfstring{$\mathcal{O}(\alpha_s$)}{O(as)}}
The goal of this section is to derive factorization formulas for $\Pi_\mu$ in QCD at the one-loop level. We utilize the diagrammatic factorization method, expanding the correlator $\Pi_{\mu,b\bar{q}}$, the short-distance function $T$, and the partonic DA of the  $B$ meson $\Phi_{b\bar{q}}$ in perturbation theory. This is presented schematically as follows:
\begin{align}
\Pi_{\mu, b \bar{q}} & =\Pi_{\mu, b \bar{q}}^{(0)}+\Pi_{\mu, b \bar{q}}^{(1)}+\ldots=\Phi_{b \bar{q}} \otimes T_\mu \notag\\
& =\Phi_{b \bar{q}}^{(0)} \otimes T^{(0)}_\mu+\left[\Phi_{b \bar{q}}^{(0)} \otimes T^{(1)}_\mu+\Phi_{b \bar{q}}^{(1)} \otimes T^{(0)}_\mu\right]+\dots.\label{eq:fac}
\end{align}
The convolution in the variable $\omega'$ defined in Eq.(\ref{eq:fac}) is denoted by $\otimes$, with the superscripts indicating the order of $\alpha_s$. The determination of the hard-scattering kernel at $\mathcal{O}(\alpha_s)$ is based on the matching condition
\begin{equation}
    \begin{aligned}
    \Phi^{(0)}_{b\bar{q}}\otimes T^{(1)}_\mu=\Pi^{(1)}_{\mu,b\bar{q}}-\Phi^{(1)}_{b\bar{q}}\otimes T^{(0)}_\mu.\label{eq:19}
    \end{aligned}
\end{equation}
One important aspect of proving the factorization of $\Pi_{\mu,b\bar{q}}$, where the second term represents the subtraction of the infrared (soft) contributions, is to establish that the hard-scattering kernel $T$ can only arise from hard and/or hard-collinear regions at the leading power in $\Lambda/m_b$. This can be achieved by a complete cancellation of the soft contributions in $\Pi_{\mu,b\bar{q}}^{(1)}$ and $\Phi_{b\bar{q}}^{(1)}\otimes T^{(0)}$. Furthermore, it needs to be shown that there is no leading contribution to the correlation function from the collinear region (with momentum scaling $l_\mu \sim (1,\lambda^2,\lambda)$) at the leading power since the  $B$ meson LCDA can only capture the soft QCD dynamics of $\Pi_{\mu,b\bar{q}}$. Following the reference \cite{Wang:2015vgv} we will utilize the method of regions to analyze the master formula of $T^{(1)}$ in Eq.(\ref{eq:19}) diagram by diagram, in order to obtain the hard coefficient function ($C$) and the jet function ($J$) concurrently. This will allow us to establish the factorization formula

\begin{equation}
\Pi_{\mu, b \bar{q}}=\Phi_{b \bar{q}} \otimes T_\mu=C \cdot J_\mu \otimes \Phi_{b \bar{q}}.
\end{equation}
Our calculation strategy is outlined as follows: (i) We first identify the leading regions of the scalar integral for each diagram; (ii) Next, we simplify the Dirac algebra in the numerator for the identified leading region and evaluate the relevant integrals using the method of regions; (iii) We then compute the contributions from the hard and hard-collinear regions using the light-cone projector of the  $B$ meson in momentum space; (iv) We establish the equivalence between the soft subtraction term and the correlation function in the soft region; (v) Finally, we sum up the contributions from the hard and hard-collinear regions separately. 

\subsection{Weak vertex correction}
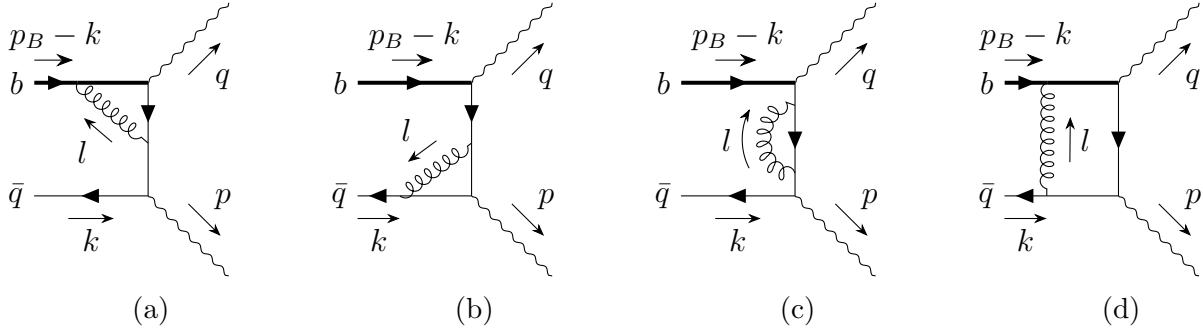
\begin{figure}[t]
    \centering
    %weak vertex correction
    \begin{subfigure}{0.24\linewidth}
    \begin{tikzpicture}
    \begin{feynman}
        \vertex (a) {\(b\)};
        \vertex [right=1.73cm of a] (b);
        \vertex [below=of b] (c);
        \vertex [left=of c] (d) {\(\bar{q}\)};
        \vertex [above right=of b] (e) ;
        \vertex [below right=of c] (f) ;
        \vertex [right=0.8cm of a] (g);
        \vertex [below=0.8cm of b] (h);
        \diagram* {
            (a) -- [fermion, line width=1.5pt,momentum={[arrow shorten=0]\(p_B-k\)}] (g) -- [line width=1.5pt] (b),
            (b) -- [fermion] (h) -- (c),
            (c) -- [fermion,reversed momentum={[arrow shorten=0.3]\(k\)}] (d),
            (b) -- [photon,momentum'={[arrow shorten=0.3]\(q\)}] (e),
            (c) -- [photon,momentum={[arrow shorten=0.3]\(p\)}] (f),
            (g) -- [gluon,reversed momentum'={[arrow shorten=0.3]\(l\)}] (h),
        };
    \end{feynman}
    \end{tikzpicture}
    \caption{}\label{fig2a}
    \end{subfigure}
    \hfill
    %scalar vertex correction
    \begin{subfigure}{0.24\linewidth}
    \begin{tikzpicture}
    \begin{feynman}
        \vertex (a) {\(b\)};
        \vertex [right=1.73cm of a] (b);
        \vertex [below=of b] (c);
        \vertex [left=of c] (d) {\(\bar{q}\)};
        \vertex [above right=of b] (e) ;
        \vertex [below right=of c] (f) ;
        \vertex [right=0.8cm of d] (g);
        \vertex [below=0.8cm of b] (h);
        \diagram* {
            (a) -- [fermion, line width=1.5pt,momentum={[arrow shorten=0.3]\(p_B-k\)}] (b),
            (b) -- [fermion] (h) -- (c),
            (c) -- (g) -- [fermion,reversed momentum={[arrow shorten=0]\(k\)}] (d),
            (b) -- [photon,momentum'={[arrow shorten=0.3]\(q\)}] (e),
            (c) -- [photon,momentum={[arrow shorten=0.3]\(p\)}] (f),
            (g) -- [gluon,reversed momentum={[arrow shorten=0.3]\(l\)}] (h),
        };
    \end{feynman}
    \end{tikzpicture}
    \caption{}\label{fig2b}
    \end{subfigure}
    \hfill
    %self-energy correction
    \begin{subfigure}{0.24\linewidth}
    \begin{tikzpicture}
    \begin{feynman}
        \vertex (a) {\(b\)};
        \vertex [right=1.73cm of a] (b);
        \vertex [below=of b] (c);
        \vertex [left=of c] (d) {\(\bar{q}\)};
        \vertex [above right=of b] (e);
        \vertex [below right=of c] (f);
        \vertex [below=0.3 of b] (g);
        \vertex [above=0.3 of c] (h);
        \diagram* {
            (a) -- [fermion, line width=1.5pt,momentum={[arrow shorten=0.3]\(p_B-k\)}] (b),
            (b) -- [fermion] (c),
            (c) -- [fermion,reversed momentum={[arrow shorten=0.3]\(k\)}] (d),
            (b) -- [photon,momentum'={[arrow shorten=0.3]\(q\)}] (e),
            (c) -- [photon,momentum={[arrow shorten=0.3]\(p\)}] (f),
            (h) -- [gluon,half left, momentum={[arrow shorten=0.3]\(l\)}] (g),
        };
    \end{feynman}
    \end{tikzpicture}
    \caption{}\label{fig2c}
    \end{subfigure}
    \hfill
    %box diagram
    \begin{subfigure}{0.24\linewidth}
    \begin{tikzpicture}
    \begin{feynman}
        \vertex (a) {\(b\)};
        \vertex [right=1.73cm of a] (b);
        \vertex [below=of b] (c);
        \vertex [left=of c] (d) {\(\bar{q}\)};
        \vertex [above right=of b] (e) ;
        \vertex [below right=of c] (f) ;
        \vertex [right=0.8cm of a] (g);
        \vertex [right=0.8cm of d] (h);
        \diagram* {
            (a) -- [fermion, line width=1.5pt,momentum={[arrow shorten=0]\(p_B-k\)}] (g) -- [line width=1.5pt] (b),
            (b) -- [fermion] (c),
            (c) -- (h) -- [fermion,reversed momentum={[arrow shorten=0]\(k\)}] (d),
            (b) -- [photon,momentum'={[arrow shorten=0.3]\(q\)}] (e),
            (c) -- [photon,momentum={[arrow shorten=0.3]\(p\)}] (f),
            (g) -- [gluon,reversed momentum={[arrow shorten=0.3]\(l\)}] (h),
        };
    \end{feynman}
    \end{tikzpicture}
    \caption{}\label{fig2d}
    \end{subfigure}
    \caption{$\mathcal{O}(\alpha_s)$  correction diagrams for $\Pi_\mu$.}
    \label{fig:2}
\end{figure}

The contribution of weak vertex correction at the one-loop order of QCD (the diagram in Figure \ref{fig:2}(a)) is 
\begin{align}
    \Pi_{\mu,\mathrm{weak}}^{A,(1)}
    =\,&\dfrac{ g_s^2 C_F \mu^{2\epsilon} }{2(\bar{n}\cdot p-\omega)}\int\dfrac{\md^D l}{(2\pi)^D}\dfrac{1}{[(p-k+l)^2+\ii 0][(m_b v+l)^2-m_b^2+\ii 0][l^2+\ii 0]}\notag\\
    & \bar{q}(k) \slashed{\bar{n}}\gamma_\rho(\slashed{p}-\slashed{k}+\slashed{l})\gamma_\mu \gamma_5(m_b\slashed{v}-\slashed{k}+\slashed{l}+m_b)\gamma_\rho b(v).
\end{align}
Hereafter, we ignore the index $b\bar{q}$ in $\Pi_{b\bar{q}}$ and $D=4-2\epsilon$. According to the scaling behaviors
\begin{equation}
    n\cdot p\sim m_b, \quad \bar{n}\cdot p\sim \Lambda, \quad k_\mu\sim\Lambda,
\end{equation}
where $\Lambda$ is a hadronic scale of order $\Lambda_{\mathrm{QCD}}$ we obtain that the leading-power contributions of scalar integral
\begin{equation}
    I_1=\int\dfrac{\md^D l}{(2\pi)^D}\dfrac{1}{[(p-k+l)^2+\ii 0][(m_b v+l)^2-m_b^2+\ii 0][l^2+\ii 0]}
\end{equation}
come from hard, hard-collinear, semi-hard, and soft regions with the power $I_1\sim \lambda^0$, which implies that only the leading power terms in the numerator contribute. The semi-hard region contribution vanishes due to the fact (see Appendix B of \cite{Becher:2014oda}) that
\begin{equation}
    \begin{aligned}
        \int\dfrac{d^D l}{(2\pi)^D}\dfrac{1}{l^2(\bar{n}\cdot l)(v\cdot l)}=\dfrac{1}{D-4}\int\dfrac{d^D l}{(2\pi)^D}\dfrac{\partial}{\partial l^\mu}\dfrac{l^\mu}{l^2(\bar{n}\cdot l)(v\cdot l)}=0
    \end{aligned},
\end{equation}
under dimensional regularization.

After applying the light-cone projector, one obtains the spinor structure in the hard region
\begin{align}
    &\mathrm{Tr}\left\{\dfrac{1+\slashed{v}}{2}\phi_{b\bar{q}}^+\slashed{n}\gamma_5\slashed{\bar{n}}\gamma_\rho\left(\dfrac{n\cdot p}{2}\slashed{\bar{n}}+\slashed{l}\right)\gamma_\mu\gamma_5(m_b\slashed{v}+\slashed{l}+m_b)\gamma_\rho\right\}\notag\\
    &=4\phi_{b\bar{q}}^+\big[m_b \,n\cdot p\, \bar{n}^\mu+2 m_b l^\mu+m_b \bar{n}^\mu \,n\cdot l-m_b n^\mu \,\bar{n}\cdot l\notag\\
    &\quad+(D-2)n^\mu l^2-2(D-2)\bar{n}^\mu \,\bar{n}\cdot l\, n\cdot l-2(D-2)n^\mu (\bar{n}\cdot l)^2\big],
\end{align}
and the spinor structure in the hard-collinear region
\begin{align}
    &\mathrm{Tr}\left\{\dfrac{1+\slashed{v}}{2}\phi_{b\bar{q}}^+ \slashed{n}\gamma_5\slashed{\bar{n}}\gamma_\rho\left(\dfrac{n\cdot p}{2}\slashed{\bar{n}}+\dfrac{n\cdot p}{2}\slashed{\bar{n}}\right)\ii \sigma_{\mu\nu}q^\nu\gamma_5\left(m_b\slashed{v}+\dfrac{n\cdot l}{2}\slashed{\bar{n}}+m_b\right)\gamma_\rho\right\}\notag\\
    &=8 (m_b \bar{n}^\mu\,n\cdot l+m_b\,n\cdot p\, \bar{n}^\mu)\phi_{b\bar{q}}^+.
    \end{align}
For tensor current, similarly:
\begin{align}
    \Pi_{\mu,\mathrm{weak}}^{T,(1)}=\,&\dfrac{ g_s^2 C_F \mu^{2\epsilon}}{2(\bar{n}\cdot p-\omega)}\int\dfrac{\md^D l}{(2\pi)^D}\dfrac{1}{[(p-k+l)^2+\ii 0][(p_B-k+l)^2-m_b^2+\ii 0][l^2+\ii 0]}\notag\\
    &\bar{q}(k) \slashed{\bar{n}}\gamma_\rho(\slashed{p}-\slashed{k}+\slashed{l})\ii\sigma_{\mu\nu}q^\nu\gamma_5(m_b\slashed{v}-\slashed{k}+\slashed{l}+m_b)\gamma_\rho b(v).
\end{align}

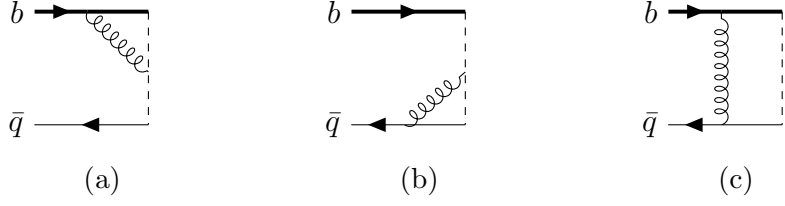
\begin{figure}[t]
    \centering
    %weak vertex Wilson line correction
    \begin{subfigure}{0.24\linewidth}
    \begin{tikzpicture}
    \begin{feynman}
        \vertex (a) {\(b\)};
        \vertex [right=1.73cm of a] (b);
        \vertex [below=of b] (c);
        \vertex [left=of c] (d) {\(\bar{q}\)};
        \vertex [left=0.8cm of b] (e);
        \vertex [below=0.8cm of b] (f);        \diagram* {
            (a) -- [fermion, line width=1.5pt] (e) -- [line width=1.5pt] (b),
            (b) -- [dashed] (c),
            (c) -- [fermion] (d),
            (e) -- [gluon] (f),
        };
    \end{feynman}
    \end{tikzpicture}
    \caption{\quad\quad\quad\quad}
    \end{subfigure}
    \begin{subfigure}{0.24\linewidth}
    %scalar vertex Wilson line correction
    \begin{tikzpicture}
    \begin{feynman}
        \vertex (a) {\(b\)};
        \vertex [right=1.73cm of a] (b);
        \vertex [below=of b] (c);
        \vertex [left=of c] (d) {\(\bar{q}\)};
        \vertex [left=0.8cm of c] (e);
        \vertex [below=0.8cm of b] (f);        
        \diagram* {
            (a) -- [fermion, line width=1.5pt] (b),
            (b) -- [dashed] (c),
            (c) -- (e) -- [fermion] (d),
            (e) -- [gluon] (f),
        };
    \end{feynman}
    \end{tikzpicture}
    \caption{\quad\quad\quad\quad}
    \end{subfigure}
    \begin{subfigure}{0.24\linewidth}
    %box Wilson line correction
    \begin{tikzpicture}
    \begin{feynman}
        \vertex (a) {\(b\)};
        \vertex [right=1.73cm of a] (b);
        \vertex [below=of b] (c);
        \vertex [left=of c] (d) {\(\bar{q}\)};
        \vertex [left=0.8cm of c] (e);
        \vertex [left=0.8cm of b] (f);        
        \diagram* {
            (a) -- [fermion,line width=1.5pt] (f) -- [line width=1.5pt] (b),
            (b) -- [dashed] (c),
            (c) -- (e) -- [fermion] (d),
            (e) -- [gluon] (f),
        };
    \end{feynman}
    \end{tikzpicture}
    \caption{\quad\quad\quad\quad}
    \end{subfigure}
    \caption{One-loop diagrams for  $B$ meson DA $\Phi_{b\bar{q}}^{\alpha\beta}(\omega')$}
    \label{fig:3}
\end{figure}

Following the same procedure, the spinor structure in the hard region is 
\begin{align}
    &\mathrm{Tr}\left\{\dfrac{1+\slashed{v}}{2}\phi_{b\bar{q}}^+\slashed{n}\gamma_5\slashed{\bar{n}}\gamma_\rho\left(\dfrac{n\cdot p}{2}\slashed{\bar{n}}+\slashed{l}\right) \ii\sigma_{\mu\nu}q^\nu\gamma_5 (m_b\slashed{v}+\slashed{l}+m_b)\gamma_\rho\right\}\notag\\
    &=4 \phi_{b\bar{q}}^+(n\cdot p\, m_b+m_b\, n\cdot l+n\cdot p\, \bar{n}\cdot l+\epsilon\, l^2)[\bar{n}\cdot q\,n_{\mu}-n\cdot q\,\bar{n}_\mu].
\end{align}
The spinor structure in the hard collinear region is
\begin{align}
    &\mathrm{Tr}\left\{\dfrac{1+\slashed{v}}{2}\phi_{b\bar{q}}^+\slashed{n}\gamma_5\slashed{\bar{n}}\gamma_\rho\left(\dfrac{n\cdot p}{2}\slashed{\bar{n}}+\dfrac{n\cdot l}{2}\slashed{\bar{n}}\right)\ii\sigma_{\mu\nu}q^\nu\gamma_5\left(m_b \slashed{v}+\dfrac{n\cdot l}{2}\slashed{\bar{n}}+m_b\right)\gamma_\rho\right\}\notag\\
    &=-4 \phi_{b\bar{q}}^+ m_b(n\cdot l+n \cdot p)[\bar{n}\cdot q\,n_{\mu}-n\cdot q\,\bar{n}_\mu].
    \end{align}
Substituting the integral results from the Appendix A in \cite{Wang:2015vgv} yields
\begin{align}
    \Pi_{\mu,\mathrm{weak}}^{A,(1),\mathrm{h}}=\,&  -\dfrac{\alpha_s C_F}{8\pi} \tilde{f}_B(\mu) m_B \dfrac{\phi^+_{b\bar{q}} (\omega)}{\bar{n} \cdot p -\omega} \left\{ \bar{n}_\mu \left[\dfrac{1}{\epsilon^2} + \dfrac{2}{\epsilon} \ln \dfrac{\mu}{n \cdot p} + \dfrac{1}{\epsilon} + 2 \ln^2\dfrac{\mu}{n \cdot p}  \right. \right. \notag\\
    & \left. \left.  +2\ln \dfrac{\mu}{m_b} - \ln^2 r - 2\mathrm{Li_2} \left(-\dfrac{\bar{r}}{r} \right) +\dfrac{2-r}{r-1} \ln r +3 + \dfrac{\pi^2}{12} \right] +n_\mu \left(\dfrac{1}{r-1}\right) \left(1+\dfrac{r}{\bar{r}} \ln r \right)\right\} ,\\
    \Pi^{A,(1),\mathrm{hc}}_{\mu,\mathrm{weak}} =\,& \dfrac{\alpha_s C_F}{8\pi} \tilde{f}_B(\mu) m_B \dfrac{\phi^+_{b\bar{q}} (\omega)}{\bar{n} \cdot p-\omega} \bar{n}_\mu \left[ \dfrac{2}{\epsilon^2} +\dfrac{2}{\epsilon} \left( \ln \dfrac{\mu^2}{n\cdot p(\omega - \bar{n} \cdot p)} +1 \right) \right. \notag\\
    & \left. +\ln^2 \dfrac{\mu^2}{n\cdot p(\omega - \bar{n} \cdot p)}+2\ln \dfrac{\mu^2}{n\cdot p(\omega - \bar{n} \cdot p)}-\dfrac{\pi^2}{6}+4 \right],\\
    \Pi^{T,(1),\mathrm{h}}_{\mu,\mathrm{weak}} =\,& \dfrac{ \alpha_s C_F}{8\pi} \dfrac{\tilde{f}_B m_B}{\bar{n} \cdot p-\omega} \phi^+_{b\bar{q}}(\omega) 
    \left[\bar{n}\cdot q\,n_{\mu}-n\cdot q\,\bar{n}_\mu  \right] \notag\\
    & \left[ \dfrac{1}{2\epsilon^2} +\dfrac{1}{\epsilon} \left(\ln \dfrac{\mu}{n\cdot p}+1 \right) +2\ln\dfrac{\mu}{m_b}+\ln^2 \dfrac{\mu}{n\cdot p} \right. \notag\\
    & \left. -\mathrm{Li_2}\left(-\dfrac{\bar{r}}{r}\right)+\dfrac{2r-1}{\bar{r}} \ln r  -\dfrac{1}{2} \ln^2 r +\dfrac{\pi^2}{24} + 2 \right], \\
    \Pi^{T,(1),\mathrm{hc}}_{\mu,\mathrm{weak}} =\,&- \dfrac{ \alpha_s C_F}{8\pi} \dfrac{\tilde{f}_B m_B}{\bar{n} \cdot p-\omega} \phi^+_{b\bar{q}}(\omega) 
    \left[\bar{n}\cdot q\,n_{\mu}-n\cdot q\,\bar{n}_\mu  \right] \notag\\
    & \left\{ \dfrac{1}{\epsilon^2} +\dfrac{1}{\epsilon} \left[1+\ln \left(\dfrac{\mu^2}{n\cdot p (\omega-\bar{n}\cdot p)}\right) \right]  + \ln \left(\dfrac{\mu^2}{n\cdot p (\omega-\bar{n}\cdot p)}\right) \right. \notag\\
    & \left.  +\dfrac{1}{2}\ln^2 \left(\dfrac{\mu^2}{n\cdot p (\omega-\bar{n}\cdot p)}\right) -\dfrac{\pi^2}{12}+2 \right\},
\end{align}
where $r=\dfrac{n\cdot p}{m_b}, \bar{r}=1-r$.% The renormalization scale of the tensor current is denoted by $\nu$.\\
\\
The contribution from the soft region in QCD is
\begin{align}
    \Pi^{(1),\mathrm{s}}_{\mu,\mathrm{weak}}=\,&\dfrac{g_s^2 C_F\mu^{2\epsilon}}{2(\bar{n}\cdot p-\omega)}\int\dfrac{\md^D l}{(2\pi)^D}\dfrac{1}{[\bar{n}\cdot (p-k+l)+i0][v\cdot l+i0][l^2+i0]}\notag\\
    &\bar{q}(k)\slashed{\bar{n}}\left\{\gamma_\mu\gamma_5,i\sigma_{\mu\nu}q^\nu\gamma_5\right\}b(v).
\end{align}
According to the Feynman rules of Wilson lines, the corresponding contribution from partonic DA (the diagram in Figure \ref{fig:3}(a)) is
\begin{align}
    \Phi_{b \bar{q}, a}^{\alpha \beta,(1)}\left(\omega, \omega^{\prime}\right)=\,& i g_s^2 C_F \int \frac{d^D l}{(2 \pi)^D} \frac{1}{[\bar{n} \cdot l+i 0][v \cdot l+i 0]\left[l^2+i 0\right]} \notag\\
    & \times\left[\delta\left(\omega^{\prime}-\omega-\bar{n} \cdot l\right)-\delta\left(\omega^{\prime}-\omega\right)\right][\bar{q}(k)]_\alpha[b(v)]_\beta.
\end{align}
By convolving with the tree-level hard-scattering kernel $T^{(0)}_{\alpha\beta}$ we derive the infrared subtraction term
\begin{align}
        \Phi^{(1)}_{b\bar{q},a}\otimes T^{(0)}_\mu=\,&\dfrac{g_s^2 C_F\mu^{2\epsilon}}{2(\bar{n}\cdot p-\omega)}\int\dfrac{\md^D l}{(2\pi)^D}\dfrac{1}{[\bar{n}\cdot (p-k+l)+i0][v\cdot l+i0][l^2+i0]}\notag\\
    &\bar{q}(k)\slashed{\bar{n}}\left\{\gamma_\mu\gamma_5,i\sigma_{\mu\nu}q^\nu\gamma_5\right\}b(v),
\end{align}
which cancels the soft region contribution from the weak vertex correction.
\subsection{Scalar vertex correction}
For axial current, the scalar vertex correction in QCD (the diagram in Figure \ref{fig:2}(b)) is
\begin{align}
    \Pi_{\mu,\mathrm{scalar}}^{A,(1)}=\,&-\dfrac{ g_s^2 C_F \mu^{2\epsilon} }{n\cdot p(\bar{n}\cdot p-\omega)}\int\dfrac{\md^D l}{(2\pi)^D}\dfrac{1}{[(l-k)^2+\ii 0][(p-l)^2+\ii 0][l^2+\ii 0]}\notag\\
    &\bar{q}(k)\gamma_\rho\slashed{l}(\slashed{p}-\slashed{l})\gamma_\rho(\slashed{p}-\slashed{k})\gamma_\mu\gamma_5 b(v)\notag\\
    =\,&-\dfrac{ g_s^2 C_F}{n\cdot p(\bar{n}\cdot p-\omega)} \bar{q}(k)\gamma_\rho\gamma_\alpha\gamma_\beta\gamma_\rho(\slashed{p}-\slashed{k})\gamma_\mu\gamma_5 b(v)\dfrac{\ii }{(4\pi)^2}I_2^{\alpha\beta},
    \end{align}
where $I_2^{\alpha\beta}$ is defined in Appendix A of \cite{Wang:2015vgv}. According to the scaling behaviors we can find that the leading power contributions of the scalar integral
\begin{equation}
    I_2=\int\dfrac{\md^D l}{(2\pi)^D}\dfrac{1}{[(l-k)^2+\ii 0][(p-l)^2+\ii 0][l^2+\ii 0]}
\end{equation}
come from the hard-collinear and soft regions with $I_2\sim \lambda^{-1}$. Further calculations indicate that the contribution from the soft region can only generate a scaleless integral that vanishes under dimensional regularization. The hard region contribution also leads to a scaleless integral
\begin{align}
        I_2^h&=\int\dfrac{d^D l}{(2\pi)^D}\dfrac{1}{[l^2+i0][l^2-\bar{n}\cdot l n\cdot p+i0][l^2+i0]}\notag\\
        &=\int^1_0 dx\int_0^x dy\int \dfrac{d^D l'}{(2\pi)^D}\dfrac{1}{[l'^2+i0]^3},\\
        l'^\mu&=l^\mu-\dfrac{1}{2}y \,n\cdot p\, \bar{n}^\mu.
\end{align}
Therefore we can ignore it safely.

The integral result $I_2^{\alpha\beta}$ contains five mutually independent Lorentz tensors, and they will contract with the indices in $ \Pi_{\mu,scalar}^{A,(1)}$. Applying the light-cone projector, four tensor structures will contribute to $ \Pi_{\mu,scalar}^{A,(1)}$ as follows
\begin{align}
    p_\alpha p_\beta:&\q 2D\,\bar{n}\cdot p\, (n\cdot p)^2 \phi_{b\bar{q}}^+ \bar{n}^\mu,\notag\\
    k_\alpha p_\beta:&\q 4 (n\cdot p)^2 \omega \phi_{b\bar{q}}^+ \bar{n}^\mu,\notag\\
    p_\alpha k_\beta:&\q 2(D-2)(n\cdot p)^2\omega\phi_{b\bar{q}}^+\bar{n}^\mu,\notag\\
    g_{\alpha\beta}:&\q 2D^2 \,n\cdot p\,\phi_{b\bar{q}}^+ \bar{n}^\mu.
\end{align}
For tensor current
\begin{align}
    \Pi_{\mu,\mathrm{scalar}}^{T,(1)}=\,&-\dfrac{ g_s^2 C_F \mu^{2\epsilon}}{n\cdot p(\bar{n}\cdot p-\omega)}\int\dfrac{\md^D l}{(2\pi)^D}\dfrac{1}{[(l-k)^2+\ii 0][(p-l)^2+\ii 0][l^2+\ii 0]}\notag\\
    &\bar{d}(k)\gamma_\rho\slashed{l}(\slashed{p}-\slashed{l})\gamma_\rho(\slashed{p}-\slashed{k})\ii\sigma_{\mu\nu}q^\nu\gamma_5 b(v)\notag\\
    =\,&-\dfrac{ g_s^2 C_F}{n\cdot p(\bar{n}\cdot p-\omega)}\bar{q}(k)\gamma_\rho\gamma_\alpha\gamma_\beta\gamma_\rho(\slashed{p}-\slashed{k})\ii\sigma_{\mu\nu}q^\nu\gamma_5b(v) \dfrac{\ii }{(4\pi)^2}I_2^{\alpha\beta}.
    \end{align}
The spinor structure is
\begin{align}
    p_\alpha p_\beta&:\quad -D \,\bar{n}\cdot p\, (n\cdot p)^2\phi_{b\bar{q}}^+\left[\bar{n}\cdot q\,n_{\mu}-n\cdot q\,\bar{n}_\mu  \right],\notag\\
    k_\alpha p_\beta&:\quad -2(n\cdot p)^2\omega\phi_{b\bar{q}}^+\left[\bar{n}\cdot q\,n_{\mu}-n\cdot q\,\bar{n}_\mu  \right],\notag\\
    p_\alpha k_\beta&:\quad (2-D)(n\cdot p)^2 \omega \phi_{b\bar{q}}^+ \left[\bar{n}\cdot q\,n_{\mu}-n\cdot q\,\bar{n}_\mu  \right],\notag\\
    g_{\alpha\beta}&:\quad -D^2 \,n\cdot p\,\phi_{b\bar{q}}^+ \left[\bar{n}\cdot q\,n_{\mu}-n\cdot q\,\bar{n}_\mu  \right].
\end{align}
Substituting the integral results from the Appendix A in \cite{Wang:2015vgv} yields
\begin{align}
    \Pi^{A,(1),\mathrm{hc}}_{\mu,\mathrm{scalar}}=\,& \dfrac{\alpha_s C_F}{4\pi} \dfrac{\tilde{f}_B m_B}{\bar{n}\cdot p -\omega} \phi^+_{b\bar{q}}(\omega) \bar{n}_\mu \left[ \left(1-\dfrac{\bar{n}\cdot p}{\omega} \ln \dfrac{\bar{n}\cdot p - \omega}{\bar{n}\cdot p}\right)\dfrac{1}{\epsilon} + \ln \left(-\dfrac{\mu^2}{p^2}\right)  \right. \notag\\
    & \left.  - \dfrac{\bar{n}\cdot p}{\omega} \ln \dfrac{\bar{n}\cdot p-\omega}{\bar{n}\cdot p} \ln \left(-\dfrac{\mu^2}{p^2}\right) +\dfrac{\bar{n}\cdot p}{2\omega} \ln^2 \dfrac{\bar{n}\cdot p-\omega}{\bar{n}\cdot p}- \ln \dfrac{\bar{n}\cdot p-\omega}{\bar{n}\cdot p}+1 \right],\\
    \Pi^{T,(1),\mathrm{hc}}_{\mu,\mathrm{scalar}} =\,&- \dfrac{ \alpha_s C_F}{8\pi} \dfrac{\tilde{f}_B m_B}{\bar{n}\cdot p -\omega} \phi^+_{b\bar{q}}(\omega)  \left[\bar{n}\cdot q\,n_{\mu}-n\cdot q\,\bar{n}_\mu  \right] \notag\\
    &\left[ \left(1-\dfrac{\bar{n}\cdot p}{\omega} \ln \dfrac{\bar{n}\cdot p - \omega}{\bar{n}\cdot p}\right)\dfrac{1}{\epsilon} \right. \left. + \ln \left(-\dfrac{\mu^2}{p^2}\right) - \dfrac{\bar{n}\cdot p}{\omega} \ln \dfrac{\bar{n}\cdot p-\omega}{\bar{n}\cdot p} \ln \left(-\dfrac{\mu^2}{p^2}\right)\right. \notag\\
    &\left.+\dfrac{\bar{n}\cdot p}{2\omega} \ln^2 \dfrac{\bar{n}\cdot p-\omega}{\bar{n}\cdot p}- \ln \dfrac{\bar{n}\cdot p-\omega}{\bar{n}\cdot p}+1 \right].
\end{align}
The contribution from the soft region is
\begin{align}
    \Pi_{\mu, \mathrm{scalar}}^{(1), \mathrm{s}}=\,&-\frac{g_s^2 C_F}{2(\bar{n} \cdot p-\omega)} \int \frac{d^D l}{(2 \pi)^D} \frac{1}{[\bar{n} \cdot(p-l)+i 0]\left[(l-k)^2+i 0\right]\left[l^2+i 0\right]}\notag\\
    &\bar{q}(k)\slashed{\bar{n}}\slashed{l}\slashed{\bar{n}}\left\{\gamma_\mu\gamma_5,i\sigma_{\mu\nu}q^\nu\gamma_5\right\}b(v).
\end{align}
The contribution from the corresponding DA (the diagram in Figure \ref{fig:3}(b)) is
\begin{align}
    \Phi_{b \bar{q}, b}^{\alpha \beta,(1)}\left(\omega, \omega^{\prime}\right)=\,& i g_s^2 C_F \int \frac{d^D l}{(2 \pi)^D} \frac{1}{[\bar{n} \cdot (l-k)+i 0][l^2+i 0]\left[(l-k)^2+i 0\right]} \notag\\
    & \times\left[\delta\left(\omega^{\prime}-\omega-\bar{n} \cdot l\right)-\delta\left(\omega^{\prime}-\omega\right)\right][\bar{q}(k)\slashed{\bar{n}}\slashed{l}]_\alpha[b(v)]_\beta.
\end{align}
By convolving with the tree-level hard-scattering kernel, we derive the infrared subtraction term
\begin{align}
    \Phi^{(1)}_{b\bar{q},b}\otimes T^{(0)}_\mu=\,&-\frac{g_s^2 C_F}{2(\bar{n} \cdot p-\omega)} \int \frac{d^D l}{(2 \pi)^D} \frac{1}{[\bar{n} \cdot(p-l)+i 0]\left[(l-k)^2+i 0\right]\left[l^2+i 0\right]}\notag\\
    &\bar{q}(k)\slashed{\bar{n}}\slashed{l}\slashed{\bar{n}}\left\{\gamma_\mu\gamma_5,i\sigma_{\mu\nu}q^\nu\gamma_5\right\}b(v),
\end{align}
which cancels the contribution from the soft region of one-loop correction to the scalar vertex. 
\subsection{Wave function renormalization}
For the axial current, the self-energy correction to the intermediate quark propagator in QCD (the diagram in Figure \ref{fig:2}(c)) is
\begin{align}
\Pi_{\mu, \mathrm{w f c}}^{(1)}=\,& \frac{g_s^2 C_F \mu^{2\epsilon}}{(n \cdot p)^2(\bar{n} \cdot p-\omega)^2} \int \frac{d^D l}{(2 \pi)^D} \frac{1}{\left[(p-k+l)^2+i 0\right]\left[l^2+i 0\right]} \notag\\
& \bar{q}(k) (\slashed{p}-\slashed{k}) \gamma_\rho(\slashed{p}-\slashed{k}+\slashed{l}) \gamma^\rho (\slashed{p}-\slashed{k}) \gamma_\mu\gamma_5 b(v).
\end{align}
It's apparent that there is no soft or collinear divergence and we can calculate it straightforwardly
\begin{equation}
    \begin{aligned}
    \Pi^{A,(1)}_{\mu,\mathrm{wfc}} &= -\dfrac{\alpha_s C_F}{8\pi}  \dfrac{\tilde{f}_B m_B}{\bar{n}\cdot p-\omega} \phi^+_{b\bar{q}}(\omega) \bar{n}_\mu \left[ \dfrac{1}{\epsilon} +\ln \left(\dfrac{\mu^2}{n\cdot p ( \omega -\bar{n}\cdot p)}\right)+1 \right].
    \end{aligned}
\end{equation}
Also, the massless quark will not contribute to hard scattering kernel \cite{Wang:2015vgv}. The wave function renormalization of $b$ quark in QCD is
\begin{equation}
    \begin{aligned}
        \Pi^{A,(1)}_{\mu,\mathrm{bwf}} &= -\dfrac{\alpha_s C_F}{8\pi} \left[\dfrac{3}{\epsilon} + 3\ln \dfrac{\mu^2}{m^2_b} +4\right] \Pi^{A,(0)}_\mu,\label{eq:PiAbwf}
    \end{aligned}
\end{equation}
while in heavy quark effective theory, the wave function renormalization gives a scaleless integral
\begin{equation}
    \Phi_{b\bar{q},\mathrm{bwf}}^{(1)}\otimes T^{(0)}_\mu=0.\label{eq:phibqoT0}
\end{equation}
Combine the results of Eq.(\ref{eq:PiAbwf}) and Eq.(\ref{eq:phibqoT0}) we obtain
\begin{equation}
\Pi_{\mu, \mathrm{b w f}}^{(1)}-\Phi_{b \bar{q}, \mathrm{b w f}}^{A,(1)} \otimes T^{(0)}_\mu=-\frac{\alpha_s C_F}{8 \pi}\left[\frac{3}{\epsilon}+3 \ln \frac{\mu^2}{m_b^2}+4\right] \Pi_\mu^{A,(0)}.
\end{equation}
Similarly, for the tensor current
    \begin{align}
    \Pi^{T,(1)}_{\mu,\mathrm{wfc}} &= \dfrac{ \alpha_s C_F}{16\pi}  \dfrac{\tilde{f}_B m_B}{\bar{n}\cdot p-\omega} \phi^+_{b\bar{q}}(\omega) \left[\bar{n}\cdot q\,n_{\mu}-n\cdot q\,\bar{n}_\mu  \right] \left[ \dfrac{1}{\epsilon} +\ln \left(\dfrac{\mu^2}{n\cdot p  (\omega -\bar{n}\cdot p)}\right)+1 \right],\\
    \Pi^{T,(1)}_{\mu,\mathrm{bwf}}&-\Phi_{b \bar{q}, \mathrm{b w f}}^{T,(1)} \otimes T^{(0)} = -\dfrac{\alpha_s C_F}{8\pi} \left[\dfrac{3}{\epsilon} + 3\ln \dfrac{\mu^2}{m^2_b} +4\right] \Pi^{T,(0)}_\mu.
    \end{align}
\subsection{Box diagram}
The box diagram contribution (the diagram in Figure \ref{fig:2}(d)) for the axial current is
    \begin{align}
    \Pi^{A,(1)}_{\mu,\mathrm{box}} =\,&  g^2_s C_F \mu^{2\epsilon}\int \dfrac{\md^D l}{(2\pi)^D} \dfrac{-1}{[(m_b v+l)^2 -m_b^2+\ii 0 ] [(p-k+l)^2 +\ii 0][(k-l)^2 +\ii 0][l^2+\ii 0]} \notag\\
    & \bar{q}(k) \gamma_\rho (\slashed{k}-\slashed{l})(\slashed{p}-\slashed{k}+\slashed{l}) \gamma_\mu \gamma_5 (m_b \slashed{v}+\slashed{l}+m_b) \gamma^\rho b(v).
    \end{align}
The scaling behavior of the scalar integral
\begin{equation}
    I_4=\int \dfrac{\md^D l}{(2\pi)^D} \dfrac{1}{[(m_b v+l)^2 -m_b^2+\ii 0 ] [(p-k+l)^2 +\ii 0][(k-l)^2 +\ii 0][l^2+\ii 0]}
\end{equation}
is $I_4\sim \lambda^{-1}(\lambda^{-2})$ in the hard-collinear and semi-hard (soft) regions. However, the semi-hard contribution corresponds to a scaleless integral. Moreover, the numerator for the hard-collinear region is power-suppressed which is different from the case in $B\rightarrow\pi$ \cite{Wang:2015vgv} because there is no polarization structure for scalar meson interpolating current. 

The soft contribution of the box diagram is
\begin{align}
        \Pi^{(1),\mathrm{s}}_{\mu,\mathrm{box}}=\,&\dfrac{g_s^2 C_F \mu^{2\epsilon}}{2}\int\dfrac{\md^D l}{(2\pi)^D}\dfrac{-1}{[v\cdot l+i0][\bar{n}\cdot(p-k+l)+i0][(k-l)^2+i0][l^2+i0]}\notag\\
        &\bar{q}(k)\slashed{v}(\slashed{k}-\slashed{l})\slashed{\bar{n}}\left\{\gamma_\mu\gamma_5,i\sigma_{\mu\nu}q^\nu\gamma_5\right\}b(v).
\end{align}
Now we calculate the corresponding infrared subtraction term. The contribution from the diagram Figure \ref{fig:3}(c) is
\begin{align}
    \Phi_{b \bar{q}, c}^{\alpha \beta,(1)}\left(\omega, \omega^{\prime}\right)=\,& -i g_s^2 C_F \int \frac{d^D l}{(2 \pi)^D} \frac{1}{\left[(l-k)^2+i 0\right][v \cdot l+i 0]\left[l^2+i 0\right]} \notag\\
    & \times \delta\left(\omega^{\prime}-\omega+\bar{n} \cdot l\right)[\bar{q}(k)\slashed{v}(\slashed{l}-\slashed{k})]_\alpha[b(v)]_\beta.
\end{align}
By convolving with the tree-level hard-scattering kernel we derive the infrared subtraction term as follows
\begin{align}
        \Phi_{b\bar{q},c}^{(1)}\otimes T^{(0)}_\mu=\,&\dfrac{g_s^2 C_F \mu^{2\epsilon}}{2}\int\dfrac{\md^D l}{(2\pi)^D}\dfrac{-1}{[v\cdot l+i0][\bar{n}\cdot(p-k+l)+i0][(k-l)^2+i0][l^2+i0]}\notag\\
        &\bar{q}(k)\slashed{v}(\slashed{k}-\slashed{l})\slashed{\bar{n}}\left\{\gamma_\mu\gamma_5,i\sigma_{\mu\nu}q^\nu\gamma_5\right\}b(v),
\end{align}
which again cancels the soft contribution from the box diagram.
\subsection{The hard-scattering kernel at \texorpdfstring{$\mathcal{O}(\alpha_s)$}{O(as)}}
From the matching condition Eq.(\ref{eq:19}), one can derive the hard-scattering kernel at one-loop level
\begin{align}
\Phi_{b \bar{q}}^{(0)} \otimes T^{(1)}_\mu=\,& {\left[\Pi_{\mu, \mathrm{w e a k}}^{(1)}+\Pi_{\mu, \mathrm{scalar}}^{(1)}+\Pi_{\mu, \mathrm{w f c}}^{(1)}+\Pi_{\mu, \mathrm{b o x}}^{(1)}+\Pi_{\mu, \mathrm{b w f}}^{(1)}+\Pi_{\mu, \mathrm{d w f}}^{(1)}\right] } \notag\\
& -\left[\Phi_{b \bar{q}, a}^{(1)}+\Phi_{b \bar{q}, b}^{(1)}+\Phi_{b \bar{q}, c}^{(1)}+\Phi_{b \bar{q}, \mathrm{b w f}}^{(1)}+\Phi_{b \bar{q}, \mathrm{d w f}}^{(1)}\right] \otimes T^{(0)}_\mu \notag\\
=\,& {\left[\Pi_{\mu, \mathrm{w e a k}}^{(1), \mathrm{h}}+\left(\Pi_{\mu, \mathrm{b w f}}^{(1)}-\Phi_{b \bar{q}, \mathrm{b w f}}^{(1)}\otimes T_\mu^{(0)}\right)\right] } \notag\\
& +\left[\Pi_{\mu, \mathrm{w e a k}}^{(1), \mathrm{h c}}+\Pi_{\mu, \mathrm{scalar}}^{(1), \mathrm{h c}}+\Pi_{\mu, \mathrm{w f c}}^{(1), \mathrm{h c}}\right],
\end{align}
where the terms in the first square brackets and the second brackets correspond to the hard coefficients and the jet functions, respectively. Finally, we establish the factorization formulae of correlation functions defined in Eq.(\ref{eq:3}) at $\mathcal{O}(\alpha_s)$ and leading power in $\Lambda/m_b$
\begin{align}
    \Pi^A=\,&-\dfrac{1}{2} \tilde{f}_B(\mu) m_B \sum_{k=\pm} C^{A,(k)} (n\cdot p ,\mu) \int_{0}^{\infty} \dfrac{\md \omega}{\omega-\bar{n}\cdot p} J^{A,(k)} \left(\dfrac{\mu^2}{n\cdot p \,\omega},\dfrac{\omega}{\bar{n}\cdot p}\right) \phi^{k}_B (\omega,\mu),\notag\\
    \widetilde{\Pi}^A=\,&-\dfrac{1}{2} \tilde{f}_B(\mu) m_B \sum_{k=\pm} \tilde{C}^{A,(k)} (n\cdot p ,\mu) \int_{0}^{\infty} \dfrac{\md \omega}{\omega-\bar{n}\cdot p} \tilde{J}^{A,(k)} \left(\dfrac{\mu^2}{n\cdot p\,\omega},\dfrac{\omega}{\bar{n}\cdot p}\right) \phi^{k}_B (\omega,\mu),\notag\\
    \Pi^T=\,& \dfrac{1}{4} \tilde{f}_B(\mu) m_B\sum_{k=\pm} C^{T,(k)} (n\cdot p ,\mu) \int_{0}^{\infty} \dfrac{\md \omega}{\omega-\bar{n}\cdot p} J^{T,(k)} \left(\dfrac{\mu^2}{n\cdot p \,\omega},\dfrac{\omega}{\bar{n}\cdot p}\right) \phi^{k}_B (\omega,\mu).\quad\quad
\end{align}
The hard coefficient functions for axial current are
\begin{align}
		C^{A,(-)} =\,& \tilde{C}^{A,(-)} = 1 ,\notag\\
		C^{A,(+)} =\,& \dfrac{\alpha_s C_F}{4\pi} \dfrac{1}{\bar{r}} \left[\dfrac{r}{\bar{r}} \ln r +1 \right],  \notag\\
		\tilde{C}^{A,(+)} =\,& 1- \dfrac{\alpha_s C_F}{4\pi} \left[2 \ln^2 \dfrac{\mu}{n\cdot p} + 5 \ln \dfrac{\mu}{m_b } - \ln^2 r - 2 \mathrm{Li_2} \left(-\dfrac{\bar{r}}{r} \right) + \dfrac{2-r}{r-1}\ln r +\dfrac{\pi^2}{12} +5 \right] ,
        \label{eq:CA}
\end{align}
and jet functions are
\begin{align}
    J^{A,(-)} =\,& \tilde{J}^{A,(-)} =  0, \notag\\
    J^{A,(+)} =\,&  1, \notag\\
    \tilde{J}^{A,(+)}
    =\,& 1+\dfrac{\alpha_s C_F}{4\pi}\left[\ln^2 \dfrac{\mu^2}{n\cdot p (\omega-\bar{n}\cdot p)} +3\ln \dfrac{\mu^2}{n\cdot p (\omega-\bar{n}\cdot p)} \right. \notag\\
    & \left. - \dfrac{2\bar{n}\cdot p}{\omega} \ln \dfrac{\bar{n}\cdot p -\omega}{\bar{n}\cdot p} \ln \dfrac{\mu^2}{n\cdot p (\omega-\bar{n}\cdot p)} - \dfrac{\bar{n}\cdot p}{\omega} \ln^2 \dfrac{\bar{n}\cdot p -\omega}{\bar{n}\cdot p} -\dfrac{\pi^2}{6} +5 \right].
\end{align}

The hard functions for axial current are the same as $ B\rightarrow \pi $ transition. The crucial discrepancy of jet functions compared with $B\rightarrow \pi$ attributes to the difference of interpolating currents, which leads to distinct contributions of Figure \ref{fig:2}(b) and (d) to the jet functions. Through a detailed numerical analysis, we have found that the effects of Figure \ref{fig:2}(b) and (d) to form factors are comparable in magnitude. There is an additional scale-dependent term in jet functions due to the fact that scalar current is not conserved.
\\
The hard functions for tensor current are
\begin{align}
    C^{T,(-)} =\,& 1,\notag\\
    C^{T,(+)} =\,& 1-\dfrac{\alpha_s C_F}{4\pi } \bigg[ 2 \ln^2 \dfrac{\mu}{n\cdot p} + 7 \ln \dfrac{\mu}{m_b} -2\mathrm{Li_2}\left(-\dfrac{\bar{r}}{r} \right) \notag\\
    &+\dfrac{4r-2}{1-r} \ln r - \ln^2 r +\dfrac{\pi^2}{12} + 6 \bigg],
\end{align}
and the jet functions are
\begin{align}
    J^{T,(-)} =\,& 0 ,\notag\\
    J^{T,(+)} 
    =\,& 1+\dfrac{\alpha_s C_F}{4\pi}\left[\ln^2 \dfrac{\mu^2}{n\cdot p (\omega-\bar{n}\cdot p)} +3\ln \dfrac{\mu^2}{n\cdot p (\omega-\bar{n}\cdot p)} \right. \notag\\
    & \left. - \dfrac{2\bar{n}\cdot p}{\omega} \ln \dfrac{\bar{n}\cdot p -\omega}{\bar{n}\cdot p} \ln \dfrac{\mu^2}{n\cdot p (\omega-\bar{n}\cdot p)} - \dfrac{\bar{n}\cdot p}{\omega} \ln^2 \dfrac{\bar{n}\cdot p -\omega}{\bar{n}\cdot p} -\dfrac{\pi^2}{6} +5 \right] . 
    \label{eq:JT}
\end{align}
The tensor current is also not conserved and there is an additional scale-dependent term in the hard coefficient function $C^{T,(+)}$ compared to the axial current part.  The hard coefficient functions are consistent with results obtained by soft-collinear effective theory from matching QCD$\to $SCET$_\mathrm{I}$ \cite{Beneke:2004rc, Bauer:2000yr}. We also observe that the jet functions of the axial and tensor currents are the same, which is also consistent with SCET. In the theoretical framework of SCET, because the leading power contributions only come from the light-cone components of the momentum transfer $q^\nu$, the axial and tensor currents have the same SCET$_\mathrm{I}$ operator basis at leading power \cite{Beneke:2004rc}, which implies that the jet functions obtained by matching SCET$_\mathrm{I}$ operators to SCET$_\mathrm{II}$ operators are the same.

To verify the factorization-scale dependence of the correlation functions, we list the following evolution equations
\begin{align}
    \dfrac{\md }{\md \ln \mu} \tilde{C}^{(+)}(n\cdot p,\mu) =\,& -\dfrac{\alpha_s C_F}{4\pi} \left[\Gamma^{(0)}_{\mathrm{cusp}} \ln \dfrac{\mu}{n\cdot p}+5 \right] \tilde{C}^{(+)}(n\cdot p,\mu) \, ,\\
    \dfrac{\md}{\md \ln \mu} \left[\tilde{f}_B (\mu) \phi_B^+(\omega,\mu)\right] =\,&-\dfrac{\alpha_s C_F}{4\pi}\left[\Gamma^{(0)}_{\mathrm{cusp}} \ln\dfrac{\mu}{\omega} -5 \right]\left[\tilde{f}_B (\mu) \phi_B^+(\omega,\mu)\right]  \notag\\
	& -\dfrac{\alpha_s C_F}{4\pi}\int_{0}^{\infty} \md \omega' \omega \Gamma_+ (\omega,\omega';\mu) \left[\tilde{f}_B (\mu) \phi^+_B(\omega',\mu)\right] \, ,\\ 
	\dfrac{\md}{\md \ln \mu}\tilde{J}^{(+)} \left( \dfrac{\mu^2}{n\cdot p \omega},\dfrac{\omega}{\bar{n}\cdot p} \right) =\,& \dfrac{\alpha_s C_F}{4\pi} \left[\Gamma^{(0)}_{\mathrm{cusp}} \ln \dfrac{\mu^2}{n\cdot p \,\omega}\right] \tilde{J}^{(+)} \left( \dfrac{\mu^2}{n\cdot p \ \omega},\dfrac{\omega}{\bar{n}\cdot p} \right) \notag\\
	& +\dfrac{\alpha_s C_F}{4\pi} \int_{0}^{\infty} \md \omega' \omega \Gamma_+(\omega,\omega';\mu) \tilde{J}^{(+)} \left( \dfrac{\mu^2}{n\cdot p \,\omega},\dfrac{\omega'}{\bar{n}\cdot p} \right), 
    \label{eq:evolutionforscalein}
\end{align}
where the function $\Gamma_+$ is given by \cite{Lange:2003ff}
\begin{equation}
	\Gamma_+ (\omega,\omega';\mu) = -\Gamma^{(0)}_{\mathrm{cusp}}  \left[\dfrac{\theta(\omega'-\omega)}{\omega'(\omega'-\omega)} + \dfrac{\theta(\omega-\omega')}{\omega(\omega-\omega')} \right]_\oplus
\end{equation}
at one-loop order, and $\Gamma^{(0)}_{\mathrm{cusp}}=4$ due to the geometry of Wilson line \cite{Wang:2015vgv}.
The scale dependence of the correlation functions at the one-loop order then simply implies
\begin{equation}
\frac{\md}{\md \ln \mu}[\Pi^{(1)}, \widetilde{\Pi}^{(1)}]+(\gamma_S+\gamma_T)[\Pi^{(0)},\widetilde{\Pi}^{(0)}]=0,
\end{equation}
which is straightforward to verify with the anomalous dimensions of the scalar current and tensor current
    \begin{equation}
        \gamma_S=-6\dfrac{\alpha_s C_F}{4\pi},\q\q\gamma_T=2\dfrac{\alpha_s C_F}{4 \pi}.
    \end{equation}
Observing the equations (\ref{eq:CA})$\sim$(\ref{eq:JT}), and (\ref{eq: fBdef}), we notice that it is impossible to eliminate the large logarithms of order $\mathrm{ln}(m_b/\Lambda_{QCD})$ simultaneously in the hard functions, jet functions, $\tilde{f}_B(\mu)$, and  $B$ meson distribution amplitudes by selecting a single value of $\mu$. To address this issue, the resummation of these logarithms to all orders of $\alpha_s$ can be accomplished by solving the three renormalization group equations presented above. We will choose $\mu$ at the scale of hard-collinear $\mu_{hc}\sim \sqrt{m_b \Lambda}$. The evolution functions resulting from the running of the renormalization scale from hard scale $\mu_{h1}\sim n\cdot p$ to $\mu_{hc}$ in $\tilde{C}^{(+)}(n\cdot p,\mu)$, and from $\mu_{h2}\sim m_b$ to $\mu_{hc}$ in $\tilde{f_B}(\mu)$ are
\begin{align}
\tilde{C}^{(+)}(n \cdot p, \mu) & =U_1\left(n \cdot p, \mu_{h 1}, \mu\right) \tilde{C}^{(+)}\left(n \cdot p, \mu_{h 1}\right) \notag\\
\tilde{f}_B(\mu) & =U_2\left(\mu_{h 2}, \mu\right) \tilde{f}_B\left(\mu_{h 2}\right).
\end{align}
In order to perform the next-to-leading log (NLL) resummation of the large logarithms in the hard coefficient functions $\tilde{C}^{(+)}$ and $\tilde{f}_B$, it is necessary to extend the renormalization group equation (\ref{eq:evolutionforscalein}) to a more generalized form
\begin{align}
    \frac{\md}{\md \ln \mu} \tilde{C}^{(+)}(n \cdot p, \mu)&=\left[-\Gamma_{\text {cusp }}\left(\alpha_s\right) \ln \frac{\mu}{n \cdot p}+\gamma\left(\alpha_s\right)\right] \tilde{C}^{(+)}(n \cdot p, \mu),\notag\\
    \dfrac{\md}{\md\ln\mu} \tilde{f}_B(\mu)&=\tilde{\gamma}(\alpha_s)\tilde{f}_B(\mu),
\end{align}
where the $\Gamma_{\mathrm{cusp}}$, $\gamma(\alpha_s)$ and $\tilde{\gamma}(\alpha_s)$ can be expanded as
\begin{align}
    \Gamma_{\text {cusp }}\left(\alpha_s\right) & =\frac{\alpha_s C_F}{4 \pi}\left[\Gamma_{\text {cusp }}^{(0)}+\left(\frac{\alpha_s}{4 \pi}\right) \Gamma_{\text {cusp }}^{(1)}+\left(\frac{\alpha_s}{4 \pi}\right)^2 \Gamma_{\text {cusp }}^{(2)}+\ldots\right], \notag\\
    \gamma\left(\alpha_s\right) & =\frac{\alpha_s C_F}{4 \pi}\left[\gamma^{(0)}+\left(\frac{\alpha_s}{4 \pi}\right) \gamma^{(1)}+\ldots\right], \notag\\
    \tilde{\gamma}\left(\alpha_s\right) & =\frac{\alpha_s C_F}{4 \pi}\left[\tilde{\gamma}^{(0)}+\left(\frac{\alpha_s}{4 \pi}\right) \tilde{\gamma}^{(1)}+\ldots\right]. 
\end{align}
The explicit formulas for $\Gamma_{\mathrm{cusp}}^{(i)}$, $\gamma^{(i)}$, and $\tilde{\gamma}^{(i)}$ are available in \cite{Beneke:2011nf}, and the solutions of renormalization group equations with NLL accuracy are given by
\begin{align}
    U_1\left(n\cdot p, \mu_{h1}, \mu\right) =\,&\exp \left(\int_{\alpha_s\left(\mu_{h1}\right)}^{\alpha_s(\mu)} d \alpha_s\left[\frac{\gamma\left(\alpha_s\right)}{\beta\left(\alpha_s\right)}+\frac{\Gamma_{\text {cusp }}\left(\alpha_s\right)}{\beta\left(\alpha_s\right)}\left(\ln \frac{n\cdot p}{\mu_{h1}}-\int_{\alpha_s\left(\mu_{h1}\right)}^{\alpha_s\left(\mu\right)} \frac{d \alpha_s^{\prime}}{\beta\left(\alpha_s^{\prime}\right)}\right)\right]\right) \notag\\
    =\,& \exp \left(-\frac{\Gamma_0}{4 \beta_0^2}\left(\frac{4 \pi}{\alpha_s\left(\mu_{h1}\right)}\left[\ln r'-1+\frac{1}{r'}\right]-\frac{\beta_1}{2 \beta_0} \ln ^2 r' \right.\right.\notag\\
    &\left.\left.+\left(\frac{\Gamma_1}{\Gamma_0}-\frac{\beta_1}{\beta_0}\right)[r'-1-\ln r']\right)\right)\times\left(\frac{n\cdot p}{\mu_{h1}}\right)^{-\frac{\Gamma_0}{2 \beta_0} \ln r'} r'^{-\frac{\gamma_0}{2 \beta_0}} \notag\\
    & \times\left[1-\frac{\alpha_s\left(\mu_{h1}\right)}{4 \pi} \frac{\Gamma_0}{4 \beta_0^2}\left(\frac{\Gamma_2}{2 \Gamma_0}[1-r']^2+\frac{\beta_2}{2 \beta_0}\left[1-r'^2+2 \ln r'\right]\right.\right.\notag\\
    &\left.-\frac{\Gamma_1 \beta_1}{2 \Gamma_0 \beta_0}\left[3-4 r'+r'^2+2 r' \ln r'\right]+\frac{\beta_1^2}{2 \beta_0^2}[1-r'][1-r'-2 \ln r']\right) \notag\\
    &\left.+\frac{\alpha_s\left(\mu_{h1}\right)}{4 \pi}\left(\ln \frac{n\cdot p}{\mu_{h1}}\left(\frac{\Gamma_1}{2 \beta_0}-\frac{\Gamma_0 \beta_1}{2 \beta_0^2}\right)+\frac{\gamma_1}{2 \beta_0}-\frac{\gamma_0 \beta_1}{2 \beta_0^2}\right)[1-r']+\mathcal{O}\left(\alpha_s^2\right)\right],\\
    U_2\left(\mu_{h 2}, \mu\right)  =\,&\operatorname{exp}\left[\int_{\alpha_s\left(\mu_{h 2}\right)}^{\alpha_s(\mu)} d \alpha_s \frac{\tilde{\gamma}\left(\alpha_s\right)}{\beta\left(\alpha_s\right)}\right] \notag\\
     =\,&z^{-\frac{\tilde{\gamma}_0}{2 \beta_0} C_F}\left[1+\frac{\alpha_s\left(\mu_{h 2}\right) C_F}{4 \pi}\left(\frac{\tilde{\gamma}^{(1)}}{2 \beta_0}-\frac{\tilde{\gamma}^{(0)} \beta_1}{2 \beta_0^2}\right)(1-z)+\mathcal{O}\left(\alpha_s^2\right)\right],
\end{align}
where $z=\alpha_s(\mu)/\alpha_s(\mu_{h2})$, $r'=\alpha_s(\mu)/\alpha_s(\mu_{h1})$.

For the scale dependence of the strong coupling constant, we evaluate it with the help of RunDec \cite{Chetyrkin:2000yt, Herren:2017osy, Schmidt:2012az}. 

Finally, the renormalization group improved correlation functions at NLL accuracy are
\begin{align}
        \Pi^A =\,& -\dfrac{1}{2} m_B \left[ U_2 (\mu_{h_2},\mu) \tilde{f}_B(\mu_{h_2})\right] C^{A,(+)}(n\cdot p,\mu) \phi^+_B(\omega,\mu) ,   \notag\\
        \widetilde{\Pi}^A =\,& -\dfrac{1}{2} m_B \left[ U_1(n\cdot p,\mu_{h_1},\mu ) U_2 (\mu_{h_2},\mu)  \right] \tilde{f}_B(\mu_{h_2}) \tilde{C}^{A,(+)} (n\cdot p,\mu_{h_1}) \notag\\
        & \times \int_{0}^{\infty} \dfrac{\md \omega}{\omega - \bar{n}\cdot p} \tilde{J}^{A,(+)} \left( \dfrac{\mu^2}{n\cdot p\, \omega},\dfrac{\omega}{\bar{n}\cdot p} \right) \phi^+_B(\omega,\mu) ,\notag\\
        \Pi^T =\,& \dfrac{1}{4} m_B \left[ U_1(n\cdot p,\mu_{h_1},\mu ) U_2 (\mu_{h_2},\mu)  \right] \tilde{f}_B(\mu_{h_2}) C^{T,(+)} (n\cdot p,\mu_{h_1}) \notag\\
        & \times \int_{0}^{\infty} \dfrac{\md \omega}{\omega - \bar{n}\cdot p} J^{T,(+)} \left( \dfrac{\mu^2}{n\cdot p\, \omega},\dfrac{\omega}{\bar{n}\cdot p} \right) \phi^+_B(\omega,\mu).
\end{align}
\section{The LCSR for \texorpdfstring{$B\to S$}{B -> S} form factors at \texorpdfstring{$\mathcal{O}(\alpha_s)$}{O(as)}}
Applying Borel transformation on both sides of partonic level and hadronic level correlation functions, we derive the one-loop level final expressions for $B\rightarrow S$ form factors
\begin{align}
        f^+_{BS} (q^2) =\,& \dfrac{m_B}{m_S \bar{f}_S}\me^{m^2_S/(n\cdot p\, \omega_M )}\left[U_2 (\mu_{h_2},\mu) \tilde{f}_B(\mu_{h_2}) \right] \notag\\
        &\times \int_{0}^{\omega_s} \md \omega' \me^{-\omega'/{\omega_M}} \bigg[ U_1(n\cdot p,\mu_{h_1},\mu ) \tilde{C}^{A,(+)} (n\cdot p,\mu_{h_1}) \phi_{B,\mathrm{eff}}^+(\omega',\mu)  \notag\\
		&  + \dfrac{n\cdot p-m_B}{m_B} C^{A,(+)} (n\cdot p,\mu) \phi^+_{B}(\omega',\mu) \bigg]  ,\notag\\
        f^-_{BS} (q^2) =\,&\dfrac{n\cdot p}{m_S \bar{f}_S}\me^{m^2_S/(n\cdot p\, \omega_M )}  \left[U_2 (\mu_{h_2},\mu) \tilde{f}_B(\mu_{h_2}) \right] \int_{0}^{\omega_s} \md \omega' \me^{-\omega'/{\omega_M}} \left[ C^{A,(+)} (n\cdot p,\mu) \phi^+_{B}(\omega',\mu) \right],\notag\\
        f^T_{BS} (q^2) =\,& \dfrac{m_B+m_S}{2m_S \bar{f}_S}\me^{m^2_S/(n\cdot p\, \omega_M )} \left[U_2 (\mu_{h_2},\mu) \tilde{f}_B(\mu_{h_2}) \right] \left[U_1(n\cdot p,\mu_{h_1},\mu ) C^{T,(+)} (n\cdot p,\mu_{h_1})  \right] \notag\\
        &\times\int_{0}^{\omega_s} \md \omega' \me^{-\omega'/{\omega_M}} \phi_{B,\mathrm{eff}}^+(\omega',\mu),
\end{align}
where the effective DA is
\begin{align}
        \phi^+_{B,\mathrm{eff}} =\,& \phi^+_B(\omega',\mu) + \dfrac{\alpha_s C_F}{4\pi} \left\{ \int_{0}^{\omega'} \md \omega \left[ \dfrac{2}{\omega-\omega'} \left(\ln \dfrac{\mu^2}{n\cdot p \,\omega'} - 2 \ln \dfrac{\omega'-\omega}{\omega'} \right)_\oplus \right.\right. \notag\\
        & \left. \left.+\left(\dfrac{3}{\omega-\omega'}\right)_\oplus+ \dfrac{2}{\omega} \ln \dfrac{\omega'-\omega}{\omega'} \right] \phi^+_B(\omega) - \int_{\omega'}^{\infty} \md \omega \left[ \ln^2 \dfrac{\mu^2}{n\cdot p \,\omega'} + 3 \ln \dfrac{\mu^2}{n\cdot p \,\omega'}\right. \right. \notag\\
        & \left.\left. -2 \ln \dfrac{\mu^2}{n\cdot p \,\omega'}\ln \dfrac{\omega-\omega'}{\omega} + \dfrac{\pi^2}{6}+5 \right] \dfrac{\md}{\md \omega} \phi^+(\omega,\mu) \right\} .
\end{align}

\section{Numerical analysis}

In this section, we investigate the phenomenological applications of the form factors. We present the shape of the form factors, the differential decay widths, branching ratios, and several angular observables. We begin by discussing the LCDA of  $B$ mesons and then analyze the choice of Borel parameter values, followed by providing the numerical values and errors of the form factors in the large recoil region. Using the $z$-parameterization of the form factors and constraints from the strong coupling constant, we extend the form factors from the small $q^2$ region to the entire kinematically allowed region.

\subsection{Theory inputs}

\begin{table}%[H]
    \caption{The numerical values of the input parameters used in the LCSR calculations for determining the exclusive form factors of $B \rightarrow S$, as well as the subsequent phenomenological analysis for the semileptonic decay observables of bottom mesons.}
     \footnotesize
    \centering
    \begin{tabular}{|c|c c||c|c c|}\hline\hline
     Parameter & Value & Ref. & Parameter & Value & Ref.\\
     \hline\hline
     $G_F$ & $1.166379\times 10^{-5}\, \mathrm{GeV^{-2}}$ & \cite{PhysRevD.66.010001} & $\alpha_s^{(5)}(m_Z)$ & $0.1179\pm 0.0009$ & \cite{PhysRevD.66.010001}\\
     $m_\mu$ & $106.658\, \mathrm{MeV}$ & \cite{PhysRevD.66.010001} & $m_\tau$ & $1776.86\pm 0.12\, \mathrm{MeV}$ & \cite{PhysRevD.66.010001}\\
     \hline\hline
     $\mu=\mu_{hc}$ & $1.5\pm 0.5\,\mathrm{GeV}$ & & $\mu_0$ & $1\,\mathrm{GeV}$ & \\
     $\mu_{h1}$ & $[m_b/2,2 m_b]$ & & $\mu_{h2}$ & $[m_b/2,2 m_b]$ & \\
     $\overline{m}_b(\overline{m}_b)$ & $4.18\pm 0.03\, \mathrm{GeV}$ & \cite{PhysRevD.66.010001} &  &  &  \\
     $m_{B_d}$ & $5279.66\pm 0.12\, \mathrm{MeV}$ & \cite{PhysRevD.66.010001} & $\tau_{B_d}$ & $1.519\pm 0.004\, \mathrm{ps}$ & \cite{PhysRevD.66.010001}\\
     $m_{B_s}$ & $5366.92\pm 0.10\, \mathrm{MeV}$ & \cite{PhysRevD.66.010001} & $\tau_{B_s}$ & $1.527\pm0.011\, \mathrm{ps}$ & \cite{PhysRevD.66.010001}\\
     $f_{B_d}|_{N_f=2+1+1}$ & $190.0\pm1.3\, \mathrm{MeV}$ & \cite{FlavourLatticeAveragingGroupFLAG:2021npn} & $f_{B_s}|_{N_f=2+1+1}$ & $230.3\pm1.3\, \mathrm{MeV}$ & \cite{FlavourLatticeAveragingGroupFLAG:2021npn}\\
     $f_{B1}$ & $288\substack{+25\\-24}\,\mathrm{MeV}$ & \cite{Pullin:2021ebn} & $f_{B_{s1}}$ & $341\substack{+20\\-24}\,\mathrm{MeV}$ & \cite{Pullin:2021ebn}\\
     $f_{B_1}^T$ & $267\substack{+21\\-22}\,\mathrm{MeV}$ & \cite{Pullin:2021ebn} & $f_{B_{s1}}^T$ & $318\substack{+18\\-22}\,\mathrm{MeV}$ & \cite{Pullin:2021ebn}\\
     $m_{B_{s1}(1^+)}$ & $5828.70\pm 0.20\, \mathrm{MeV}$ & \cite{PhysRevD.66.010001} & $m_{B_{d1}(1^+)}$ & $5726.1\pm 1.3\, \mathrm{MeV}$ & \cite{PhysRevD.66.010001}\\
     \hline\hline
     $s_{0\mathrm{min}}$ & $\{4.2, 5.1, 5.1, 6.2\}\ \mathrm{GeV^2}$ &  & $M^2_{\mathrm{min}}$ & $\{4.2, 4.2, 4.2, 4.2\}\ \mathrm{GeV^2}$ & \\
     $s_{0\mathrm{cen}}$ & $\{4.5, 5.4, 5.4, 6.5\}\,\mathrm{GeV^2}$ &  & $M^2_{\mathrm{cen}}$ & $\{4.5, 4.5, 4.5, 4.5\}\, \mathrm{GeV^2}$ & \\
     $s_{0\mathrm{max}}$ & $\{4.8, 5.7, 5.7, 6.8\}\, \mathrm{GeV^2}$ &  & $M^2_{\mathrm{max}}$ & $\{4.8, 4.8, 4.8, 4.8\}\, \mathrm{GeV^2}$ & \\
     \hline\hline
     $\lambda_{B_d}(\mu_0)$ & $350\pm 150\, \mathrm{MeV}$ & \cite{Beneke:2020fot, Wang:2021yrr, Shen:2020hfq, Wang:2018wfj, Wang:2016qii, Janowski:2021yvz} &  & $\{0.7, 6.0\}$ &  \\
      &  &  & $\{\hat{\sigma}_1(\mu_0),\hat{\sigma}_2(\mu_0)\}$ & $\{0.0, \pi^2/6\}$ & \cite{Beneke:2020fot, Shen:2020hfq}  \\
     $\lambda_{B_s}(\mu_0)$ & $400\pm 150\, \mathrm{MeV}$ & \cite{Beneke:2020fot, Shen:2020hfq} &   & $\{-0.7, -6.0\}$ &  \\
     \hline\hline
     $\overline{m}_u(2\mathrm{GeV})$ & $2.20\pm 0.08\, \mathrm{MeV}$ & \cite{PhysRevD.66.010001} & $\overline{m}_d(2\mathrm{GeV})$ & $4.69\pm 0.05\, \mathrm{MeV}$ & \cite{PhysRevD.66.010001}\\
     $\overline{m}_s(2\mathrm{GeV})$ & $93.0\pm 0.6 \, \mathrm{MeV}$ & \cite{PhysRevD.66.010001} & & & \\
     $m_{a_0(1450)}$ & $1474\pm 19\, \mathrm{MeV}$ & \cite{PhysRevD.66.010001} & $m_{K_0^*(1430)}$ & $1425\pm 50\, \mathrm{MeV}$ & \cite{PhysRevD.66.010001}\\
     $m_{f_0(1500)}$ & $1506\pm 6\, \mathrm{MeV}$ & \cite{PhysRevD.66.010001} & $\bar{f}_{a_0(1450)}(\mu_0)$ & $460\pm 50\, \mathrm{MeV}$ & \cite{Cheng:2005nb,PhysRevD.87.114001} \\
     $\bar{f}_{K_0^*(1450)}(\mu_0)$ & $445\pm 50\, \mathrm{MeV}$ & \cite{Cheng:2005nb,PhysRevD.87.114001} & $\bar{f}_{f_0(1500)}(\mu_0)$ & $490\pm 50\, \mathrm{MeV}$ & \cite{Cheng:2005nb,PhysRevD.87.114001}\\
     \hline\hline
    \end{tabular} 
   \label{tab:1}
\end{table}

\begin{figure}[htbp]
    \centering
    \begin{subfigure}[b]{0.4\linewidth}
    \includegraphics[width=\linewidth]{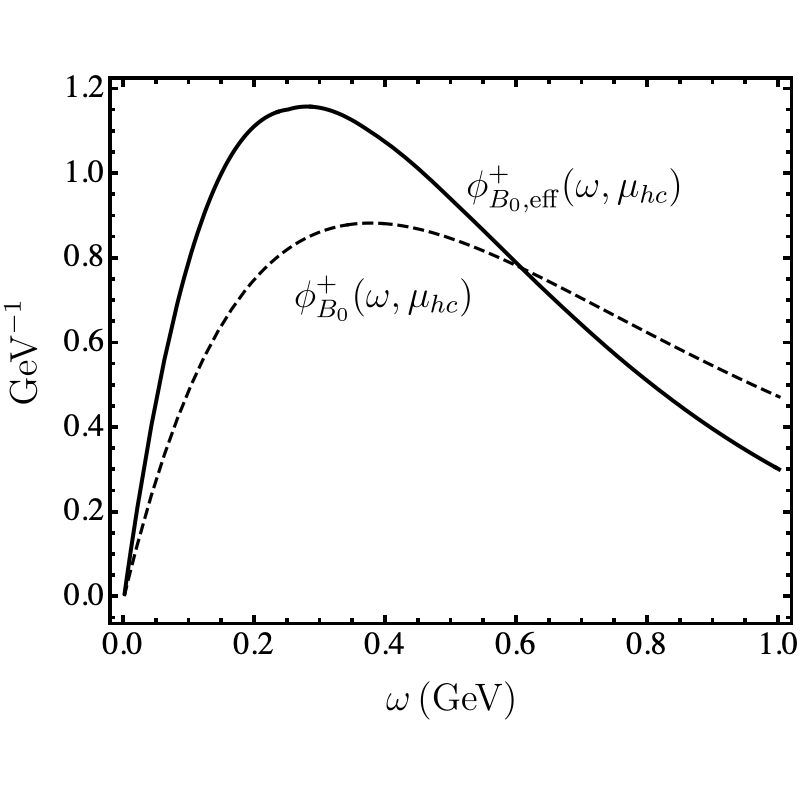}
    \end{subfigure}
    \begin{subfigure}[b]{0.4\linewidth}
    \includegraphics[width=\linewidth]{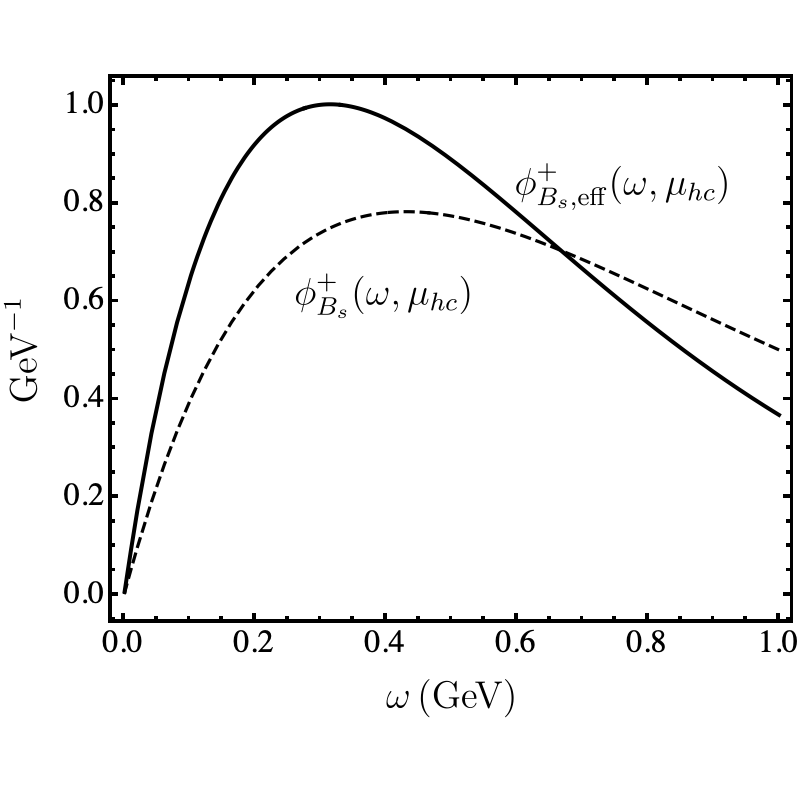}
    \end{subfigure}
    \caption{Three parameters modeled  $B$ meson LCDA with input parameters $\lambda_{B_0}(\mu_0)=350\,\mathrm{MeV}$ (left panel), $\lambda_{B_s}(\mu_0)=450\,\mathrm{MeV}$ (right panel), $\{\hat{\sigma}_1(\mu_0),\hat{\sigma}_2(\mu_0)\}=\{0.0, \pi^2/6\}$ and corresponding effective LCDA $\phi_{B_0,\mathrm{eff}}$ and $\phi_{B_s,\mathrm{eff}}$ at one-loop order.}
    \label{fig:LCDA}
\end{figure}

In the following, we will apply the viable phenomenological models for the DAs of $B$ meson within the framework of heavy quark effective theory, which are expected to satisfy nontrivial constraints stemming from the equations of motion, as well as the appropriate asymptotic behavior at small quark momentum,   derived from the conformal symmetry analysis. Specifically, the three-parameter model \cite{Beneke:2018wjp} (refer to \cite{Feldmann:2022uok} for an alternative parametrization involving an expansion based on associated Laguerre polynomials) is used for DAs of the  $B$ meson
\begin{equation}
    \begin{aligned}
        \phi_B^+(\omega,\mu)=\,&U_{\phi}(\mu,\mu_0)\dfrac{1}{\omega^{p+1}}\dfrac{\Gamma(\beta)}{\Gamma(\alpha)}\mathcal{G}(\omega;0,2,1),
    \end{aligned}
\end{equation}
where the function $\mathcal{G}(\omega;0,2,1)$ is Meijer G function \cite{luke1969special}
\begin{equation}
    \begin{aligned}
        \mathcal{G}(\omega;l,m,n)\equiv G_{23}^{21}\left(\dfrac{\omega}{\omega_0}\Big|\substack{1,\beta+l\\p+m,\alpha,p+n}\right),
    \end{aligned}
\end{equation}
the parameters $p=\dfrac{\Gamma^{(0)}_{\mathrm{cusp}}}{2\beta_0}\mathrm{ln}[\alpha_s(\mu)/\alpha(\mu_0)]$, and the evolution factor at one-loop order is \cite{Braun:2017liq, Braun:2015pha}
\begin{align}
        U_{\phi}(\mu,\mu_0)&=\mathrm{exp}\Bigg\{-\dfrac{\Gamma_{\mathrm{cusp}}^{(0)}}{4\beta_0^2}\bigg(\dfrac{4\pi}{\alpha_s(\mu_0)}\bigg[\mathrm{ln}\,r''-1+\dfrac{1}{r''}\bigg]\notag\\
        -\dfrac{\beta_1}{2\beta_0}&\mathrm{ln}^2\, r''+\left(\dfrac{\Gamma_{\mathrm{cusp}}^{(1)}}{\Gamma_{\mathrm{cusp}}^{(0)}}-\dfrac{\beta_1}{\beta_0}\right)[r''-1-\mathrm{ln}\,r'']\bigg)\Bigg\}\left(\me^{2\gamma_E}\mu_0\right)^{\dfrac{\Gamma_{\mathrm{cusp}}^{(0)}}{2\beta_0}\mathrm{ln}\,r''}{r^{\prime\prime}}^{\dfrac{\gamma_{t2}^{(0)}}{2\beta_0}},
\end{align}
where $r''=\alpha_s(\mu)/\alpha_s(\mu_0)$.
\\
For the leading-twist  $B$ meson DAs, the following inverse moments are defined for convenience
\begin{align}
        \dfrac{1}{\lambda_B(\mu)}=\,&\int_0^\infty \dfrac{d\omega}{\omega}\phi_B^+(\omega,\mu),\notag\\
        \dfrac{\hat{\sigma}_n(\mu)}{\lambda_B(\mu)}=\,&\int_0^\infty\dfrac{d\omega}{\omega}\mathrm{ln}^n\dfrac{\mathrm{e}^{-\gamma_E}\lambda_B(\mu)}{\omega}\phi_B^+(\omega,\mu).
\end{align}
By employing the standard definitions of inverse logarithmic moments for the leading-twist DA of the  $B$ meson, we can readily express these essential non-perturbative parameters in relation to the three shape parameters in our model as follows \cite{Beneke:2018wjp, Shen:2020hfq, Beneke:2020fot}
\begin{align}
        \lambda_B(\mu_0)=\,&\dfrac{\alpha-1}{\beta-1}\omega_0,\notag\\
        \hat{\sigma}_1(\mu_0)=\,&\psi(\beta-1)-\psi(\alpha-1)+\mathrm{ln}\dfrac{\alpha-1}{\beta-1},\notag\\
        \hat{\sigma}_2(\mu_0)=\,&\hat{\sigma}_1^2(\mu)+\psi'(\alpha-1)-\psi'(\beta-1)+\dfrac{\pi^2}{6},
\end{align}
where $\psi(x)$ is the digamma  function.
The employed values for $\lambda_B$, $\hat{\sigma}_1$ and $\hat{\sigma}_2$ are shown in Table \ref{tab:1}.
The shape of the LCDA and $\phi_{B,\mathrm{eff}}(\omega')$ is shown in Figure \ref{fig:LCDA}.

\begin{figure}[htbp]
\centering
\begin{subfigure}[b]{0.4\linewidth}
\includegraphics[width=\linewidth]{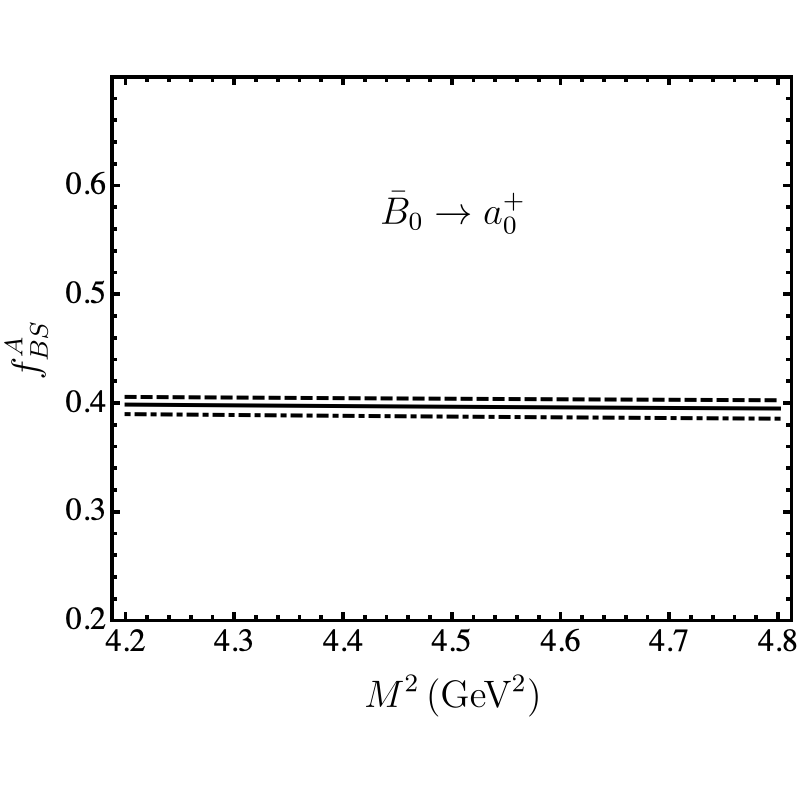}
\end{subfigure}
\begin{subfigure}[b]{0.4\linewidth}
\includegraphics[width=\linewidth]{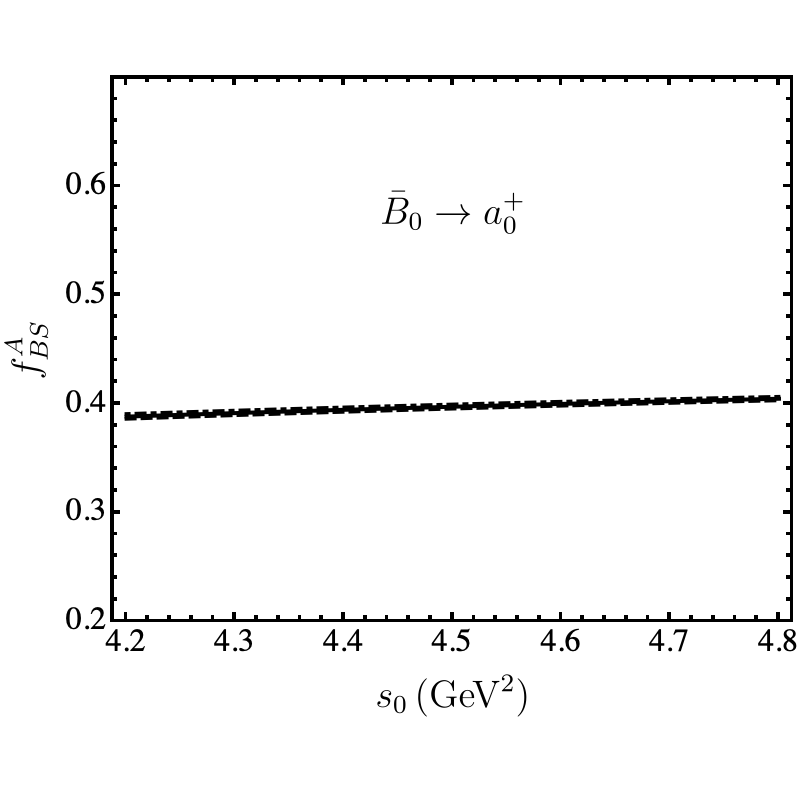}
\end{subfigure}
\caption{The Borel parameter dependence (left panel) and effective threshold dependence (right panel) of the renormalization group-improved $\bar{B}_0\rightarrow a_0^+(1450)\ell\bar{\nu}_\ell$ form factor, calculated using LCSR up to NLL accuracy, are presented. The solid, dotted, and dashed lines correspond to $s_0=s_{0\mathrm{min}}, s_{0\mathrm{cen}}, s_{0\mathrm{max}}$ (left panel) and $M^2=M^2_{\mathrm{min}}, M^2_{\mathrm{cen}}, M^2_{\mathrm{max}}$ (right panel), respectively.}
\label{fig:4}
\end{figure}

Regarding the choice of Borel parameter and effective threshold, we follow the criteria stated in \cite{Wang:2015vgv}. The final range of values for them we obtained is shown in Table~\ref{tab:1}. From Figure \ref{fig:4}, we can observe that the form factors vary smoothly with the change in Borel parameters and effective thresholds in this region. The choice of the effective threshold coincides with \cite{Lu:2006fr}. In principle, the effective threshold is not a property of the scalar meson but is dependent on the LCSR and the specific process. However, due to our use of the same unitarity relation, the consistency of our obtained effective thresholds with the results in \cite{Lu:2006fr} can serve as a rough verification.

Furthermore, we will consider the variation range of the matching scales $\mu_{h1}$ and $\mu_{h2}$ to be from $m_b/2$ to $2\,m_b$. This approach is widely employed in the realm of exclusive heavy-hadron decay processes \cite{Beneke:2018wjp, Shen:2020hfq, Beneke:2020fot}.

\subsection{Numerical results of form factors}

Substituting these parameters into the form factor expression obtained from the LCSR, we obtain the value of the form factors at $q^2 = 0$ shown in Table \ref{tab:ffs}.

\begin{table}%[H]
    \caption{Numerical results for the $B\rightarrow S$ transition form factors at the large recoil point $q^2=0$.}
    \centering
    \begin{tabular}{|c|c|c|c|c|}
    \hline\hline
     Processes & Methods & $f_+(q^2=0)$ & $f_-(q^2=0)$ & $f_T(q^2=0)$ \\
     \hline\hline
     $\bar{B}_0\rightarrow a_0^+(1450)$ & This work & 0.40(10) & -0.39(10) & 0.54(13) \\
     & LCSR\cite{Wang:2008da} & 0.52 & -0.44 & 0.66 \\
     & LCSR\cite{PhysRevD.83.025024} & 0.53 & -0.53 &  \\
     & pQCD\cite{Li:2008tk} & 0.68 &  & 0.92 \\
     \hline\hline
     $\bar{B}_0\rightarrow \bar{K}_0^*(1430)$ & This work & 0.43(10) & -0.42(10) & 0.58(13) \\
     & LCSR\cite{Wang:2008da} & 0.49 & -0.41 & 0.60 \\
     & LCSR\cite{PhysRevD.83.025024} & 0.49 & -0.49 & 0.69 \\
     & QCDSR\cite{Yang:2005bv} & 0.31 & -0.31 & -0.26 \\
     & pQCD\cite{Li:2008tk} & 0.60 &  & 0.78 \\
     \hline\hline
     $\bar{B}_s\rightarrow K_0^{*+}(1430)$ & This work & 0.47(11) & -0.46(11) & 0.64(15) \\
     & LCSR\cite{Wang:2008da} & 0.42 & -0.34 & 0.52 \\
     & LCSR\cite{PhysRevD.83.025024} & 0.44 & -0.44 &  \\
     & QCDSR\cite{Aliev:2007rq} & 0.24 &  &  \\
     & pQCD\cite{Li:2008tk} & 0.56 &  & 0.72 \\
     \hline\hline
     $\bar{B}_s\rightarrow f_0(1500)$ & This work & 0.45(9) & -0.44(9) & 0.61(13) \\
     & LCSR\cite{Wang:2008da} & 0.43 & -0.37 & 0.56 \\
     & LCSR\cite{PhysRevD.83.025024} & 0.41 & -0.41 & 0.59 \\
     & pQCD\cite{Li:2008tk} & 0.60 &  & 0.82 \\
     \hline\hline
    \end{tabular} 
    \label{tab:ffs}
\end{table}

\begin{figure}%[htbp]
\centering
\begin{subfigure}[b]{0.4\linewidth}
\includegraphics[width=\linewidth]{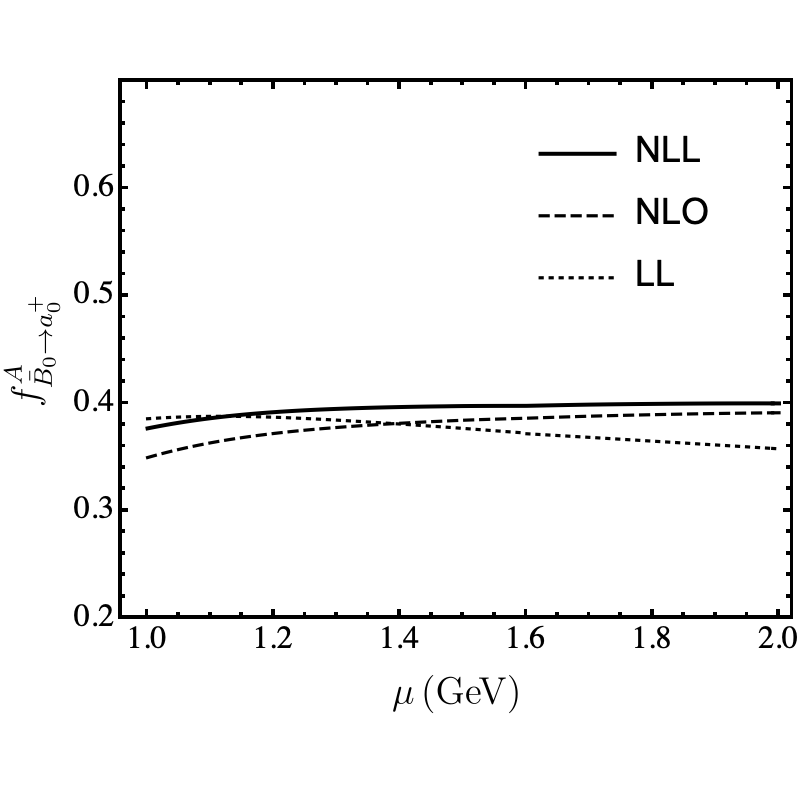}
\end{subfigure}
\begin{subfigure}[b]{0.4\linewidth}
\includegraphics[width=\linewidth]{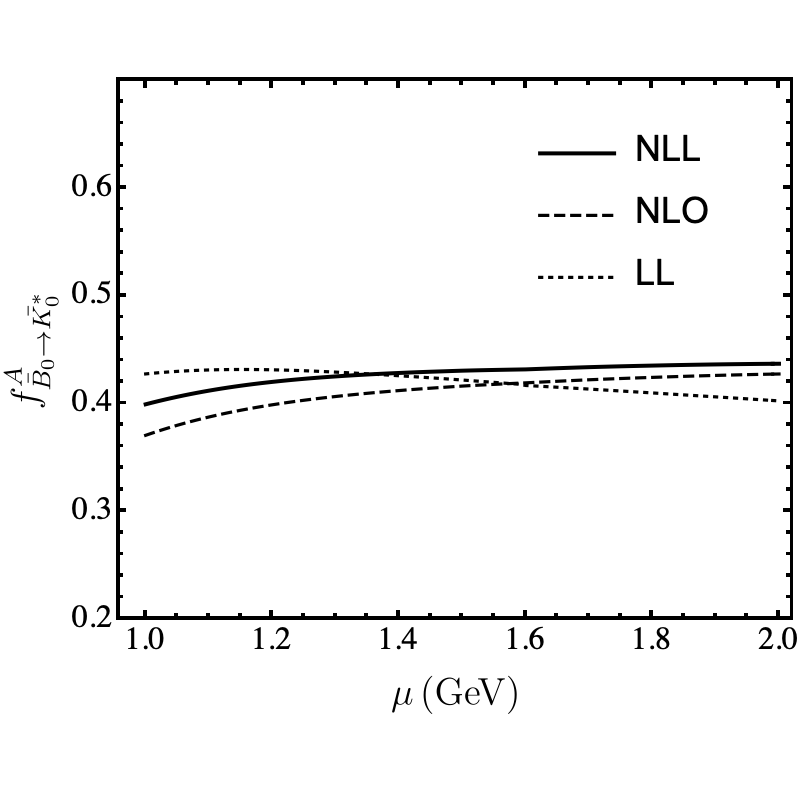}
\end{subfigure}
\begin{subfigure}[b]{0.4\linewidth}
\includegraphics[width=\linewidth]{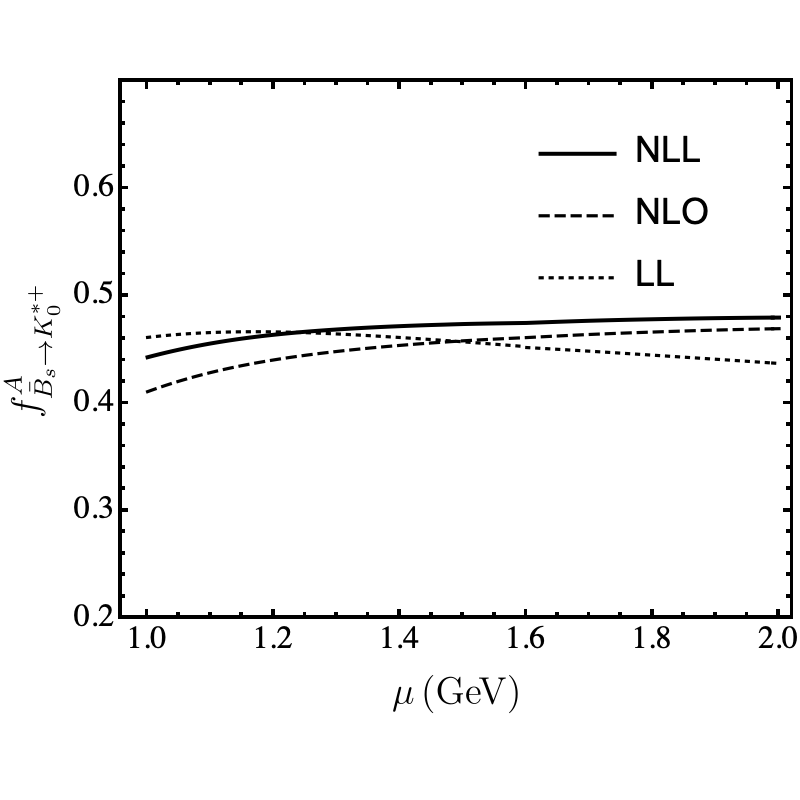}
\end{subfigure}
\begin{subfigure}[b]{0.4\linewidth}
\includegraphics[width=\linewidth]{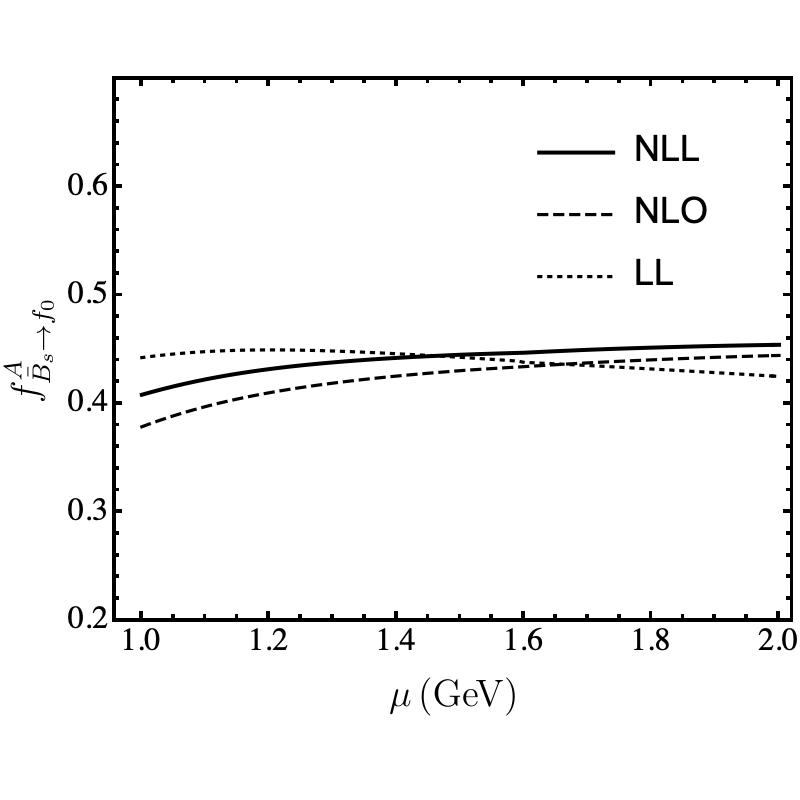}
\end{subfigure}
\caption{Factorization scale dependence of the form factor $f_{BS}^A$ for $B\rightarrow S$ transition.}
\label{fig:fpscale}
\end{figure}

Now we investigate the dependence of the factorization scale at leading logarithm (LL) and NLL accuracies. The result at LL accuracy can be obtained by setting the precision of the evolution function of the renormalization equation to be at $\mathcal{O}(1)$. Due to the cancellation of the one-loop corrections between the hard functions and the jet functions, the NLL corrections relative to the LL accuracy results are only about 5\%.  From Figure \ref{fig:fpscale}, it is evident that the high precision of NLL reduces the scale dependence of the result, thereby enhancing the reliability of the LCSR method. 

The predictions of the form factors obtained from LCSR are relatively more accurate at small $q^2$, but there is a large error at large $q^2$ \cite{Khodjamirian:2006st}. To obtain the numerical values of the form factors in the entire kinematically allowed region, i.e., the shape of the form factors, it is widely recognized that the analyticity, crossing symmetry, unitarity, and the asymptotic behaviors of form factors \cite{Hill:2005ju, Boyd:1997qw, Bourrely:2005hp, Bourrely:2008za} impose strong constraints on the $q^2$ behavior of the form factors. Using these constraints, we derive the BCL parameterization based on the $z$-series expansion \cite{Bourrely:2008za}
\begin{align}
    f_{BS}^{A,T}&=\dfrac{1}{1-q^2/m_{B_{q1}}^2}\sum_{k=0}^{k=N-1}b_k^{A,T}\left[z(q^2,t_0)^k-(-1)^{k-N}\dfrac{k}{N}z(q^2,t_0)^N\right],\notag\\
    f_{BS}^{P}&=\dfrac{1}{1-q^2/m_{B_q}^2}\sum_{k=0}^{k=N-1}b_k^{P} z(q^2,t_0)^k ,
\end{align}
where $m_{B_{q1}}$ and $m_{B_q}$ correspond to the masses of the lowest resonances of  $B$ mesons with $J^P=1^+$ and $J^P=0^-$, respectively. Through our concrete fitting procedure, we have found that $N=3$ is sufficient to provide satisfactory fitting results. It is worth mentioning that the BCL parameterization satisfies the scaling behavior $f^A_{BS}(q^2)\sim 1/q^2$ at $|q^2|\rightarrow\infty$ predicted by perturbative QCD \cite{Akhoury:1994tnu}.  We have summarized the masses of these resonances in Table \ref{tab:1}. 

The variable $z$ is defined as a conformal transformation \cite{Bourrely:2008za}
\begin{equation}
    \begin{aligned}
        z(q^2,t_0)=\dfrac{\sqrt{t_+-q^2}-\sqrt{t_+-t_0}}{\sqrt{t_+-q^2}+\sqrt{t_+-t_0}},
    \end{aligned}
\end{equation}
where $t_+=(m_B+m_S)^2$ is the threshold parameter for $B\rightarrow S$ process. The $z$-parameterization maps the entire cut-$q^2$ plane to a unit disk. The value of the free parameter $t_0<t_+$ determines the mapping of $q^2$ to the origin when $q^2$ takes the value of $t_0$. To minimize the range of $z$ under kinematic constraints, the value of $t_0$ can be chosen as follows
\begin{equation}
    t_0=t_+-\sqrt{t_+\left[t_+-(m_B-m_S)^2\right]}.
\end{equation}

We proceed to determine the values of the unspecified parameters in the model by fitting them to accurately predicted data points. Since lattice QCD (LQCD) can provide relatively accurate predictions of the form factors near large $q^2$, the usual approach is to perform a joint fit of the LCSR and lattice data points. As there are currently no LQCD results available for the $B\rightarrow S$ form factors, we are only able to constrain the error of the form factors at large $q^2$ using the strong coupling constant. This constant (for the axial form factor, $g_{B_{q1}BS}$) is related to the form factors through the dispersion relation
\begin{equation}
    \begin{aligned}
        f_{BS}^A(q^2)=\dfrac{g_{B_{q1}BS}f_{B_{q1}}}{2m_{B_{q1}}(1-q^2/m_{B_{q1}}^2)}+\dfrac{1}{\pi}\int_{s_A}^\infty ds \dfrac{\im\ f_{BS}^A(s)}{s-q^2}.
    \end{aligned}
\end{equation}
By calculating the residue of $f_{BS}^A(q^2)$ at the pole $m_{B_{q1}}^2$, we derive the constraint from the strong coupling constant for the axial form factor
\begin{equation}
    g_{B_{q1}BS}=\dfrac{2m_{B_{q1}}}{f_{B_{q1}}}\lim_{q^2\rightarrow m_{B_{q1}}^2}\left[(1-q^2/m_{B_{q1}}^2)f^A_{BS}(q^2)\right].
\end{equation}
The strong coupling constraints for the other form factors are similar with the dispersion relations presented in Appendix \ref{appendix:a}. Since no literature currently provides the strong couplings we need, we calculate them ourselves. According to \cite{Khodjamirian:2020mlb}, we show the scheme for calculating the strong couplings with LCSR in Appendix \ref{appendix:b}.

\begin{figure}%[htbp]
\centering
\begin{subfigure}[b]{0.4\linewidth}
\includegraphics[width=\linewidth]{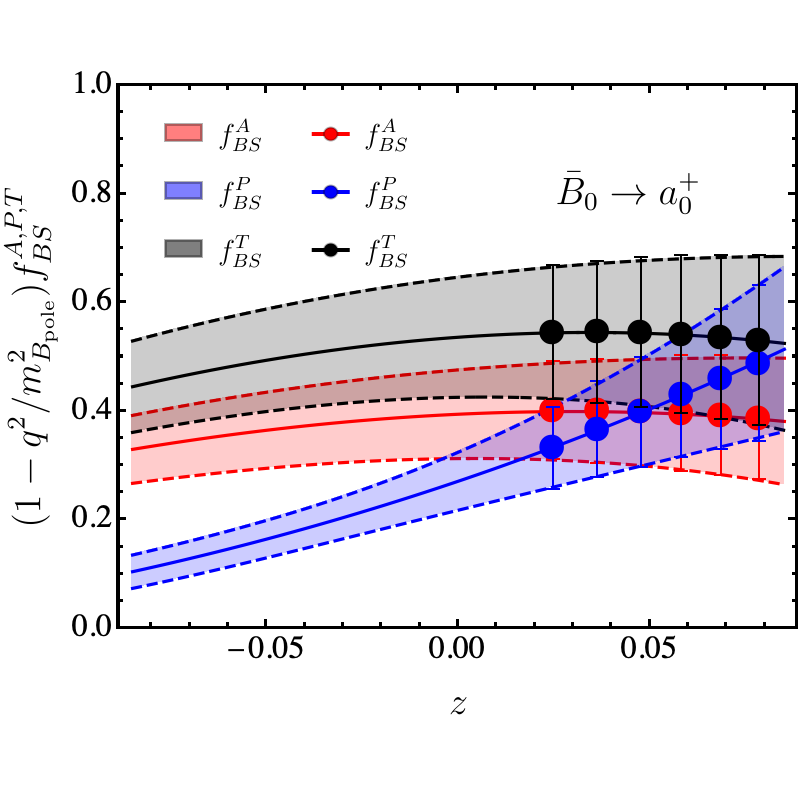}
\end{subfigure}
\begin{subfigure}[b]{0.4\linewidth}
\includegraphics[width=\linewidth]{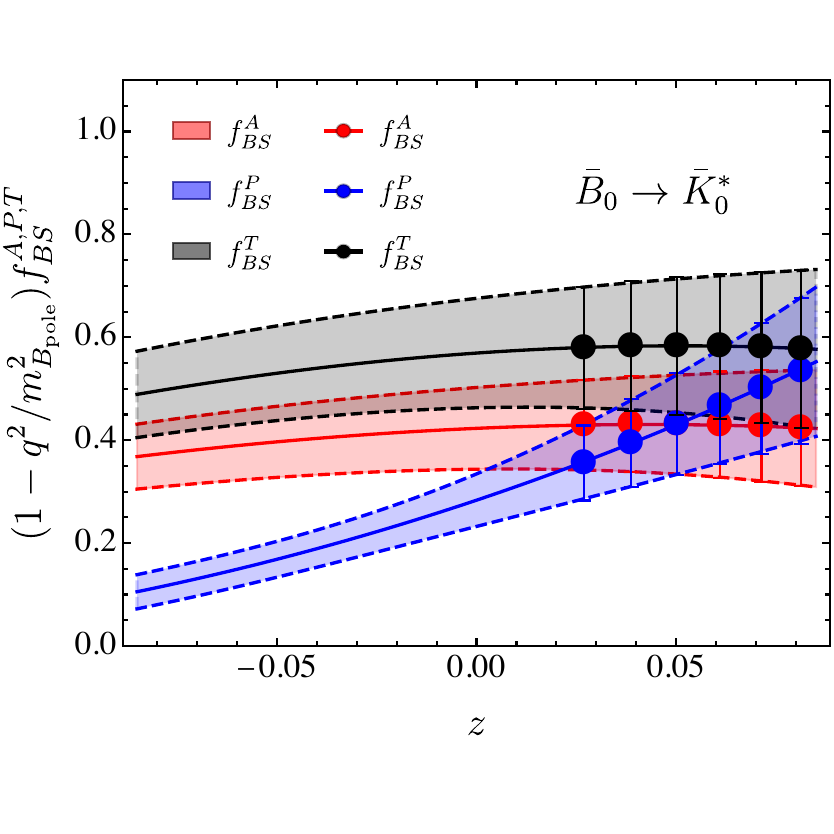}
\end{subfigure}
\begin{subfigure}[b]{0.4\linewidth}
\includegraphics[width=\linewidth]{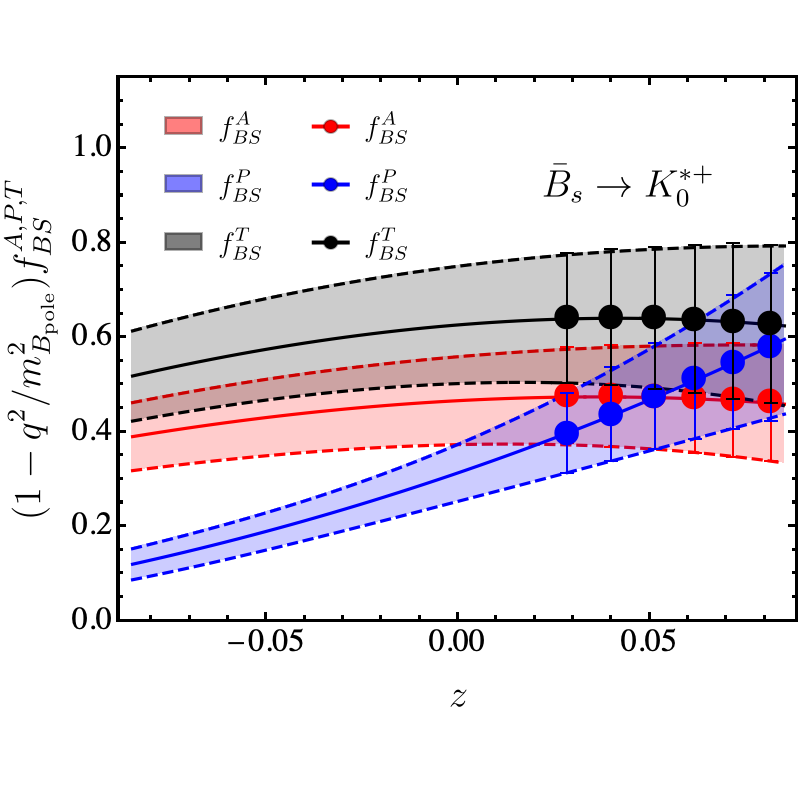}
\end{subfigure}
\begin{subfigure}[b]{0.4\linewidth}
\includegraphics[width=\linewidth]{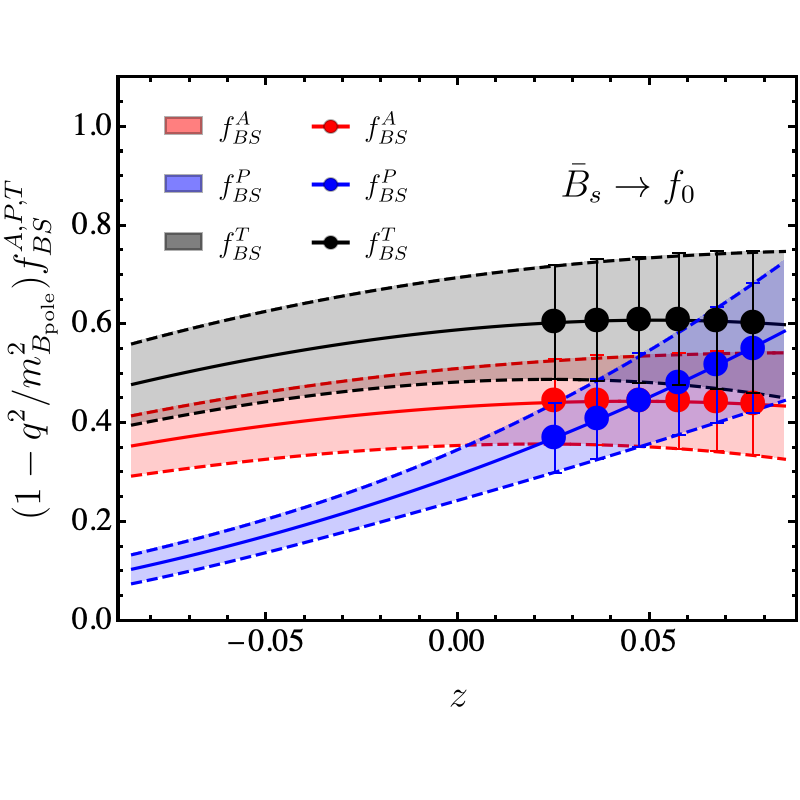}
\end{subfigure}
\caption{The shape of form factors $f_{BS}^A(z)$, $f_{BS}^P(z)$ and $f_{BS}^T(z)$ from fitting the BCL parameters for $B\rightarrow S$. }
\label{fig:bclfitz}
\end{figure}

\begin{figure}%[htbp]
\centering
\begin{subfigure}[b]{0.4\linewidth}
\includegraphics[width=\linewidth]{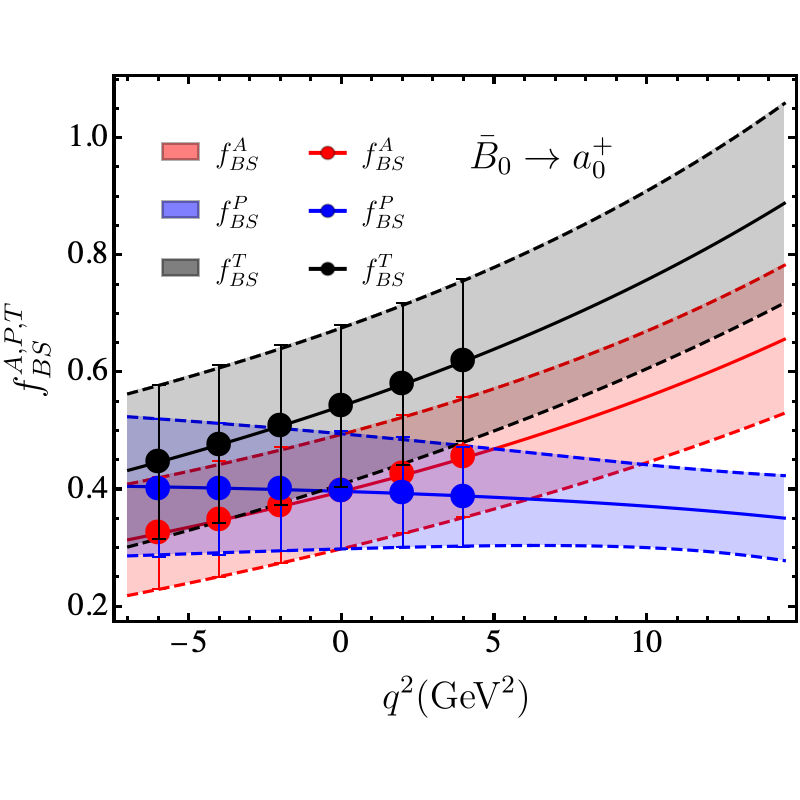}
\end{subfigure}
\begin{subfigure}[b]{0.4\linewidth}
\includegraphics[width=\linewidth]{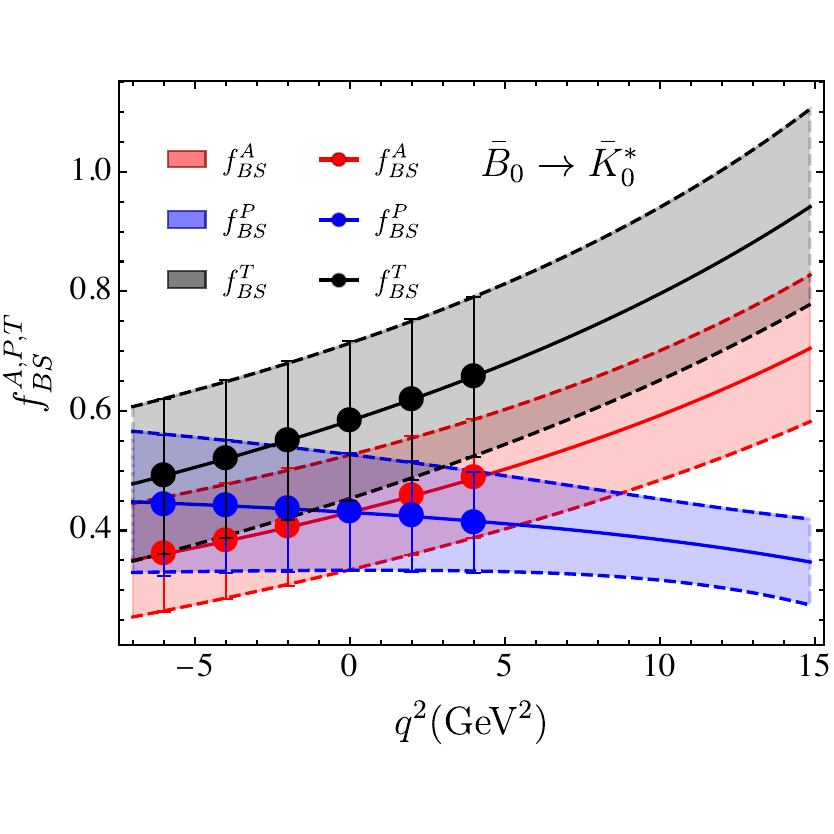}
\end{subfigure}
\begin{subfigure}[b]{0.4\linewidth}
\includegraphics[width=\linewidth]{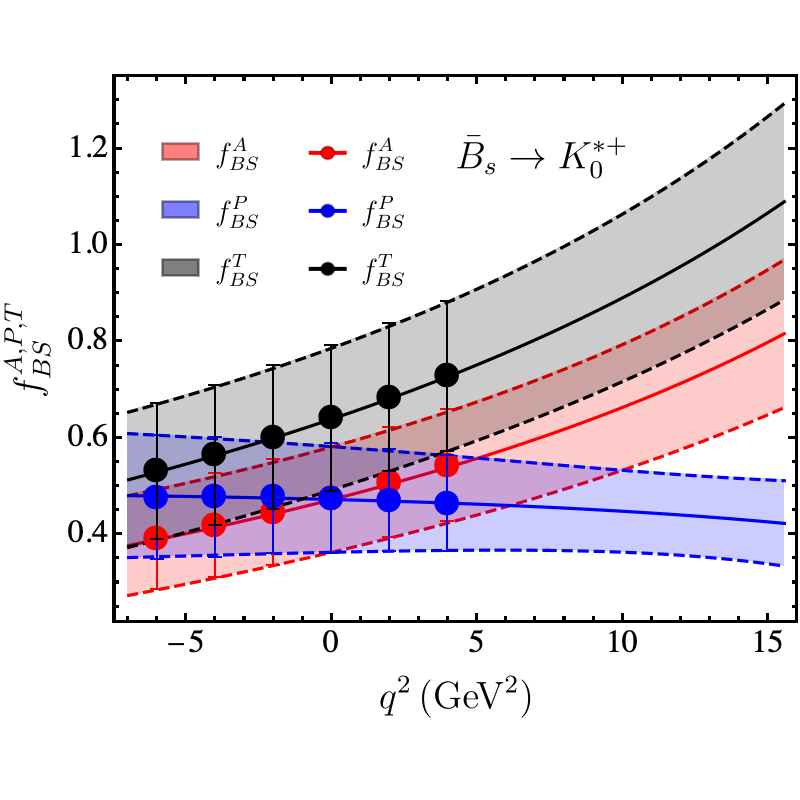}
\end{subfigure}
\begin{subfigure}[b]{0.4\linewidth}
\includegraphics[width=\linewidth]{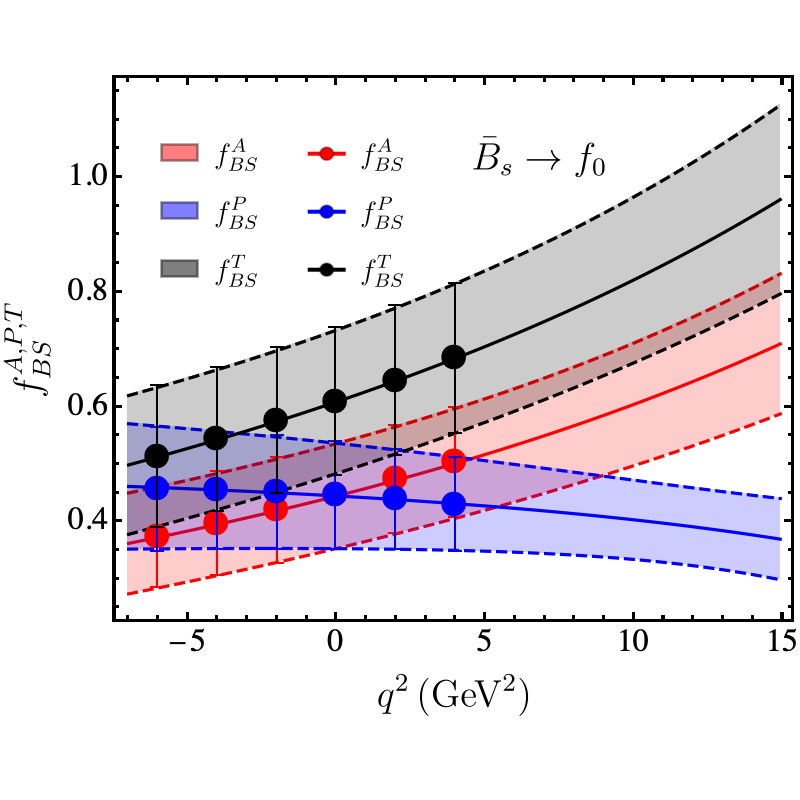}
\end{subfigure}
\caption{The shape of form factors $f_{BS}^A(q^2)$, $f_{BS}^P(q^2)$ and $f_{BS}^T(q^2)$ from fitting the BCL parameters for $B\rightarrow S$. }
\label{fig:bclfitqq}
\end{figure}

We now proceed to perform the $\chi^2$ fitting of the BCL parameters. Using the six data points obtained from LCSR corresponding to $q^2=-6,-4,-2,0,2,4\,\mathrm{GeV^2}$, and taking into account the constraints from the strong coupling constant, a total of seven data points are used. It should be noted that the fitting is subject to the condition $f_{BS}^A(0) = f_{BS}^P(0)$. For the LCSR calculations, there exist strong correlations among the form factors. To preserve the information on correlation during the fitting of the BCL parameters, we performed a combined fitting of the three form factors. The combined fitting allows us to obtain fitting results for the nine BCL parameters that exhibit strong correlations. These strong correlations impose stringent constraints on the shape of the form factors. Consequently, in the future, when combining the results from LCSR calculations with LQCD data, we can provide form factor results with smaller uncertainties in the large recoil region \cite{Cui:2022zwm}. The dependencies of the form factors on the variable $z$ and the momentum transfer $q^2$ obtained from the combined fitting are shown in Figure \ref{fig:bclfitz} and Figure \ref{fig:bclfitqq}, respectively. The numerical results of the combined fitting for 9 BCL parameters are presented in Tables \ref{tab:bclpara1} -- \ref{tab:bclpara4},     corresponding to the fitting results for the four respective processes. The values of $\chi^2$ corresponding to our fitting results are all around 0.05, indicating that the parameterization form we adopted is in good agreement with the results obtained from LCSR.

\begin{table}[H]
     \caption{Fitting results of the BCL parameters for the  $\bar{B}_0\rightarrow a_0^+(1450)$ transition form factors.}
   \footnotesize
    \centering
    \begin{tabular}{|c|c||c c c c c c c c c|}
    \hline\hline
    \multicolumn{2}{|c||}{$\bar{B}_0\rightarrow a_0^+(1450)$ Form Factors} & \multicolumn{9}{|c|}{Correlation Matrix}\\
    \hline
    Parameters & Values & $b_0^A$ & $b_1^A$ & $b_2^A$ & $b_0^P$ & $b_1^P$ & $b_2^P$ & $b_0^T$ & $b_1^T$ & $b_2^T$ \\
    \hline\hline
 $b_0^A$ & 0.398(98) & 1.000 & 0.636 & 0.634 & 0.990 & 0.646 & 0.644 & 0.991 & 0.640 & 0.653 \\
 $b_1^A$ & 0.828(1200) & \text{} & 1.000 & 1.000 & 0.646 & 0.563 & 0.563 & 0.648 & 0.559 & 0.565 \\
 $b_2^A$ & -13.568(6664) & \text{} & \text{} & 1.000 & 0.644 & 0.562 & 0.562 & 0.646 & 0.559 & 0.565 \\
 $b_0^P$ & 0.544(133) & \text{} & \text{} & \text{} & 1.000 & 0.638 & 0.636 & 0.990 & 0.640 & 0.653 \\
 $b_1^P$ & 0.928(1203) & \text{} & \text{} & \text{} & \text{} & 1.000 & 1.000 & 0.648 & 0.560 & 0.566 \\
 $b_2^P$ & -13.031(6683) & \text{} & \text{} & \text{} & \text{} & \text{} & 1.000 & 0.647 & 0.559 & 0.565 \\
 $b_0^T$ & 0.398(98) & \text{} & \text{} & \text{} & \text{} & \text{} & \text{} & 1.000 & 0.641 & 0.653 \\
 $b_1^T$ & 6.007(1054) & \text{} & \text{} & \text{} & \text{} & \text{} & \text{} & \text{} & 1.000 & 1.000 \\
 $b_2^T$ & 13.232(8838) & \text{} & \text{} & \text{} & \text{} & \text{} & \text{} & \text{} & \text{} & 1.000 \\
    \hline\hline
    \end{tabular} 
    \label{tab:bclpara1}
\end{table}

\begin{table}[H]
    \caption{Fitting results of the BCL parameters for the  $\bar{B}_0\rightarrow \bar{K}_0^*(1430)$ transition form factors.}
    \footnotesize
    \centering
    \begin{tabular}{|c|c||c c c c c c c c c|}
    \hline\hline
    \multicolumn{2}{|c||}{$\bar{B}_0\rightarrow \bar{K}_0^*(1450)$ Form Factors} & \multicolumn{9}{|c|}{Correlation Matrix}\\
    \hline
    Parameters & Values & $b_0^A$ & $b_1^A$ & $b_2^A$ & $b_0^P$ & $b_1^P$ & $b_2^P$ & $b_0^T$ & $b_1^T$ & $b_2^T$ \\
    \hline\hline
 $b_0^A$ & 0.430(96) & 1.000 & 0.633 & 0.577 & 0.990 & 0.637 & 0.595 & 0.991 & 0.616 & 0.634 \\
 $b_1^A $& 0.793(1116) & \text{} & 1.000 & 0.993 & 0.644 & 0.573 & 0.552 & 0.644 & 0.561 & 0.569 \\
 $b_2^A$ & -8.882(5324) & \text{} & \text{} & 1.000 & 0.590 & 0.547 & 0.529 & 0.590 & 0.535 & 0.543 \\
 $b_0^P$ & 0.583(130) & \text{} & \text{} & \text{} & 1.000 & 0.632 & 0.589 & 0.990 & 0.618 & 0.636 \\
 $b_1^P$ & 0.929(1113) & \text{} & \text{} & \text{} & \text{} & 1.000 & 0.996 & 0.640 & 0.560 & 0.568 \\
 $b_2^P$ & -8.641(5365) & \text{} & \text{} & \text{} & \text{} & \text{} & 1.000 & 0.598 & 0.540 & 0.548 \\
 $b_0^T$ & 0.430(97) & \text{} & \text{} & \text{} & \text{} & \text{} & \text{} & 1.000 & 0.618 & 0.636 \\
 $b_1^T$ & 6.058(953) & \text{} & \text{} & \text{} & \text{} & \text{} & \text{} & \text{} & 1.000 & 1.000 \\
 $b_2^T$ & 14.648(7353) & \text{} & \text{} & \text{} & \text{} & \text{} & \text{} & \text{} & \text{} & 1.000 \\
    \hline\hline
    \end{tabular} 
    \label{tab:bclpara2}
\end{table}

\begin{table}[H]
    \caption{Fitting results of the BCL parameters for the  $\bar{B}_s\rightarrow K_0^{*+}(1430)$ transition form factors.}
    \footnotesize
    \centering
    \begin{tabular}{|c|c||c c c c c c c c c|}
    \hline\hline
    \multicolumn{2}{|c||}{$\bar{B}_s\rightarrow K_0^{*+}(1430)$ Form Factors} & \multicolumn{9}{|c|}{Correlation Matrix}\\
    \hline
    Parameters & Values & $b_0^A$ & $b_1^A$ & $b_2^A$ & $b_0^P$ & $b_1^P$ & $b_2^P$ & $b_0^T$ & $b_1^T$ & $b_2^T$ \\
    \hline\hline
 $b_0^A$ & 0.473(109) & 1.000 & 0.590 & 0.531 & 0.990 & 0.599 & 0.553 & 0.991 & 0.580 & 0.587 \\
 $b_1^A$ & 0.916(1005) & \text{} & 1.000 & 0.993 & 0.604 & 0.496 & 0.473 & 0.604 & 0.485 & 0.488 \\
 $b_2^A$ & -12.218(5674) & \text{} & \text{} & 1.000 & 0.547 & 0.468 & 0.448 & 0.547 & 0.458 & 0.461 \\
 $b_0^P$ & 0.640(147) & \text{} & \text{} & \text{} & 1.000 & 0.590 & 0.543 & 0.990 & 0.582 & 0.589 \\
 $b_1^P$ & 1.035(1002) & \text{} & \text{} & \text{} & \text{} & 1.000 & 0.996 & 0.601 & 0.485 & 0.488 \\
 $b_2^P$ & -11.918(5701) & \text{} & \text{} & \text{} & \text{} & \text{} & 1.000 & 0.555 & 0.463 & 0.466 \\
 $b_0^T$ & 0.473(110) & \text{} & \text{} & \text{} & \text{} & \text{} & \text{} & 1.000 & 0.578 & 0.585 \\
 $b_1^T$ & 5.869(850) & \text{} & \text{} & \text{} & \text{} & \text{} & \text{} & \text{} & 1.000 & 1.000 \\
 $b_2^T$ & 12.96(7695) & \text{} & \text{} & \text{} & \text{} & \text{} & \text{} & \text{} & \text{} & 1.000 \\
    \hline\hline
    \end{tabular} 
    \label{tab:bclpara3}
\end{table}

\begin{table}[H]
    \caption{Fitting results of the BCL parameters for the  $\bar{B}_s\rightarrow f_0(1500)$ transition form factors.}
     \footnotesize   
    \centering
    \begin{tabular}{|c|c||c c c c c c c c c|}
    \hline\hline
    \multicolumn{2}{|c||}{$\bar{B}_s\rightarrow f_0(1500)$ Form Factors} & \multicolumn{9}{|c|}{Correlation Matrix}\\
    \hline
    Parameters & Values & $b_0^A$ & $b_1^A$ & $b_2^A$ & $b_0^P$ & $b_1^P$ & $b_2^P$ & $b_0^T$ & $b_1^T$ & $b_2^T$ \\
    \hline\hline
 $b_0^A$ & 0.445(91) & 1.000 & 0.550 & 0.550 & 0.990 & 0.564 & 0.564 & 0.991 & 0.556 & 0.573 \\
 $b_1^A$ & 1.125(950) & \text{} & 1.000 & 1.000 & 0.564 & 0.475 & 0.475 & 0.565 & 0.471 & 0.478 \\
 $b_2^A$ & -11.935(5264) & \text{} & \text{} & 1.000 & 0.564 & 0.475 & 0.475 & 0.565 & 0.471 & 0.478 \\
 $b_0^P$ & 0.609(125) & \text{} & \text{} & \text{} & 1.000 & 0.554 & 0.554 & 0.990 & 0.557 & 0.574 \\
 $b_1^P$ & 1.224(952) & \text{} & \text{} & \text{} & \text{} & 1.000 & 1.000 & 0.566 & 0.471 & 0.479 \\
 $b_2^P$ & -11.383(5275) & \text{} & \text{} & \text{} & \text{} & \text{} & 1.000 & 0.566 & 0.471 & 0.479 \\
 $b_0^T$ & 0.445(92) & \text{} & \text{} & \text{} & \text{} & \text{} & \text{} & 1.000 & 0.555 & 0.572 \\
 $b_1^T$ & 6.306(834) & \text{} & \text{} & \text{} & \text{} & \text{} & \text{} & \text{} & 1.000 & 1.000 \\
 $b_2^T$ & 15.51(6995) & \text{} & \text{} & \text{} & \text{} & \text{} & \text{} & \text{} & \text{} & 1.000 \\
    \hline\hline
    \end{tabular} 
   \label{tab:bclpara4}
\end{table}

\begin{figure}[htbp]
\centering
\begin{subfigure}[b]{0.4\linewidth}
\includegraphics[width=\linewidth]{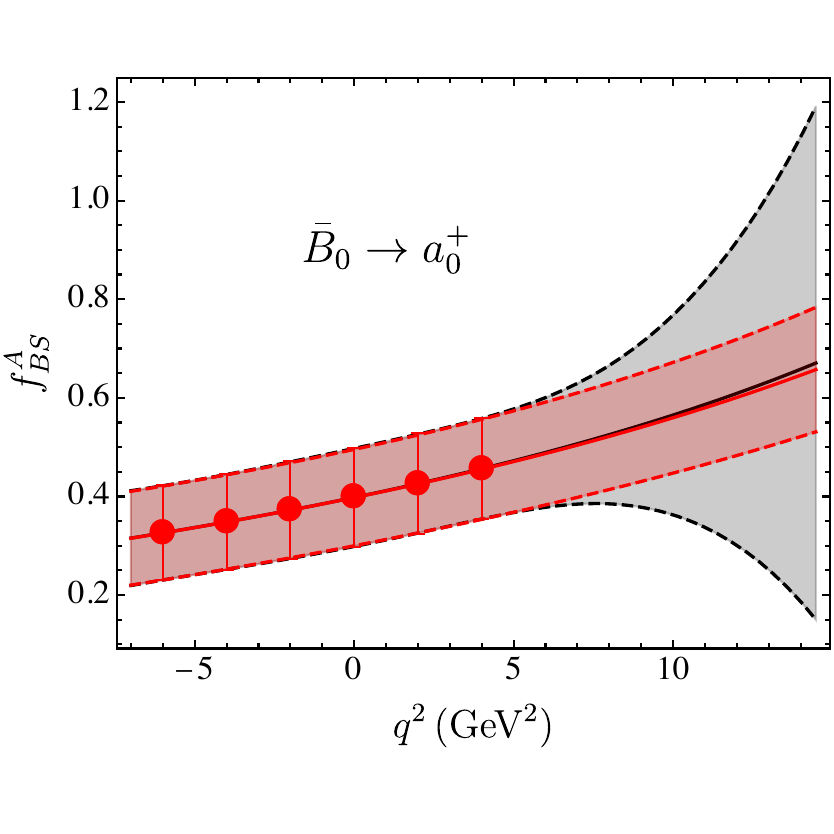}
\end{subfigure}
\caption{The shape of form factor $f_{BS}^A(q^2)$ from fitting the BCL parameters for $B\rightarrow a_0 l\bar{\nu}_l$, where the red curve and area correspond to the shape and error after adding the strong coupling constant constraint, while the black curve and grey area correspond to the fitting results of the data points given only by LCSR.}
\label{fig:BCLcontrast}
\end{figure}

The strong coupling constants provide constraints on the form factors at $q^2 = m_{B_{\mathrm{pole}}}^2$. From the fitting results shown in Figure \ref{fig:BCLcontrast}, it is evident that the constraint of the strong coupling constants effectively reduces the uncertainties of the form factors in the large $q^2$ region.

\subsection{Phenomenological analysis of \texorpdfstring{$B\to S\ell\bar{\nu}_\ell$}{B -> S l nu} observables}

In this section, we utilize the fitting results of the form factors obtained through the BCL parameterization to conduct phenomenological studies of processes $B\to S\ell\bar{\nu}_\ell$. Starting from the general results of decay angular distributions, we provide the differential decay widths, branching ratios, lepton-flavor universality ratios, and several new physics sensitive observables related to angular distributions. These observables include forward-backward asymmetries, ``flat terms" and lepton polarization asymmetries.

The effective Hamiltonian of the $b\rightarrow u$ transition is 
\begin{equation}
    \mathcal{H}_{\mathrm{eff}}(b\rightarrow u\ell\bar{\nu}_\ell)=\dfrac{G_F}{\sqrt{2}}V_{ub}\bar{u}\gamma_\mu(1-\gamma_5)b \bar{\ell}\gamma^\mu(1-\gamma_5)\nu_\ell+h.c.,
\end{equation}
from which we can derive the angular distributions for $B\rightarrow Sl\bar{\nu}_l$ in the  $B$ meson rest frame
\begin{align}
\dfrac{d \Gamma(B \rightarrow S \ell \bar{\nu}_\ell)}{d q^2 d \cos \theta_\ell}=\,& \left(\frac{q^2-m_\ell^2}{q^2}\right)^2 \frac{ \sqrt{\lambda} G_F^2\left|V_{ub}\right|^2}{256 m_B^3 \pi^3} \notag\\
 & \times\left\{\frac{m_\ell^2}{q^2}\left|\cos \theta_\ell \sqrt{\lambda} f_{BS}^A+\left(m_B^2-m_S^2\right) f_{BS}^P\right|^2+\sin ^2 \theta_\ell\,\lambda \left|f_{BS}^A\left(q^2\right)\right|^2\right\}, \notag\\
 =\,& \left(\dfrac{q^2-m_\ell^2}{q^2}\right)^2\dfrac{\sqrt{\lambda}G_F^2|V_{ub}|^2}{256 m_B^3 \pi^3}\left(a_{\theta_\ell}(q^2)+b_{\theta_\ell}(q^2)\cos\theta_\ell+c_{\theta_\ell}(q^2)\cos^2\theta_\ell\right),
\end{align}
with the three coefficient functions and $\lambda$ defined as
\begin{align}
    a_{\theta_\ell}(q^2)=\,&\lambda\left|f_{BS}^A(q^2)\right|^2+\dfrac{m_\ell^2}{q^2}(m_B^2-m_S^2)\left|f_{BS}^P(q^2)\right|^2,\\
    b_{\theta_\ell}(q^2)=\,&2\dfrac{m_\ell^2}{q^2}\sqrt{\lambda}(m_B^2-m_S^2)\mathrm{Re}\left[f_{BS}^A(q^2)f_{BS}^{P*}(q^2)\right],\\
    c_{\theta_\ell}(q^2)=\,&\left(\dfrac{m_\ell^2}{q^2}\sqrt{\lambda}-\lambda\right)\left|f_{BS}^A(q^2)\right|^2,\\
    \lambda=\,&(m_B^2-q^2-m_S^2)^2-4q^2 m_S^2.
\end{align}
The angle $\theta_\ell$ is defined as the angle between the momentum of the final-state lepton $\ell^-$ and the momentum of the final-state scalar meson.

\begin{figure}%[htbp]
\centering
\begin{subfigure}[b]{0.4\linewidth}
\includegraphics[width=\linewidth]{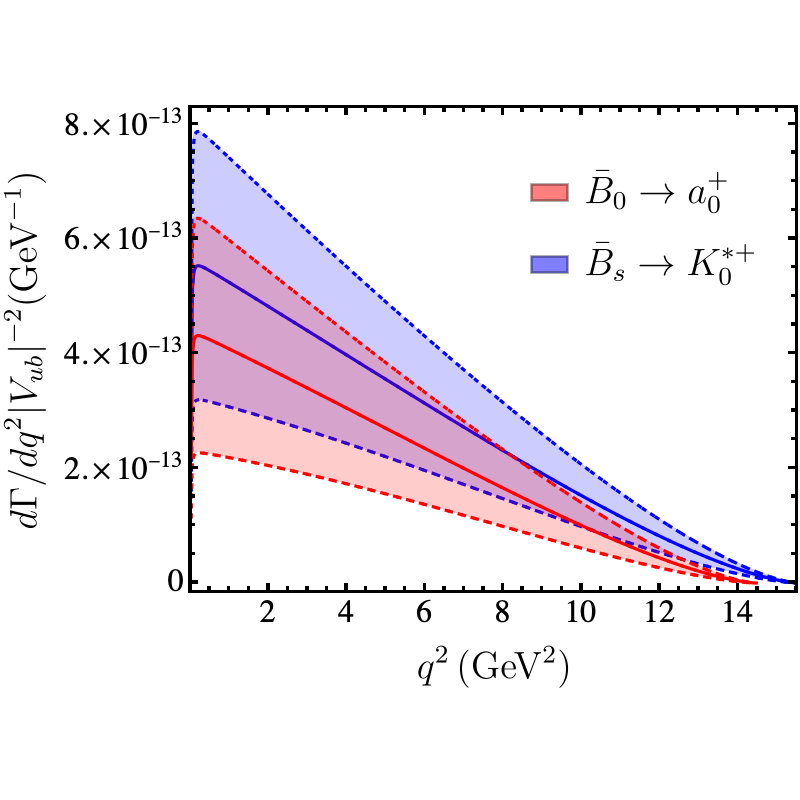}
\end{subfigure}
\begin{subfigure}[b]{0.4\linewidth}
\includegraphics[width=\linewidth]{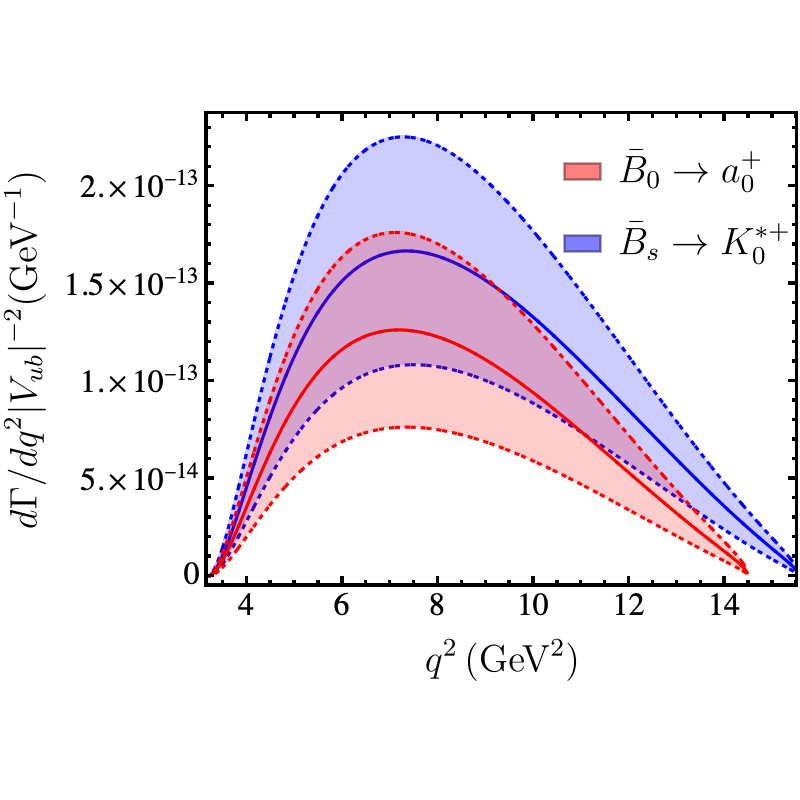}
\end{subfigure}
\caption{The differential decay width of $B\rightarrow S \ell\bar{\nu}_\ell.$}
\label{fig:dGammaSlnu}
\end{figure}

By integrating out the angle $\theta_\ell$, we can obtain the differential decay width easily
\begin{align}
    \dfrac{d \Gamma(B \rightarrow S \ell \bar{\nu}_\ell)}{d q^2}=\,&\left(\dfrac{q^2-m_\ell^2}{q^2}\right)^2\dfrac{\sqrt{\lambda}G_F^2|V_{ub}|^2}{256 m_B^3 \pi^3} \cdot2\left[a_{\theta_\ell}(q^2)+\dfrac{1}{3}c_{\theta_\ell}(q^2)\right]\notag\\
    =\,& \left(\frac{q^2-m_\ell^2}{q^2}\right)^2 \frac{\sqrt{\lambda} G_F^2\left|V_{ub}\right|^2}{384 m_B^3 \pi^3} \cdot \frac{1}{q^2} \notag\\
    & \times\left\{\left(m_\ell^2+2 q^2\right) \lambda \left|f_{BS}^A\left(q^2\right)\right|^2+3 m_\ell^2\left(m_B^2-m_S^2\right)^2 \left|f_{BS}^P\left(q^2\right)\right|^2\right\}.
\end{align}
By comparing with the experimental result, we are able to extract the CKM matrix element $\left|V_{ub}\right|$ from this observable. However, due to the current lack of experimental data, we only provide $d\mathrm{\Gamma}(B\rightarrow S \ell \bar{\nu}_\ell)/dq^2\,\left|V_{ub}\right|^{-2}$, as shown in Figure \ref{fig:dGammaSlnu}.

From the angular distribution, we can also obtain two observables relative to electroweak symmetry breaking. They are the forward-backward asymmetry and the ``flat term"
\begin{align}
    \mathcal{A}_{\mathrm{FB}}^{B\rightarrow S\ell\bar{\nu}_\ell}(q^2)=\,&\left[\dfrac{d\mathrm{\Gamma}(B\rightarrow S\ell\bar{\nu}_\ell)}{dq^2}\right]^{-1}\int_{-1}^1 d\cos\theta_\ell\, \mathrm{sgn}(\cos\theta_\ell)\dfrac{d^2\mathrm{\Gamma}(B\rightarrow S\ell\bar{\nu}_\ell)}{dq^2d\cos\theta_\ell}\notag\\
    =\,& \left[\dfrac{1}{2}b_{\theta_\ell}(q^2)\right]:\left[a_{\theta_\ell}(q^2)+\dfrac{1}{3}c_{\theta_\ell}(q^2)\right],\\
    \mathcal{F}_{\mathrm{H}}^{B\rightarrow S\ell\bar{\nu}_\ell}(q^2)=\,&1+\dfrac{2}{3} \left[\dfrac{d\mathrm{\Gamma}(B\rightarrow S\ell\bar{\nu}_\ell)}{dq^2}\right]^{-1} \dfrac{d^2}{d(\cos\theta_\ell)^2}\dfrac{d^2\mathrm{\Gamma}\left(B\rightarrow S\ell\bar{\nu}_\ell\right)}{dq^2d\cos\theta_\ell}\notag\\
    =\,& \left[a_{\theta_\ell}(q^2)+c_{\theta_\ell}(q^2)\right]:\left[a_{\theta_\ell}(q^2)+\dfrac{1}{3}c_{\theta_\ell}(q^2)\right].
\end{align}
They vanish under limit $m_\ell\rightarrow 0$.

\begin{figure}[htbp]
\centering
\begin{subfigure}[b]{0.4\linewidth}
\includegraphics[width=\linewidth]{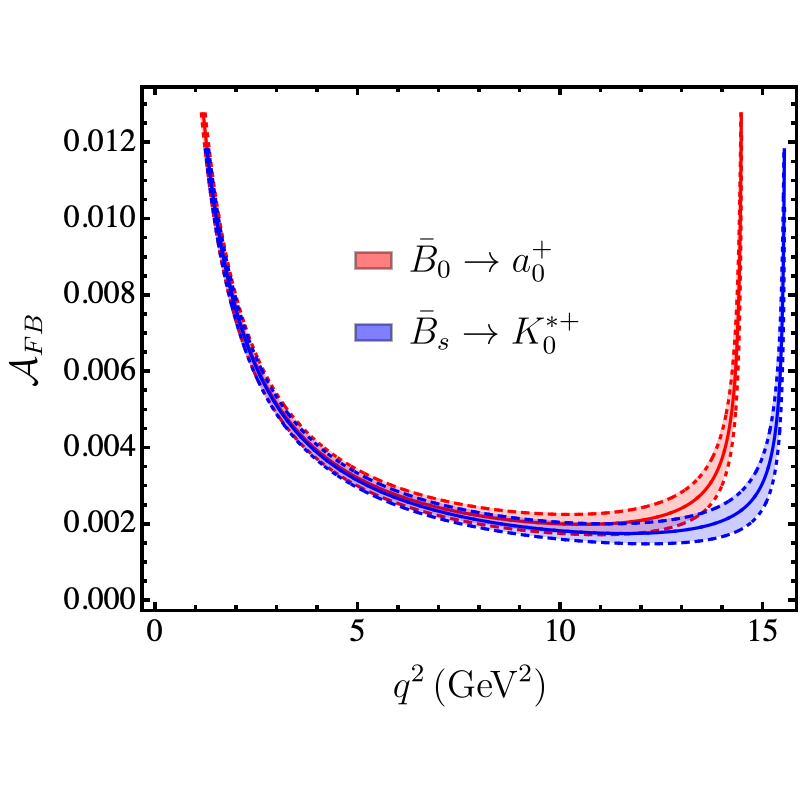}
\end{subfigure}
\begin{subfigure}[b]{0.4\linewidth}
\includegraphics[width=\linewidth]{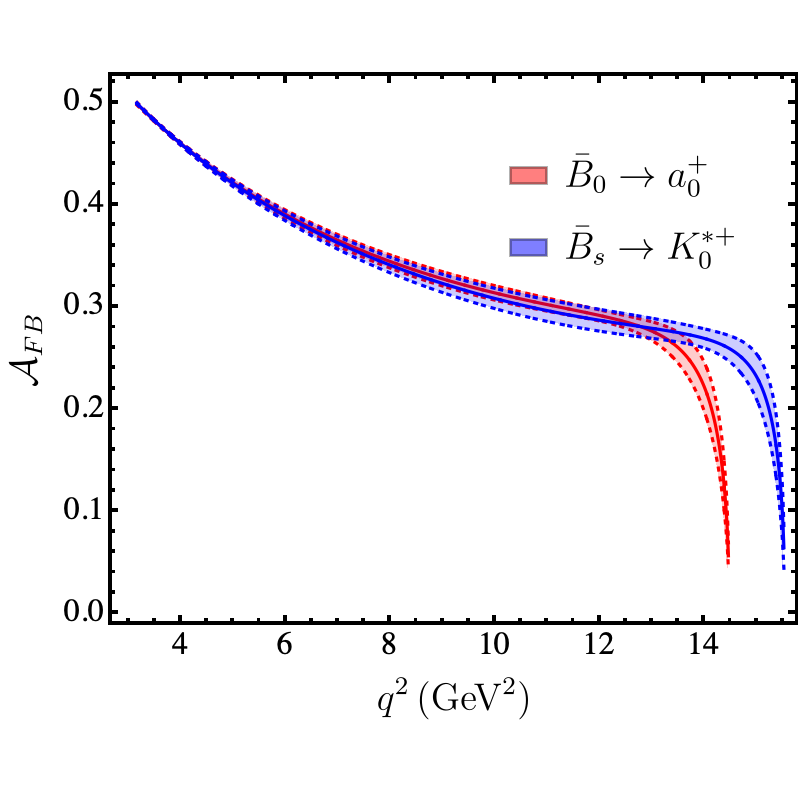}
\end{subfigure}
\begin{subfigure}[b]{0.4\linewidth}
\includegraphics[width=\linewidth]{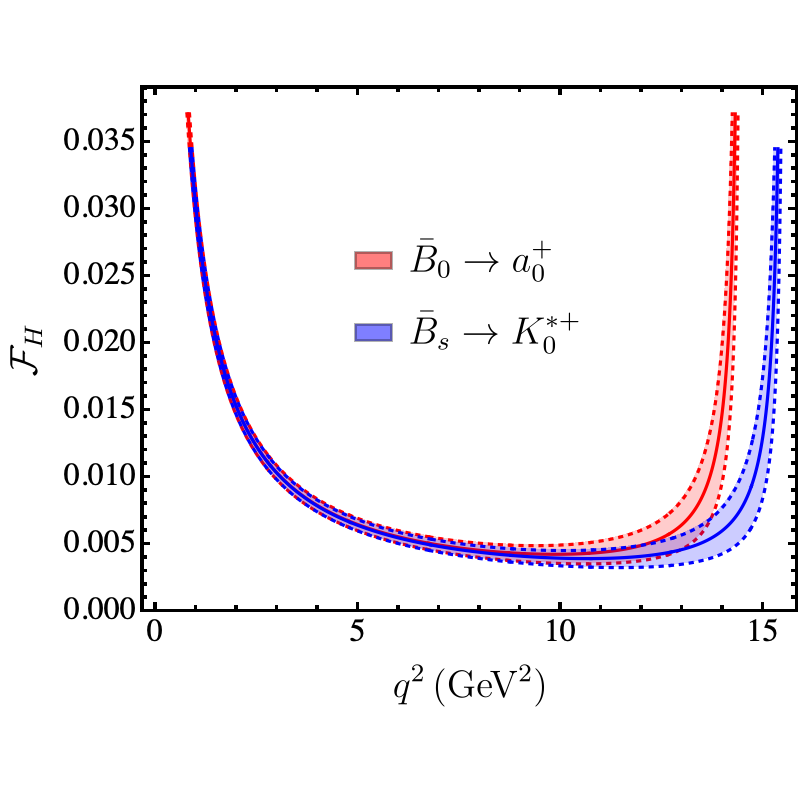}
\end{subfigure}
\begin{subfigure}[b]{0.4\linewidth}
\includegraphics[width=\linewidth]{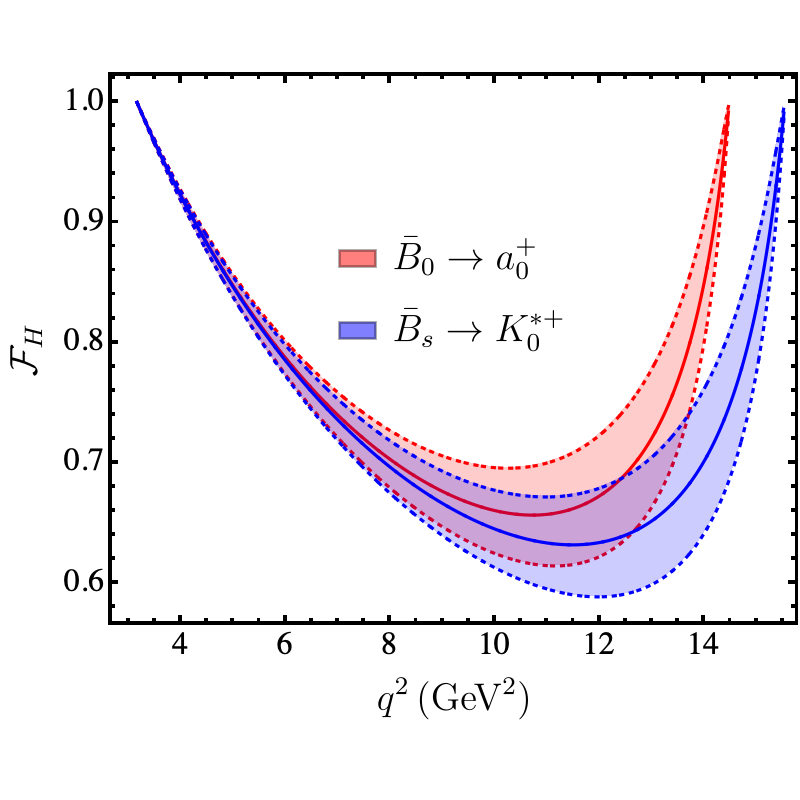}
\end{subfigure}
\begin{subfigure}[b]{0.4\linewidth}
\includegraphics[width=\linewidth]{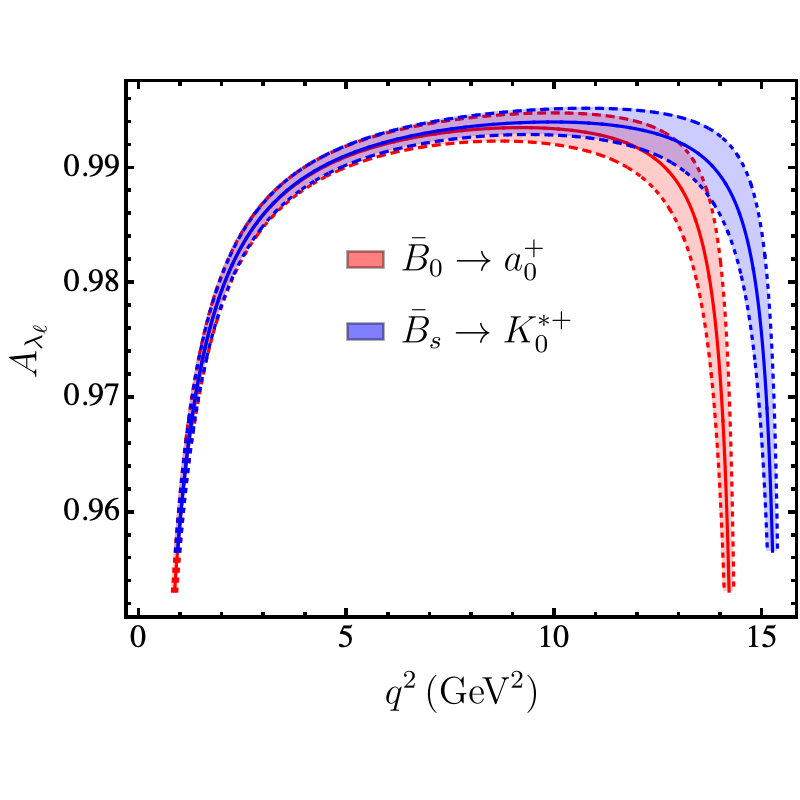}
\end{subfigure}
\begin{subfigure}[b]{0.4\linewidth}
\includegraphics[width=\linewidth]{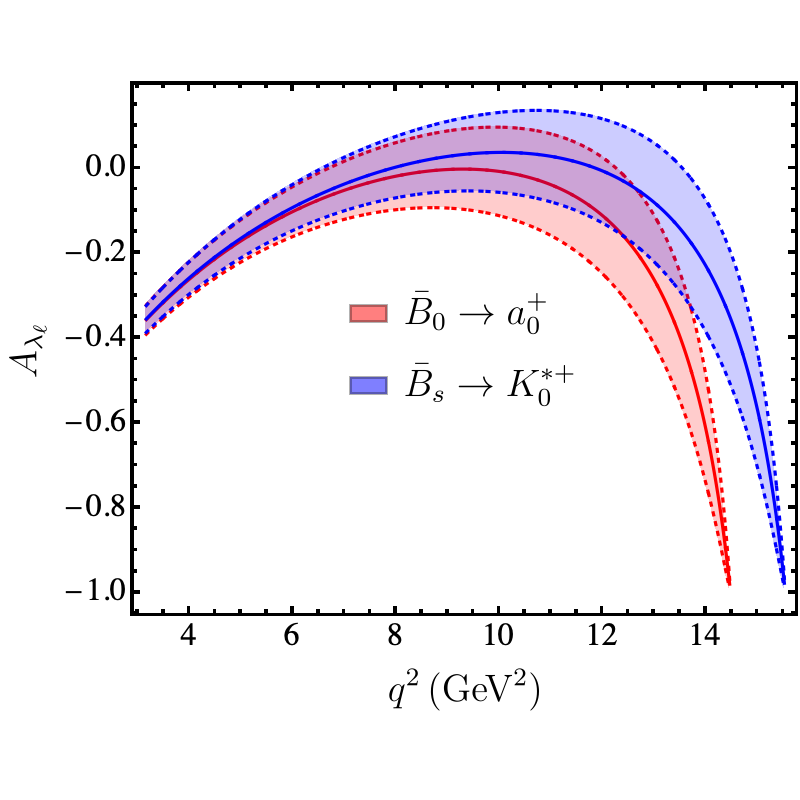}
\end{subfigure}
\caption{Theoretical predictions for three categories of integrated observables derived from the combined BCL $z$-series expansion fitting of $B\rightarrow S\ell\bar{\nu}_\ell$ form factors obtained via LCSR.}
\label{fig:observableslnu}
\end{figure}

Another angular observable we are concerned with is the polarization asymmetry associated with helicity-violating new physics interactions
\begin{align}
    \mathcal{A}_{\lambda_\ell}^{B\rightarrow S\ell\bar{\nu}_\ell}(q^2)=\,& \left[\dfrac{d\mathrm{\Gamma}(B\rightarrow Sl\bar{\nu}_\ell)}{dq^2}\right]^{-1} \left[\dfrac{d\mathrm{\Gamma}^{\lambda_\ell=-1/2}}{dq^2}-\dfrac{d\mathrm{\Gamma}^{\lambda_\ell=+1/2}}{dq^2}\right]\left(B\rightarrow S\ell\bar{\nu}_\ell\right)\notag\\
    =\,&1-\dfrac{2}{3}\left\{\left[3\left(a_{\theta_\ell}(q^2)+c_{\theta_\ell}(q^2)\right)+\dfrac{2m_\ell^2}{q^2-m_\ell^2}c_{\theta_\ell}(q^2)\right]:\left[a_{\theta_\ell}(q^2)+\dfrac{1}{3}c_{\theta_\ell}(q^2)\right]\right\},
\end{align}
The numerical results of these three observables, which are sensitive to new physics, obtained from the BCL fitting incorporating LCSR and constraints from the strong coupling constant, are presented in Figure \ref{fig:observableslnu}.

We also provide integrated observables corresponding to these measurements. These results, along with the branching ratios, are presented in Table \ref{tab:observableslnu}. Additionally, for convenience and comparison purposes, we have included the results from other relevant studies in Table \ref{tab:observableslnu}. These results need to be verified in future experiments.

\begin{figure}[htbp]
\centering
\begin{subfigure}[b]{0.4\linewidth}
\includegraphics[width=\linewidth]{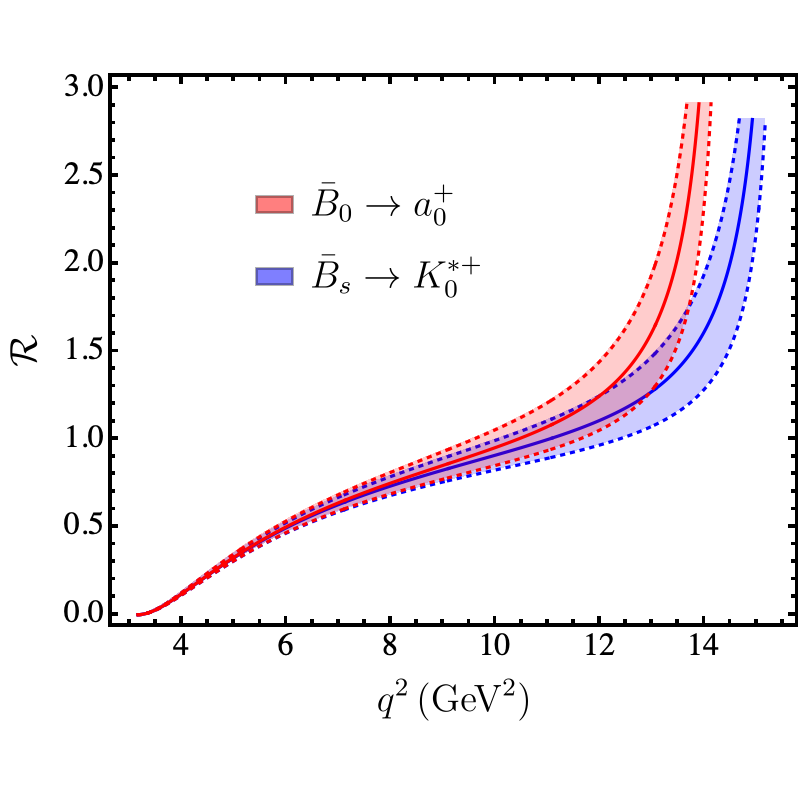}
\end{subfigure}
\caption{The Lepton flavor universality observable of $B\rightarrow S \ell\bar{\nu}_\ell.$}
\label{fig: LFUlnu}
\end{figure}

Finally, we utilize the form factors obtained through the combined fitting using BCL parameterization to calculate the lepton-flavor universality  observables
\begin{align}
    \mathcal{R}_{S}=\frac{\Gamma\left({B} \rightarrow S \tau \bar{\nu}_\tau\right)}{\Gamma\left(B \rightarrow S \mu \bar{\nu}_\nu\right)}=\frac{\int_{m_\tau^2}^{q_{\max }^2} d q^2 d \Gamma\left(B \rightarrow S \tau \bar{\nu}_\tau\right) / d q^2}{\int_{m_\mu^2}^{q_{\max }^2} d q^2 d \Gamma\left(B \rightarrow S \mu \bar{\nu}_\mu\right) / d q^2}
\end{align}
for the two considered $B\rightarrow Sl\bar{\nu}_l$ processes. The numerical results are
\begin{align}
    \mathcal{R}_{a_0(1450)}&= 0.309\pm 0.032, \notag\\
    \q\mathcal{R}_{K_0^*(1430)}&=0.337\pm 0.032.
\end{align}
The ratio of the differential decay widths associated with lepton-flavor universality is depicted in Figure \ref{fig: LFUlnu}.

\begin{table}[H]
  \caption{Theoretical predictions for branch ratios of $B\to S$ and three categories of integrated observables derived from the combined BCL $z$-series expansion fitting of $B\rightarrow S\ell\bar{\nu}_\ell$ form factors obtained via LCSR. }
     \footnotesize
    \centering
    \begin{tabular}{|c|c|c|c|c|c|}
    \hline\hline
     Processes & Methods & $\mathcal{BR}$ & $\mathcal{A}_{\mathrm{FB}}$ & $\mathcal{F}_\mathrm{H}$ & $\mathcal{A}_{\lambda_l}$\\
     \hline\hline
     $\bar{B}_0\rightarrow a_0^+(1450)\mu\bar{\nu}_\mu$ & This work & $1.00(43)\times 10^{-4}$ & $6.25(32)\times 10^{-3}$ & $1.61(13)\times 10^{-2}$ & $0.976(3)$ \\
     & LCSR\cite{Wang:2008da} & $1.8\substack{+0.9\\-0.6}\times 10^{-4}$ &   &   &  \\
     & pQCD\cite{Li:2008tk} & $3.25\substack{+2.36\\-1.36}\times 10^{-4}$  &   &   &  \\
     \hline\hline
     $\bar{B}_0\rightarrow a_0^+(1450)\tau\bar{\nu}_\tau$ & This work & $3.0(12)\times 10^{-5}$ & 0.341(3) & 0.757(28) & -0.158(85) \\
     & LCSR\cite{Wang:2008da} & $6.3\substack{+3.4\\-2.5}\times 10^{-5}$ &   &   &  \\
     & pQCD\cite{Li:2008tk} & $1.32\substack{+0.97\\-0.57}\times 10^{-4}$  &   &   & \\
     \hline\hline
     $\bar{B}_0\rightarrow \bar{K}_0^{*}(1430)\nu_\ell\bar{\nu}_\ell$ & This work & $2.50(97)\times 10^{-6}$ &  &   &  \\
     \hline\hline
     $\bar{B}_s\rightarrow K_0^{*+}(1430)\mu\bar{\nu}_\mu$ & This work & $1.60(67)\times 10^{-4}$ & $5.81(31)\times 10^{-3}$ & $1.50(13)\times 10^{-2}$ & $0.978(3)$ \\
     & LCSR\cite{Wang:2008da} & $1.3\substack{+1.2\\-0.4}\times 10^{-4}$ &   &   & \\
     & QCDSR\cite{Yang:2005bv} & $3.6\substack{+3.8\\-2.4}\times 10^{-5}$ &   &   & \\
     & pQCD\cite{Li:2008tk} & $2.45\substack{+1.77\\-1.05}\times 10^{-4}$  &   &   & \\
     \hline\hline
     $\bar{B}_s\rightarrow \bar{K}_0^{*+}(1430)\tau\bar{\nu}_\tau$ & This work & $5.2(20)\times 10^{-5}$ & 0.334(4) & 0.737(30) & -0.120(90) \\
     & LCSR\cite{Wang:2008da} & $5.2\substack{+5.7\\-1.8}\times 10^{-5}$ &   &   & \\
     & pQCD\cite{Li:2008tk} & $1.09\substack{+0.82\\-0.47}\times 10^{-4}$ &   &   & \\
     \hline\hline
     $\bar{B}_s\rightarrow {f}_0 (1500)\nu_\ell\bar{\nu}_\ell$ & This work & $2.67(101)\times 10^{-6}$ &   &   &  \\
     \hline\hline
    \end{tabular} 
     \label{tab:observableslnu}
\end{table}

\subsection{Phenomenological analysis of the \texorpdfstring{$B\to S\nu_\ell\bar{\nu}_\ell$}{B -> S nu bar(nu)} observables}

For the $B\rightarrow S \nu\bar{\nu}$ process, due to the absence of the $Z \nu \bar{\nu}$ vertex in the leading-power effective theory of electroweak interactions, there are no charm-loop effects generated by $O_{1,2}$ operators. We can easily obtain the differential decay width formula for the $B\rightarrow S \nu\bar{\nu}$ process from the effective weak Hamiltonian of the $b\rightarrow s\nu\bar{\nu}$ transition
\begin{align}
    \frac{d \Gamma\left(B \rightarrow S \nu_{\ell} \bar{\nu}_{\ell}\right)}{d q^2}= \frac{G_F^2 \alpha_{\mathrm{em}}^2}{256 \pi^5} \dfrac{\lambda^{3 / 2}}{m_B^3 \sin ^4 \theta_W}\left|V_{t b} V_{t s}^*\right|^2\left[X_t\left(\frac{m_t^2}{m_W^2}, \frac{m_H^2}{m_W^2}, \sin \theta_W, \mu\right)\right]^2
    \left|f_{B S}^{A}\left(q^2\right)\right|^2.
\end{align}
For the short-distance Wilson coefficient $X_t$, we have taken into account the NLO QCD correction and two-loop electroweak correction \cite{10.1143/PTP.65.1772, BUCHALLA1993285, Buchalla:1998ba, Misiak:1999yg,Brod:2010hi}.
\begin{figure}[htbp]
\centering
\begin{subfigure}[b]{0.4\linewidth}
\includegraphics[width=\linewidth]{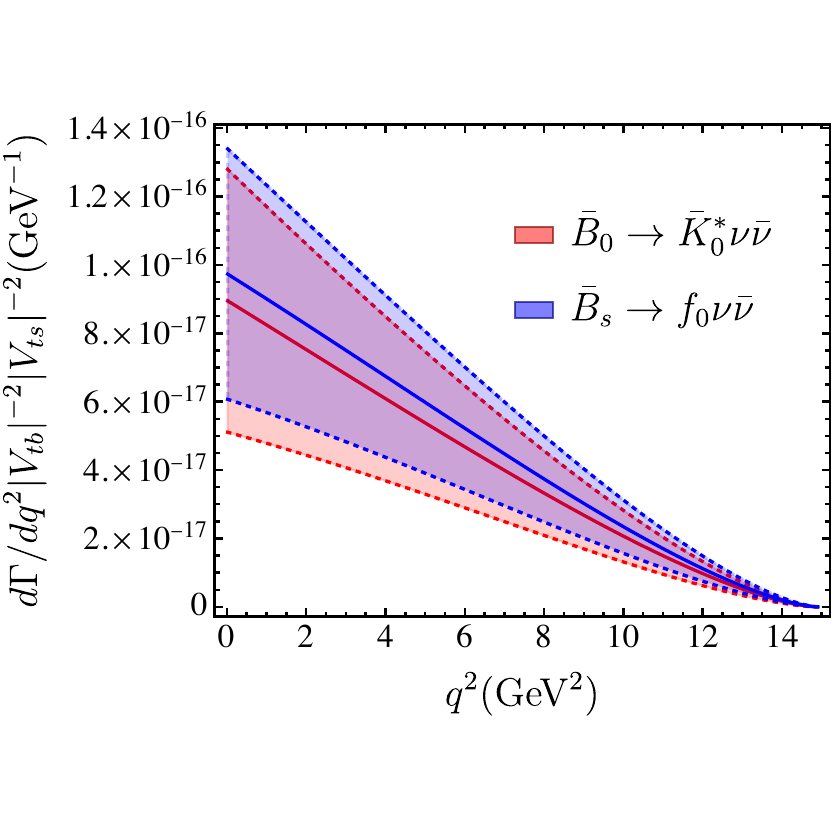}
\end{subfigure}
\caption{The differential decay width of $B\rightarrow S \nu_\ell\bar{\nu}_\ell.$}
\label{fig:dGammaSll}
\end{figure}

We are currently unable to accurately estimate the charm-loop effects in the $B\rightarrow S\ell^+{\ell^-}$ process. Particularly in the region where $q^2$ approaches the $c\bar{c}$ resonances, it is challenging to calculate the effects of multiple soft-gluon emissions with OPE \cite{Khodjamirian:2010vf}. Therefore, based solely on the obtained form factors, we cannot provide precise predictions for the relevant observables of the $B\rightarrow S\ell^+{\ell^-}$ process. In this paper, we refrain from further discussing this aspect. 

\section{Conclusion}

In this work, we have calculated next-to-leading order QCD   corrections to the $B \rightarrow S$ form factors for the first time. We first established the method of calculating the form factors using LCSR at the tree level. Subsequently, we have employed the method of regions to discuss the factorization of NLO correlation functions, where the short-distance functions include both hard and jet functions. The crucial aspect of proving factorization is the strict cancellation between the contributions from the soft regions and the infrared subtraction terms. The results of the correlation functions demonstrate that the $B \rightarrow S$ form factors are solely related to the $B$ meson DA $\phi_B^+(\omega,\mu)$. The resulting hard functions are consistent with SCET results. The jet functions corresponding to the axial current and axial tensor current are identical, which is also consistent with the results in the theoretical framework of SCET. We have also verified the scale dependence of the correlation functions and performed resummation of large logarithms using standard renormalization group methods, resulting in the renormalization group improved form factors. 

Extensive numerical analysis has been conducted on the $B\to S$ form factors. We have observed a significant cancellation effect between the contributions of the hard function and the jet function at NLO, leading to an approximately 5\% improvement in the form factor results at NLL accuracy compared to LL accuracy at the large recoil point. We have provided comparisons of results obtained using different methods, which are consistent within the error range. 

By applying the BCL parameterization combined with the constraint of the strong coupling constant, we have performed a fitting of the data points obtained for the form factors, extending the results to the entire kinematic region. The constraint of strong coupling constant effectively reduces the errors in the large $q^2$ region where LQCD data points are lacking. In the fitting process, we have taken into account the correlations between the input LCSR data points and provided the results of three sets of BCL parameters as well as their correlation matrix. It is expected that this approach will reduce the errors of the form factors in the large recoil region by combining them with LQCD data points in the future.

Utilizing the results of $B\to S$ form factors, we have extensively discussed the physical observables of semi-leptonic $B$ meson decays. For the $B \rightarrow S \ell\bar{\nu}_\ell$ processes, the differential decay widths and branching ratios are given. The corresponding integrated observables, such as lepton-flavor universality ratios, forward-backward asymmetries, “flat terms” and lepton polarization asymmetries are also given. For the $B \rightarrow S\nu_\ell\bar{\nu}_\ell$ processes, we only provide the results of the differential decay widths and branching ratios. These results can be verified in future experiments.

\section*{Acknowledgements}

We are grateful to Yu-Ming Wang for illuminating discussions. 
The work is partly supported by the National Natural Science Foundation of China with Grant No.11735010, 12275277, 12070131001 and 12075125, the National Key Research and Development Program of China under Contract No.2020YFA0406400 and the Natural Science Foundation of Tianjin with Grant No.19JCJQJC61100.

%\newpage
\appendix
\section{Dispersion relations for form factors}
\label{appendix:a}
By inserting a complete set of states in the hadronic matrix elements, we obtain
\begin{align}
        \bra{S(p)}\bar{q}\gamma_\mu\gamma_5 b\ket{\bar{B}(p_B)}=\,&\bra{B(-p_B)S(p)}\bar{q}\gamma_\mu\gamma_5 b\ket{0}\notag\\
        =\,&\dfrac{\langle B(-p_B)S(p)| B_q(-q)\rangle\langle{B_q(-q)}|\bar{q}\gamma_\mu\gamma_5 b\ket{0}}{m_B^2-q^2}\notag\\
        &+\dfrac{\bra{B(-p_B)S(p)}\overline{{B_{q1}(-q)\rangle\langle{B_{q1}(-q)}}}|\bar{q}\gamma_\mu\gamma_5b\ket{0}}{m_{B_{q1}}^2-q^2}+\mathrm{continuum},\\
        \bra{S}\bar{q}i\sigma_{\mu\nu}\gamma_5q^\nu\ket{\bar{B}}=\,&\bra{B(-p_B)S(p)}\bar{q}i\sigma_{\mu\nu}\gamma_5q^\nu\ket{0}\notag\\
        =\,&\dfrac{\bra{B(-p_B)S(p)}\overline{ B_{q1}(-q)\rangle\langle{B_{q1}(-q)}}|\bar{q}i\sigma_{\mu\nu}\gamma_5q^\nu\ket{0}}{m_{B_{q1}}^2-q^2}+\mathrm{continuum}\notag\\
        =\,&\dfrac{g_{B_{q1}BS}f_{B_{q1}}^{T}q^2 p^{\rho}i\sum_{\epsilon}\epsilon_{\rho}\epsilon_{\mu}^*}{m_{B_{q1}}^2-q^2}+\mathrm{continuum}\notag\\
        =\,&\dfrac{g_{B_{q1}BS}f_{B_{q1}}^T}{2(m_{B_{q1}}^2-q^2)}\left\{-i\left[2q^2 p_\mu+(m_B^2-m_S^2-q^2)q_\mu\right]\right\}+\mathrm{continuum},
\end{align}
where the overline notation denotes the summation of all polarization states. By comparing with the definition of the form factors we obtain the dispersion relations
\begin{align}
    f_{BS}^A=\,&\dfrac{g_{B_{q1} BS}f_{B_{q1}}}{2(m_{B_{q1}}^2-q^2)}+\mathrm{continuum},\notag\\
    f_{BS}^P=\,&\dfrac{q^2 g_{B_{q} BS}f_{B_q}}{2(m_{B_{q}}^2-q^2)}\dfrac{m_B^2-m_S^2-m_{B_q}^2}{m_B^2-m_S^2}+\mathrm{continuum},\notag\\
    f_{BS}^T=\,&\dfrac{(m_B+m_S)g_{B_{q1}BS}f_{B_{q1}}^T}{2(m_{B_{q1}}^2-q^2)}+\mathrm{continuum}.
    \label{eq: dispersion relation of FFs}
\end{align}
We have adopted the following parametrizations
\begin{align}
        \bra{B_q(p)}\bar{q}\gamma_\mu\gamma_5 b\ket{0}=\,&-i f_{B_q} p_\mu,\notag\\
        \bra{B_{q1}(p,\epsilon)}\bar{q}\gamma_\mu\gamma_5 b\ket{0}=\,&-i f_{B_{q1}}m_{B_{q1}}\epsilon^*_\mu\ \mathrm{or}\ -i f_{B_{q1}}\sqrt{p^2} \epsilon_\mu^*,\notag\\
        \bra{B_{q1}(p,\epsilon)}\bar{q}i\sigma_{\mu\nu}\gamma_5q^\nu b\ket{0}=\,&-i f_{B_{q1}}^T m_{B_{q1}}^2\epsilon_\mu^*\ \mathrm{or}\ -i f_{B_{q1}}^T p^2 \epsilon_\mu^*,\notag\\
        \braket{B(p)S(q)|B_{q}(p+q)}=\,&g_{B_qBS}\, 
        q\cdot (p+q),\notag\\
        \braket{B(p)S(q)|B_{q1}(p+q,\epsilon)}=\,&-g_{B_{q1}BS}\, q^\rho \epsilon_\rho,\notag\\
        \sum_{\epsilon}\epsilon_\mu(p) \epsilon_\nu^*(p)    =\,&-g_{\mu\nu}+\dfrac{p_\mu p_\nu}{p^2}.
\end{align}
The selection of the two kinds of parameterizations does not affect the residue's outcome but rather influences the manifestation of off-shell effects. If the second parameterization scheme is adopted, there will be an extra $q^2$ dependence on the numerator of Eq.(\ref{eq: dispersion relation of FFs}). Furthermore, it is important to note that the strong coupling constant $g_{B_qBS}$ in $f_{BS}^P$ carries the dimension of $\mathrm{GeV^{-1}}$. Do not be confused by how the strong coupling constants are parameterized. The chosen parameterization is solely for dimensional considerations and convenience in reference usage.

\section{Strong coupling constant}
\label{appendix:b}

%\subsection{LCSR}

We estimate the strong coupling constant $ g_{B_{q1} B S} $ with the tree-level LCSR by calculating the vacuum-to-scalar correlation function near the light cone in terms of OPE. 
For example, 
the strong coupling constant $g_{B_{s1} B K_0^*}$ is defined through     parametrizing the hadronic matrix element
\begin{equation}
	\langle{B_{s1}(q) K_0^*(p) | B(p+q) }\rangle = -g_{B_{s1} B K_0^*} p^\mu \epsilon_\mu^{B_{s1}},
\end{equation}
where 
$ \epsilon_\mu^{B_{s_1}} $ is the polarization vector of $ B_{s_1} $.

We start from the vacuum-to-scalar correlation function
\begin{equation}
	\begin{aligned}
		F_\mu (p,q) = \ii \int \md^4 x \me^{\ii q\cdot x} & \bra{S(p)} T\{ j_{2\mu}(x), j_1(0) \} \ket{0} = F(q^2,(p+q)^2) p_\mu +  \widetilde{F}(q^2, (p+q)^2) q_\mu , \\
		j_{2\mu} =& \bar{q}_2(x) \ii \gamma_\mu \gamma_5 b(x) , \ \ j_1(0) = (m_b + m_{q_1})\bar{b}(0) \ii \gamma_5 q_1(0).
	\end{aligned}
\end{equation}

The $ j_{2\mu} $ and $ j_1 $ are the interpolating currents for $ B_{s1} $ and $ K^*_0 $. In the above formula, the invariant amplitude $ F $ proportional to the momentum $ p_\mu $ is isolated, and the remaining kinematical structures proportional to $ q_\mu $ will not be used.

Using the decay constants defined as
\begin{equation}
	\bra{0} j_{2\mu}(0)\ket{B_{s1}(q)} = f_{B_{s1}} m_{B_{s1}} \epsilon_\mu^{(B_{s1})} , \quad \bra{0}j_1(0) \ket{B(p+q)} = m_B^2 f_B,
\end{equation}
we insert the complete set of intermediate states with $B$ and $B_{s1}$ quantum numbers and employ the double dispersion relation, then the amplitude $ F(q^2, (p+q)^2 ) $ can be written as
\begin{equation}
	F(q^2, (p+q)^2) = \dfrac{m_B^2 m_{B_{s1}} f_B f_{B_{s1}} g_{B_{s1} B K_0^* }}{(m_B^2 - (p+q)^2) (m^2_{B_{s1}} -q^2 )} + \iint_\Sigma \md s_2 \md s_1 \dfrac{\rho^h(s_1,s_2)}{(s_2^2 - (p+q)^2) (s^2_{1} -q^2 )} + \cdots.
\end{equation}
The $ \Sigma $ is used to represent the two-dimensional region $ s_1 \geq (m_B + m_{K^*_0})^2, \, s_2 \geq (m_{B_{s1}} + m_{K^*_0})^2 $. And $ \rho^h(s_1, s_2) $ denotes the hadronic spectral density of the continuum and excited states. 

At $q^2,(p+q)^2 \ll m_Q^2 $, the dispersion relation is matched to the result of the QCD calculation of $F(q^2,(p+q)^2)$. We employ the light-cone OPE in terms of scalar meson DAs. The OPE result for the correlation function in the form of a double dispersion integral
\begin{equation}
	F^{(\OPE)}(q^2, (p+q)^2) = \int_{-\infty}^{+\infty} \dfrac{\md s_2}{(s_2 - (p+q)^2)} \int_{-\infty}^{+\infty} \dfrac{\md s_1}{(s_1 - q^2)} \rho^{(\OPE)}(s_1,s_2),
\end{equation}
and the double spectral density
\begin{equation}
	\rho^{(\OPE)}(s_1,s_2) \equiv \dfrac{1}{\pi^2} \im_{s_1} \im_{s_2} F^{(\OPE)}(s_1,s_2).
\end{equation}
Then we adopt the quark-hadron duality and assume that the integral of the hadronic spectral density $\rho^h(s_1,s_2)$ taken over the two dimensional region $\Sigma$ is equal to the integral of the OPE spectral density taken over a certain region $\Sigma_0$ in the $(s_1, s_2)$ plane
\begin{equation}
	\iint_\Sigma \md s_2 \md s_1 \dfrac{\rho^h(s_1,s_2)}{(s_2^2 - (p+q)^2) (s^2_{1} -q^2 )} = \iint_{\Sigma_0} \md s_2 \md s_1 \dfrac{\rho^{(\OPE)}(s_1,s_2)}{(s_2^2 - (p+q)^2) (s^2_{1} -q^2 )}.
\end{equation}

Next, we equate the double dispersion representations of hadron and OPE result and subtract the equal integrals over the region $ \Sigma_0 $ from both sides of the equation. The remaining region is notated as
\begin{equation}
	\iint^{\Sigma_0} \md s_2 \md s_1 = \int_{-\infty}^{+\infty} \md s_2 \int_{-\infty}^{+\infty} \md s_1 - \iint_{\Sigma_0} \md s_2 \md s_1.
\end{equation}
After performing the double Borel transformation, the subtraction terms are removed and contributions of the higher-state and continuum spectrum are suppressed. The resulting LCSR give the desired strong coupling constant
\begin{equation}\label{couplingLCSR}
	g_{B_{s1} B K_0^*} = \dfrac{1}{m_B^2 m_{B_{s1}} f_B f_{B_{s1}} } \exp\left( \dfrac{m_B^2}{M_2^2} + \dfrac{m_{B_{s1}}^2}{M_1^2} \right) \iint^{\Sigma_0} \md s_2 \md s_1 \exp \left( -\dfrac{s_2}{M_2^2} - \dfrac{s_1}{M_1^2} \right) \rho^{(\OPE)} (s_1, s_2).
\end{equation} 

The OPE near the light-cone for the correlation function is valid if $q^2$ and $(p+q)^2$ are far below the heavy quark threshold $m_Q^2$. The vacuum-to-scalar meson correlation function becomes
\begin{equation}
	F_\mu(p,q) = m_Q \int \md^4 x \me^{\ii q\cdot x} \bra{S(p)} \bar{q}_2(x) \gamma_\mu \gamma_5 S_Q(x,0) \gamma_5 q_1(0) \ket{0},
\end{equation}
where $  S_Q(x,0) = -\ii \bra{0}T\{Q(x),\bar{Q}(0)\} \ket{0} $ is the heavy quark propagator expanded near the light cone. The vacuum-to-scalar meson matrix elements can be expanded in terms of the scalar meson quark-antiquark (quark-antiquark-gluon) DAs. Following the reference \cite{Wang:2008da}, the leading twist-2 DA and its contribution to the amplitude is
\begin{equation}
	\begin{aligned}
		\bra{S(p)} \bar{q}_2(x) \gamma_\mu q_1(y) \ket{0} = & p_\mu \int_{0}^{1} \md u \me^{\ii (u p\cdot x + \bar{u} p \cdot y)} \Phi_S(u,\mu), \\
		F^{(\mathrm{tw2,LO})}(q^2,(p+q)^2) =& m_Q^2 \int_{0}^{1} \dfrac{\md u}{m_Q^2 - (q+up)^2} \Phi_S(u) .
	\end{aligned}
\end{equation}

The leading order invariant amplitude with twist-2 is sufficient to meet the needs of this work, where the denominator is transformed by using the relation
\begin{equation}
	m_Q^2 - (q+up)^2 = m_Q^2 - \bar{u} q^2 - u(p+q)^2.
\end{equation}
Furthermore, we consider a Taylor expansion of the DA $\Phi_S(u)$ 
\begin{equation}
	\Phi_S(u) = \sum_{k=0}^{\infty} c_k u^k.
\end{equation}
Following the techniques from \cite{Belyaev:1994zk, Khodjamirian:2020mlb}, it is convenient to do the integrals. We take the imaginary part of the invariant amplitude and then derive the double spectral density 
\begin{equation}\label{spectral_density}
	\begin{aligned}
		\rho^{(\mathrm{tw2,LO})}(s_1,s_2) =& \dfrac{1}{\pi^2} \im_{s_1} \im_{s_2} F^{(\mathrm{tw2,LO})}(s_1,s_2) \\
		=& \dfrac{1}{\pi^2} \im_{s_1} \im_{s_2} \sum_{k=0}^{\infty} c_k \int_{0}^{1} \md u \dfrac{m_Q^2 u^k}{m_Q^2 - \bar{u}s_1 - u s_2}  \\
		=& \sum_{k=0}^{\infty} c_k \dfrac{1}{k!} \delta^{(k)}(\dfrac{s_1}{m_Q^2} - \dfrac{s_2}{m_Q^2} ) (\dfrac{s_2}{m_Q^2} - 1)^k \theta(\dfrac{s_2}{m_Q^2} - 1).
	\end{aligned}
\end{equation}

Given the spectral density, we are ready to integrate over the duality region of the LCSR. The lower boundary of the duality region is determined by the heavy quark threshold $ s_1 \geq m_Q^2, s_2 \geq m_Q^2 $. As for the upper boundary, following the discussion in reference  \cite{Belyaev:1994zk, Khodjamirian:2020mlb}, we adopt the triangle duality regions and equal Borel parameters
\begin{equation}
	s_1+s_2 \ll 2s_0, \quad M_1^2=M_2^2=2M^2
\end{equation}
and rewrite the double integral in the LCSR Eq.(\ref{couplingLCSR}) as 
\begin{equation}
	\mathcal{F}^{(\mathrm{tw2, LO})}(M^2,s_0) \equiv \int_{-\infty}^{\infty} \md s_1 \int_{-\infty}^{\infty} \md s_2 \theta(2s_0 - s_1 -s_2) \exp (-\dfrac{s_1 + s_2}{2M^2}) \rho^{(\mathrm{tw2, LO})} (s_1,s_2).
\end{equation}
Next, we substitute the double spectral density Eq.(\ref{spectral_density}) into the integral and obtain the integral result by using techniques from \cite{Belyaev:1994zk, Khodjamirian:2020mlb}
\begin{align}
        \mathcal{F}^{(\mathrm{tw2,LO})}(M^2,s_0) =& \sum_{k=0}^{\infty} c_k \dfrac{1}{k!} \int_{-\infty}^{2 s_0} \md s_1 \int_{-\infty}^{2s_0 - s_1} \md s_2 \exp (-\dfrac{s_1 + s_2}{2M^2}) \notag\\
        &\delta^{(k)}(\dfrac{s_1}{m_Q^2} - \dfrac{s_2}{m_Q^2} ) (\dfrac{s_2}{m_Q^2} - 1)^k \theta(\dfrac{s_2}{m_Q^2} - 1) \notag\\
        =& m_Q^2 \left. M^2 \left[ \exp \left(-\dfrac{m_Q^2}{M^2} -\exp \left(-\dfrac{s_0}{M^2} \right) \right) \Phi_S(u)  \right] \right|_{u=1/2}.
\end{align}
Finally, we express the strong coupling of leading order with the contribution of twist-2 by LCSR
\begin{equation}
	 g_{B_{s1} B K_0^*}^{(\mathrm{tw2,LO})} = \dfrac{1}{m_B^2 m_{B_{s1}} f_B f_{B_{s1}} } \exp\left( \dfrac{m_B^2+m_{B_{s1}}^2}{2M^2} \right) m_Q^2 \left. M^2 \left[ \exp \left(-\dfrac{m_Q^2}{M^2} - \exp \left(-\dfrac{s_0}{M^2} \right) \right) \Phi_S(u)  \right] \right|_{u=1/2}.
\end{equation}

In our work, the decay constants and masses of the axial mesons are given by reference \cite{Pullin:2021ebn} using the QCD sum rules, and the LCDA of scalar mesons come from \cite{Wang:2008da, Cheng:2005nb}.

%\newpage

\bibliography{mybibfile}
\bibliographystyle{JHEP}
\end{document}